\documentclass[a4paper,twocolumn,11pt,pra]{quantumarticle}
\pdfoutput=1

\usepackage{algorithm}
\usepackage{algpseudocode}
\usepackage[dvipsnames]{xcolor}
\usepackage{dsfont}
\usepackage{mathtools}
\usepackage{bbding}
\usepackage{textcomp}
\usepackage[normalem]{ulem}
\usepackage{nicefrac}
\usepackage[utf8]{inputenc} 	%Input what you want e.g., é, ?, a, ü
\usepackage[T1]{fontenc} 	%Output what you want e.g., é, ?, a, ü
\usepackage{graphicx}% Include figure files
\usepackage{dcolumn}% Align table columns on decimal point
\usepackage{bm}% bold math
\usepackage[english]{babel}
\usepackage{hyperref}
\usepackage{amsfonts,amssymb}
\usepackage{amsmath}
\usepackage{stmaryrd} %Extra computer-science symbols and short arrows
\usepackage{epsfig}
\usepackage{mathrsfs}
\usepackage{color,soul}
\usepackage{bbold}
\usepackage{pstricks}
\usepackage{multirow}
\usepackage{textcomp}
\usepackage{braket}
\usepackage{empheq}
\hypersetup{colorlinks,%
citecolor=purple,%
filecolor=cyan,%
linkcolor=magenta,%
urlcolor=blue}
\usepackage{rotating}
\usepackage{amsthm}
\usepackage{eurosym}
\usepackage{mwe}
\usepackage[framemethod=default]{mdframed}
\usepackage{newfloat}

\DeclareFloatingEnvironment[name={S}]{suppfigure}
\DeclareFloatingEnvironment[name={S}]{supptable}
\allowdisplaybreaks
\usepackage{enumitem}
\usepackage[numbers,sort&compress]{natbib}
\usepackage{comment}
\usepackage{cancel}
\usepackage{physics} %Many commands and macros for physics, including Dirac notation

\DeclareMathOperator{\id}{\mathds{1}}

\DeclareMathOperator{\ee}{\mathrm{e}}             %%% base of natural exponents
\DeclareMathOperator{\ii}{\mathrm{i}}             %%% the imaginary unit i^2=-1

\DeclareMathOperator{\pr}{\mathrm{Pr}}

\newcommand{\lowtilde}[1]{\underset{\widetilde{\hphantom{#1}}}{#1}}

\newcommand{\oW}{\lowtilde{\mathcal{W}}}

\newtheorem*{lemma*}{Lemma}

\newcommand{\ti}{t_{\mathrm{i}}}
\newcommand{\tf}{t_{\mathrm{f}}}
\newcommand{\D}{\mathscr{D}}
\NewDocumentCommand{\tfti}{ O{\tf} O{\ti}}{{#1\!\!\shortleftarrow\!#2}}

\newcommand{\tS}{\mathrm{S}}
\newcommand{\tE}{\mathrm{E}}
\newcommand{\tM}{\mathrm{M}}
\newcommand{\tSE}{\mathrm{SE}}
\newcommand{\tSM}{\mathrm{SM}}

\begin{document}
%\preprint{APS/123-QED}
\title{Continuous operations on non-Markovian processes}

\author{Fabio Costa}
\email{fabio.costa@su.se}
\affiliation{Nordita, KTH Royal Institute of Technology and Stockholm University, Hannes Alfv\'ens v\"ag 12, 106 91 Stockholm, Sweden.}
\author{Jing Yang}
\email{jing.yang.quantum@zju.edu.cn}
\affiliation{Institute of Fundamental and Transdisciplinary Research, Institute of Quantum Sensing, and Institute for Advanced Study in Physics, Zhejiang University, Hangzhou 310027, China}
\affiliation{Nordita, KTH Royal Institute of Technology and Stockholm University, Hannes Alfv\'ens v\"ag 12, 106 91 Stockholm, Sweden.}
\begin{abstract}
Continuous measurements are central to quantum control and sensing, yet lack a model-independent operational description that can be applied to arbitrary non-Markovian processes without specifying a microscopic measurement model. Existing multi-time frameworks, such as process matrices, allow for an arbitrary sequence of operations to be applied on a general process, but are restricted to interventions at discrete times and cannot represent measurements of finite duration. We introduce a continuous-time extension of multi-time quantum processes based on process and operation functionals, which generalize the Feynman–Vernon influence functional and yield a continuous Born rule that cleanly separates processes from operations. This framework provides a consistent representation of non-Markovian dynamics under continuous monitoring and leads to natural definitions of causality and Markovianity in continuous time, opening the way to formalising quantum causal structures in the continuum. We illustrate the formalism by analyzing continuous measurements in a generalized Caldeira–Leggett model, demonstrating its applicability to realistic non-Markovian scenarios.
\end{abstract}

\maketitle

\section{Introduction}

The analysis of open dissipative systems typically assumes quickly-vanishing system-environment correlations, leading to a memoryless reduced dynamics captured by the Gorini–Kossakowski–Sudarshan–Lindblad master equation~\cite{lindblad1976onthe, gorini1976completely, breuer2002theory}. Yet, memory effects arise ubiquitously in the presence of structured or strongly coupled environments and non-Markovian behavior is becoming increasingly relevant for emerging quantum technologies \cite{Madsen2011, Young2020, Morris2022, Svendsen2023, Vicencio2025}. Conventional descriptions through dynamical maps \cite{rivas2014quantum, breuer2016colloquium, Vacchini2021} do not capture multi-time correlations and hence only give a partial characterization of non-Markovian processes.

\begin{figure}[t]
\begin{centering}
\includegraphics[scale=0.28]{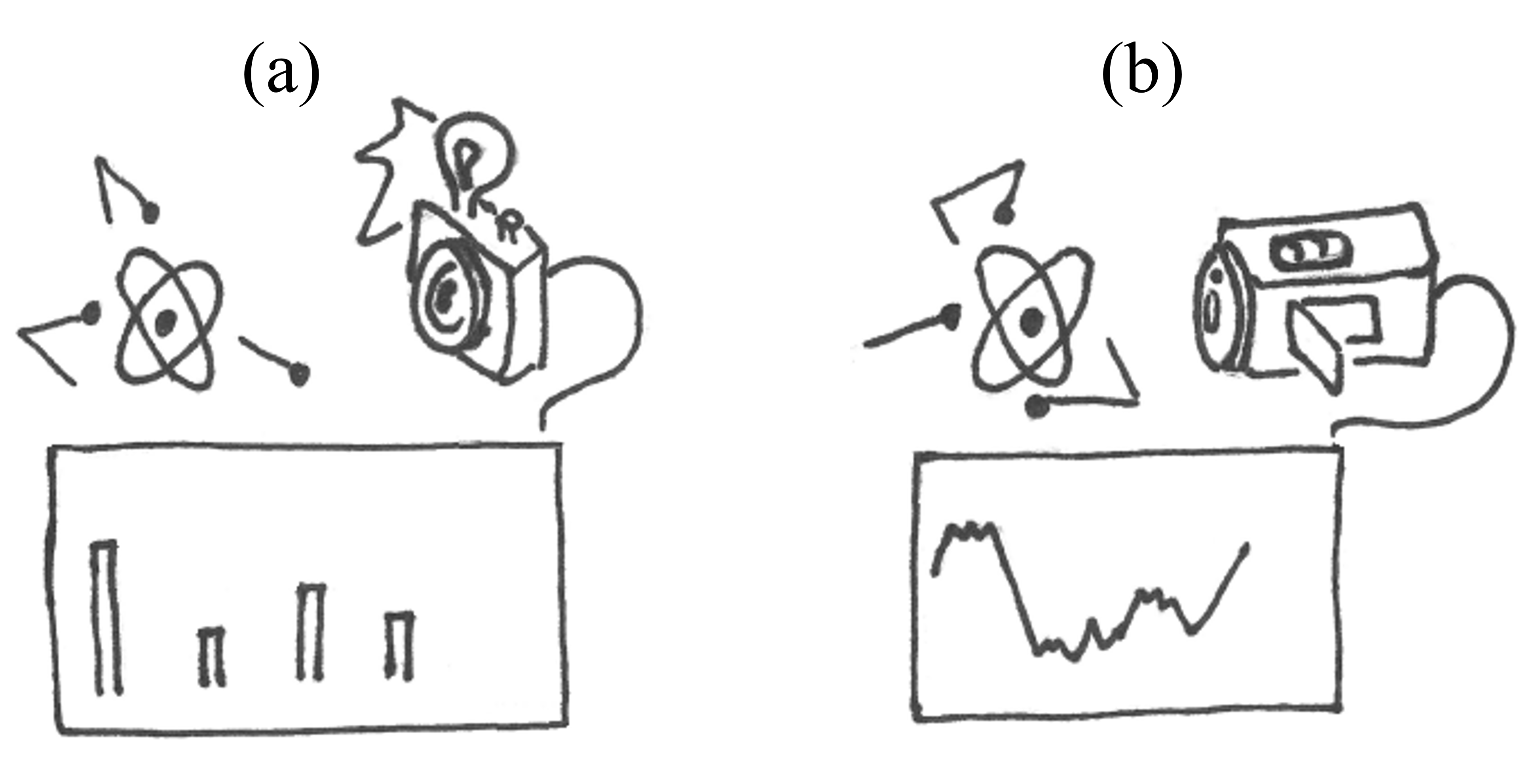}
\par\end{centering}
\caption{\protect\label{fig:illustration} (a) In its current formulation, the process matrix formalism allows modeling repeated instantaneous measurements/operations on a system (in the picture, the atom) interacting with an unobserved environment with memory (the particles). (b) This work extends the framework to continuous monitoring/finite-time operations.}
\end{figure}

A recent operational framework uses process matrices, also known as combs or process tensors, to encode all operationally accessible multi-time correlations \cite{chiribella08,chiribella09b,Pollock2018, Pollock2018a, Milz2017,Strathearn2018, Joergensen2019,link2024openquantum}. This framework represents measurement sequences and processes independently, much like observables and states, and generates  probabilities for temporal sequences of outcomes through a generalized Born rule, whose application does not require an underlying microscopic model. The approach provides a clean operational definition of Markovianity with the correct classical limit \cite{Pollock2018, Milz2020kolmogorovextension}, as well as powerful methods for reconstructing processes \cite{White2022,Giarmatzi2025multitimequantum} and characterizing memory structures \cite{Taranto2019, giarmatzi2018witnessing, Taranto2024characterising}. The formalism further connects, and draws insights from, foundational research on quantum causal models \cite{costa2016, Giarmatzi2018, Barrett2019}, indefinite causal structures \cite{oreshkov12, araujo15, chiribella09}, and quantum gravity \cite{hardy2007towards, Zych2019, Hardy2016}.

However, the framework assumes \textit{instantaneous} operations taking place at discrete instants of time, Fig.~\ref{fig:illustration}(a). This is a strong limitation for real-world applications where the measurement and the process have comparable timescales, such as superconducting circuits~\cite{blais2004cavityquantum,blais2021circuit}, or for modeling continuous measurements, Fig.~\ref{fig:illustration}(b). 
On the other hand, while continuous monitoring is common in quantum optical platforms~\cite{Wiseman2009,jacobs2014quantum,jordan2024quantum,gardiner2004quantum}, its formalization presumes Markovian noise ~\cite{Albarelli2024,mohseninia2020alwayson,mabuchi2009continuous}, leaving open the problem of continuous monitoring of non-Markovian systems.

Here, we develop a formalism for continuous-time non-Markovian quantum processes that can be monitored through continuous measurements.  Similar to the discrete case, processes and operations are modeled independently, so calibrated operations can characterize the background process without prior microscopic modeling. We derive a continuous version of the \emph{causality condition}, defining a process with fixed causal order, and of \emph{Markovianity}, which characterizes time-local dynamics without external memory. We further show that the framework reduces to the discrete-time one in the appropriate limit, and that continuous processes that satisfy, respectively, causality and Markovianity,  reduce to causal and Markovian processes in that limit, ensuring consistency of the formalism. 
As this framework represents a first continuous version of the process matrix formalism, it open the way to investigation of quantum causal structures in continuous time or spacetime.

\section{From discrete-time to continuous processes and operations}

\begin{figure}[t]
\begin{centering}
\includegraphics[scale=0.31]{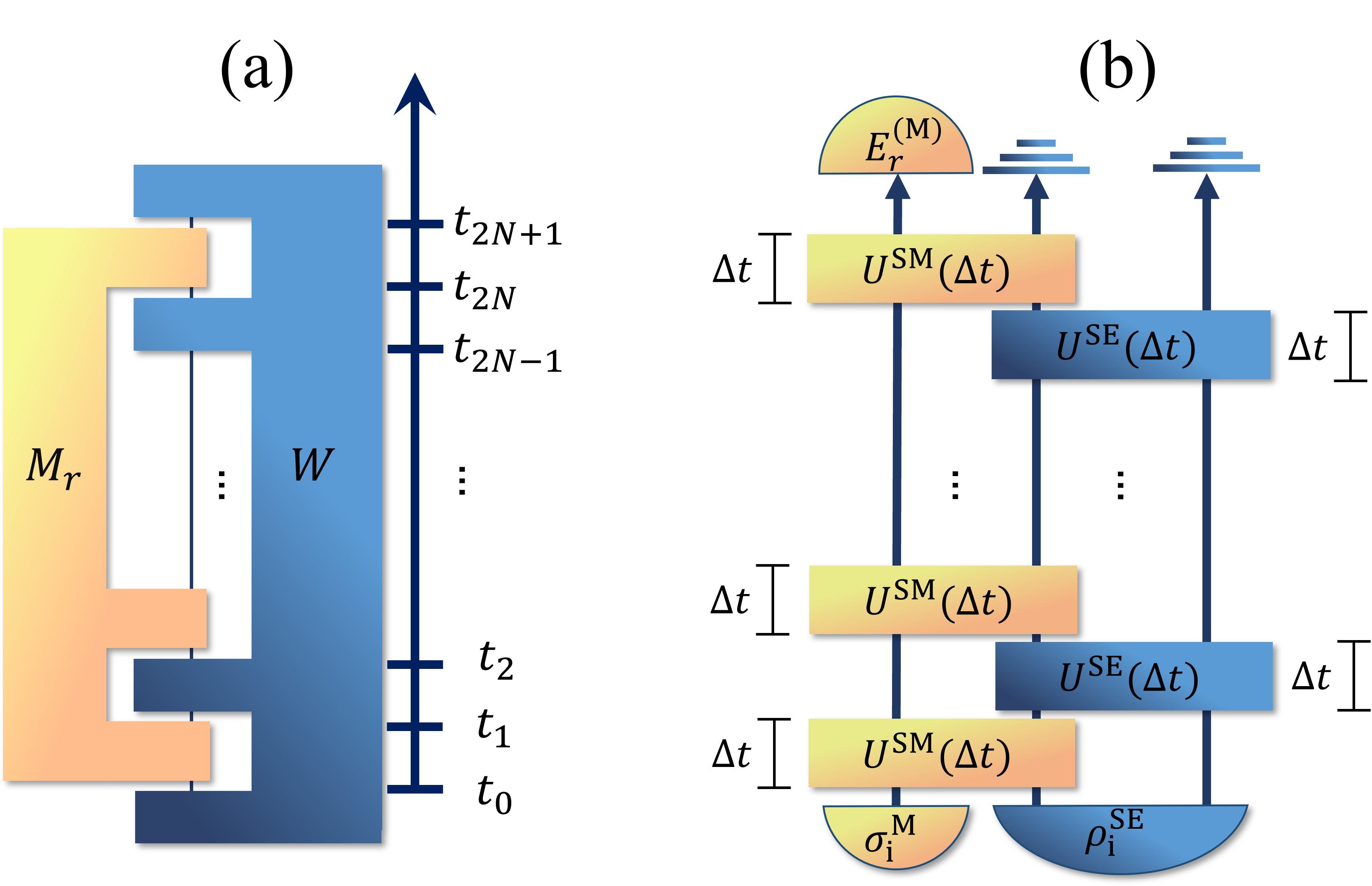}
\par\end{centering}
\caption{\protect\label{fig:Discrete-Continuous-Process} (a) Schematic illustration of discrete processes and operations. Operations act from even to odd time while the process takes over from odd to even times. (b) Trotterization of the joint evolution of the meter, system and environment. Upon taking the continuous-time limit, we obtain the representation of the continuous-time quantum processes and testers respectively.}
\end{figure}

The multi-time process formalism concerns a quantum system $S$, with Hilbert space $\mathcal{H}^S$, at a discrete (usually, finite) set of times. The system is allowed to be initially correlated with an environment and interact with it at every time step, while information about the process is obtained by repeated measurements or operations at the prescribed times, which are assumed to be instantaneous. Here, with a view to deriving a continuous-time limit, we modify the scenario by letting operations last for a finite time. Hence, we consider a scenario where process and operations alternate, with operations acting between times $t_{2k}$ and $t_{2k+1}$, $k=0,\dots, N$, while the process takes over from odd to even times, see Fig.~\ref{fig:Discrete-Continuous-Process}(a).
To model the most general multi-time correlations, the formalism introduces a copy of the system Hilbert space for each time $t_j$: $\mathcal{H}^j \cong \mathcal{H}^S$, with total Hilbert space $\mathcal{H}^{(N)} \coloneqq \otimes_{j=0}^{2N+1}\mathcal{H}^j$. A quantum process is represented by a positive semidefinite operator  $W{\in}\mathcal{L}(\mathcal{H}^{(N)})$, called a \textit{process matrix} (or \textit{comb}, or \textit{process tensor}), where $\mathcal{L}(\mathcal{H})$ denotes the space of linear operators over a Hilbert space $\mathcal{H}$. Causally-ordered process matrices are  required to satisfy a set of constraints, enforcing no causal influence from the future to the past. 

Most commonly, probing the system is modeled through independent \textit{quantum instruments} at different times, defined as collections of Completely Positive (CP) maps $\mathscr{M}^k_{r_k}{:} \mathcal{L}(\mathcal{H}^{2k}) {\rightarrow} \mathcal{L}(\mathcal{H}^{2k+1})$, where $r_k$ labels the measurement outcome (or ``readout'') and such that $\sum_{r_k}\mathscr{M}^k_{r_k}$ is completely positive and trace preserving (CPTP). The probability to observe a sequence of outcomes $r=(r_0,\dots, r_N)$ is given by the \textit{generalized Born rule} \cite{chiribella08, Shrapnel2017}
\begin{equation} \label{eq:discreteBorn}
    \pr(r) = \Tr \hat M^T_r \hat W,
\end{equation}
 where $\hat M_r {\coloneqq} \bigotimes_{k=0}^{N} \hat M^k_{r_k}$, $\hat M^k_{r_h} {\in} \mathcal{L}(\mathcal{H}^{2k} \otimes \mathcal{H}^{2k+1})$ is the Choi-Jamio{\l}kowski \cite{Choi1975, jamio72} representation of $\mathscr{M}^k_{r_k}$, and $^T$ denotes transposition in the chosen basis. Here, we need to consider more general, non-Markovian, probing schemes, which are described by \textit{quantum testers} \cite{chiribella09} (or \textit{measuring strategies} \cite{gutoski06}), defined as collections of positive semidefinite operators  $\hat{M}_r{\in} \mathcal{L}(\mathcal{H}^{(N)})$ such that $\sum_r \hat{M}_r$ is a valid (causally ordered) process matrix. 
The probability rule for this more general scenario is still given by Eq.~\eqref{eq:discreteBorn}, which can be seen as an extension of the ordinary Born rule, where $\hat{W}$ plays the role of a state and $\hat{M}_r$ that of a measurement operator. The key aspect of this expression is that processes and operations are encoded in two independent objects, and the formula is linear in both. Our goal is to obtain an expression for the continuum limit of this formula that retains these properties. 

It is convenient to connect the continuum limit with the standard Hamiltonian description of time evolution. This is facilitated by \textit{dilation theorems} \cite{gutoski06, chiribella09b}, which ensure that both process and tester can be modeled through unitary interactions with auxiliary systems, which are eventually traced out (for $\hat{W}$) or measured (for $\hat{M}_r$). More precisely, as shown in Fig.~\ref{fig:Discrete-Continuous-Process}(b), any process matrix $\hat{W}$ can be modeled by introducing an environment system $\tE$, taking an initial joint state $\hat{\rho}^{\tSE}$, letting system and environment interact through a unitary during the intervals $t_{2k+1}$ and $t_{2k+2}$, $k=0,\dots,N{-}1$, and tracing out the environment at the final time (see, e.g., Ref.~\cite{Goswami2024} for an explicit construction of the process matrix, using the ``link product'' \cite{chiribella09b}). Similarly, $\hat{M}_r$ can be dilated by introducing a meter system $\tM$, initially in some state $\sigma^\tM$, letting system and meter undergo joint unitaries between times $t_{2k}$ and $t_{2k+1}$, $k=0,\dots,N$, and measuring the  meter at the final time with a POVM (Positive Operator-Valued Measure) $\{\hat{E}^\tM_r\}_r$. Note that measurements at different times can also be modeled in this way, with different meter subsystems interacting with $\tS$ at different times and then measured independently, yielding outcomes $r=(r_0,\dots,r_N)$.

Taking the continuum limit (and under suitable regularity conditions), such alternating unitary evolution corresponds to the Trotterized evolution according to a generic time-dependent Hamiltonian~\cite{trotter1959,suzuki1991} 
\begin{equation} \label{eq:totalH}
    \hat H^{\mathrm{TOT}} = \hat{H}^{\mathrm{SE}} + \hat{H}^{\mathrm{SM}},
\end{equation}
where $\hat{H}^{\mathrm{SE}}$ acts as identity on the meter and $\hat{H}^{\mathrm{SM}}$ acts as identity on the environment (and we keep the time dependence implicit). The system's free Hamiltonian can be  included either in $\hat{H}^{\mathrm{SE}}$ or $\hat{H}^{\mathrm{SM}}$, which is not important at this point.  Even though we arrived at Eq.~\eqref{eq:totalH} from the discrete-time formalism,  we can take the existence of a model based on the above Hamiltonians, initial states, and final measurement as our \emph{constructive} definition of continuous-time measurement of a continuous-time process.

The unitary evolution operator from an initial time $\ti$ to a final time $\tf$ is given by the time-ordered exponential
\begin{equation}\label{eq:TorderedU}
   \hat{U}^\mathrm{TOT}(\tfti) = \mathcal{T}\ee^{-\ii \int_{\ti}^{\tf} \dd t \, \hat H^{\mathrm{TOT}}},
\end{equation}
and the probability to observe an outcome $r$ is given by 
\begin{multline}\label{eq:continuousprob_operators}
   \pr[r] \\
   =\Tr \left[\hat{E}^\tM_r\hat{U}^\mathrm{TOT}(\tfti)(\hat{\sigma}^\tM\otimes \hat{\rho}^\tSE )\hat{U}^{\mathrm{TOT}\dagger}(\tfti) \right]. 
\end{multline}
Here we use the square brackets in $\pr[r]$ anticipating the continuous-monitoring scenario, where the outcome is a function of time, $r(t)$, and hence the probability is a functional. However, the formula holds for any value that can be taken by the outcome $r$.

To proceed further, we restrict now to continuous-variable systems, which allows us to use coherent-state path integral methods. 
We treat all systems as one-dimensional continuous variables for notational simplicity, although extensions to multi-dimensional variables are straightforward. We denote as $\ket{\alpha}^\tS$, $\ket{\beta}^\tE$, and $\ket{\gamma}^\tM$ the coherent states in $\tS$, $\tE$, and $\tM$ respectively, with $\alpha, \beta, \gamma {\,\in \,}\mathbb{C}$, and use the short-hand $\zeta{=}(\alpha,\beta,\gamma)$. The evolution operator can be expressed through the coherent-state path integral
\begin{multline} \label{eq:coherentpathintegral}
    \matrixel{\zeta_{\mathrm{f}}}{\hat{U}^{\mathrm{TOT}}(\tfti)}{\zeta_{\mathrm{i}}} \\
    = \int_{\substack{\zeta(\ti) = \zeta_{\mathrm{i}} \\ \zeta(\tf) = \zeta_{\mathrm{f}}}}\!\! \D[\zeta] \ee^{\ii \int_{t_{\text{i}}}^{t_{\text{f}}} \dd t \left[\Im(\dot{\alpha}^*\alpha + \dot{\beta}^*\beta + \dot{\gamma}^*\gamma) - H^{\mathrm{TOT}}(\zeta)\right]},
\end{multline}
where $H^{\mathrm{TOT}}(\zeta)\coloneqq \langle\zeta|\hat H^\mathrm{TOT}|\zeta\rangle$ is the Husimi $Q$-function of the total Hamiltonian of the system, environment and meter~\cite{shankar2017quantum,nagaosa1999quantum} (see also Appendix~\ref{app:Review-Coherent-PI} for a brief review). 
Introducing the shorthand notation $O(\alpha,\,\bar{\alpha})\coloneqq\langle{\alpha|\hat{O}|\bar{\alpha}}\rangle$ and similar notations for the environment and meter's degrees of freedom, and plugging Eq.~\eqref{eq:coherentpathintegral} into Eq.~\eqref{eq:continuousprob_operators}, we arrive at ({see Appendix~\ref{app:Derivation-of-Pr-Phase}}) 
\begin{multline}\label{eq:functionalBorn}
    \pr[r] \\
   = \!\! \int \! \D[\alpha,\bar{\alpha}] \ee^{\ii \int_{t_{\text{i}}}^{t_{\text{f}}} \dd t \Im (\dot{\alpha}^*\alpha - \dot{\bar \alpha}^*\bar \alpha  )} \mathcal{M}_r[\alpha, \bar \alpha] \mathcal{W}[\alpha,\bar{\alpha}].
\end{multline}
Here, the (coherent-state) \textit{process functional} is defined as
\begin{equation}\label{eq:Wfunctional0}
  \mathcal{W}[\alpha,\bar{\alpha}]\coloneqq \widetilde{\mathcal{W}}[\alpha,\bar{\alpha}]\,\mathrm{id}(\bar{\alpha}(\tf),\alpha(\tf)), 
\end{equation}
where $\mathrm{id}(\bar{\alpha},\alpha)\coloneqq \langle\bar{\alpha}|\alpha\rangle$ and 
\begin{multline}\label{eq:Wfunctional}
    \widetilde{\mathcal{W}}[\alpha,\bar{\alpha}] 
\coloneqq \int\!\!\!\D[\beta,\bar{\beta}] \ee^{\ii \int_{t_{\text{i}}}^{t_{\text{f}}} \dd t \Im(\dot{\beta}^*\beta - \dot{\bar{\beta}}^*\bar{\beta})} 
\\
\times \mathcal{U}^{\tSE}[\alpha,\beta]\mathcal{U}^{\tSE*}[\bar{\alpha},\bar{\beta}]
\\
\times\rho^{\tSE}(\alpha(\ti),\beta(\ti), \bar{\alpha}(\ti),\bar{\beta}(\ti)) \\
\times \mathrm{id}(\bar{\beta}(\tf),\beta(\tf)) ,
\end{multline}
with
\begin{equation} \label{eq:SEUfunctional}
    \mathcal{U}^{\tSE}[\alpha,\beta] \coloneqq \ee^{-\ii\int_{t_{\text{i}}}^{t_{\text{f}}} \dd t H^\tSE(\alpha,\beta)},
\end{equation}
while the \textit{operation functional} is
\begin{multline}\label{eq:Mfunctional}
    \mathcal{M}_r[\alpha,\bar{\alpha}]
    \coloneqq \int\D[\gamma,\bar{\gamma}] \ee^{\ii \int_{t_{\text{i}}}^{t_{\text{f}}} \dd t \Im(\dot{\gamma}^*\gamma - \dot{\bar{\gamma}}^*{\bar{\gamma}})} \\
\times \mathcal{U}^{\tSM}[\alpha,\gamma]\mathcal{U}^{\tSM*}[\bar{\alpha},\bar{\gamma}] \\
\times \sigma(\gamma(\ti), \bar{\gamma}(\ti)) E_r(\bar{\gamma}(\tf),\gamma(\tf)),
\end{multline}
with 
\begin{equation} \label{eq:SMUfunctional}
    \mathcal{U}^{\tSM}[\alpha,\gamma] \coloneqq \ee^{-\ii\int_{t_{\text{i}}}^{t_{\text{f}}} \dd t H^\tSM(\alpha,\gamma)}.
\end{equation}
To fix some terminology, we say that a process functional $\mathcal{W}$ with the final $\mathrm{id}$ term, as in Eq.~\eqref{eq:Wfunctional0}, has \textit{closed future}, while a functional missing that term, as $\widetilde{\mathcal{W}}$ in Eq.~\eqref{eq:Wfunctional}, has \emph{open future}. For the special case in which the initial system-environment joint state is a product, $\hat\rho^{\mathrm{SE}}=\hat\rho^{\mathrm{S}}\otimes\hat\rho^{\mathrm{E}}$, the open-future functional can be further decomposed as 
\begin{equation}\label{eq:openpastW}
    \widetilde{\mathcal{W}}[\alpha,\bar{\alpha}] = \lowtilde{\mathcal{W}}[\alpha,\bar{\alpha}]\rho^{\mathrm{S}}(\alpha(t_{\text{i}}),\bar{\alpha}(t_{\text{i}})), 
\end{equation}
where $\lowtilde{\mathcal{W}}$ is defined as in Eq.~\eqref{eq:Wfunctional}, but with $\rho^{\tSE}$ replaced by $\rho^\mathrm{E}$, which is independent of $\alpha(\ti)$, $\bar{\alpha}(\ti)$. We refer to a functional defined in this way as having an \textit{open past}. $\lowtilde{\mathcal{W}}$ in Eq.~\eqref{eq:openpastW}, but also $\mathcal{M}_r$ in Eq.~\eqref{eq:Mfunctional}, have both open past and future, which we shorten as \textit{open boundary}, or just \textit{open}. On the other hand, we say a functional has a \textit{closed boundary} if it has closed past and future, as $\mathcal{W}$ in Eq.~\eqref{eq:Wfunctional0}. We assume a process functional has closed boundary unless otherwise specified.

The open-boundary process functional relates directly to the Feynman–Vernon (FV) influence functional \cite{feynman1963thetheory}. Indeed, 
if we decompose the system-environment Hamiltonian in free and interacting part, $\hat{H}^\tSE = \hat{H}_{0} +\hat{H}^\tSE_\mathrm{I}$, the unitary functional decomposes as $\mathcal{U}^{\mathrm{SE}}[\alpha,\beta]=\mathcal{U}_{0}[\alpha]\mathcal{V}^{\mathrm{SE}}[\alpha,\beta]$,
where $\mathcal{U}_{0}$ and  $\mathcal{V}^{\mathrm{SE}}$ are generated by $\hat{H}_{0}$ and $\hat{H}^\tSE_\mathrm{I}$, respectively. The open process functional can then be written as 
\begin{equation}
\lowtilde{\mathcal{W}}[\alpha,\,\bar{\alpha}]=\mathcal{U}_{0}[\alpha]\mathcal{U}_{0}^{*}[\bar{\alpha}]\mathcal{J}^{\mathrm{FV}}[\alpha,\,\bar{\alpha}]\label{eq:W-FV-Phase}
\end{equation}
where the (coherent-state) FV influence functional is~\cite{grabert1988quantum,weiss2012quantum,yang2020nonequilibrium} 

\begin{multline}
\mathcal{J}_{\text{FV}}[\alpha,\,\bar{\alpha}]  \coloneqq \int\mathscr{D}[\beta, \bar{\beta}]\mathrm{e}^{\ii \int_{t_{\text{i}}}^{t_{\text{f}}}\left\{ \mathrm{Im}[\dot{\beta}^{*}\beta-\dot{\bar{\beta}}^{*}\bar{\beta}]\right\} \mathrm{d}t} \\
  \times\mathcal{V}^{\mathrm{SE}}[\alpha,\,\beta]\mathcal{V}^{\mathrm{SE}*}[\bar{\alpha},\,\bar{\beta}]\rho^{\mathrm{E}}(\beta(t_{\text{i}}),\bar{\beta}(t_{\text{i}})) \\ 
  \times \mathrm{id}(\bar{\beta}(\tf),\beta(\tf)).
 \label{eq:J-FV-def}
\end{multline}
As shown in Appendix~\ref{app:Derivation-of-Pr-Phase}, the post-measurement unnormalized conditional state is given by
\begin{multline}
\label{eq:Post-Meas-state}
    \rho_r(\alpha_{\mathrm{f}},\bar{\alpha}_{\mathrm{f}})
    = \! \int_{\substack{\alpha(\tf) = \alpha_{\mathrm{f}} \\ \bar{\alpha}(\tf) = \bar{\alpha}_{\mathrm{f}}}} \!\! \D[\alpha,\bar{\alpha}]  \ee^{\ii \int_{t_{\text{i}}}^{t_{\text{f}}} \dd t \Im (\dot{\alpha}^*\alpha - \dot{\bar \alpha}^*\bar \alpha  )} \\
    \times  \mathcal{M}_r[\alpha, \bar \alpha] \widetilde{\mathcal{W}}[\alpha,\bar{\alpha}],
\end{multline}
with $\Tr \hat\rho_r {=}\pr[r]$. Therefore, our formalism generalizes FV theory in two ways: by including initial system-environment correlations through the closed-past process functional and by including conditional evolution thanks to the operation functional. 
The main novelty, however, is a continuous version of the Born rule, where (causally ordered) processes and operations are represented independently and combined in a simple formula, Eq.~\eqref{eq:functionalBorn}, which does not depend on the details of the underlying system-environment-meter model. This reproduces and greatly generalizes Cave's formalism for measurements distributed in time \cite{Caves1986, Caves1987}, which is recovered for the particular case of unitary processes, $\lowtilde{\mathcal{W}}[\alpha,\,\bar{\alpha}]=\mathcal{U}[\alpha]\mathcal{U}^{*}[\bar{\alpha}]$ and pure [i.e., with Kraus rank 1, see Eq.~\eqref{eq:Markovian-M-alpha} below and related discussion] position measurements. While the connection between process matrices and FV functionals is well-established for the discrete formalism and for numerical methods \cite{Strathearn2018, Joergensen2019}, and a formal continuous limit of combs has been linked to quantum stochastic processes \cite{Milz2020kolmogorovextension}, a formalism to model continuous operations on non-Markovian processes was hitherto missing.

\section{Properties of the process functionals}

We have constructed continuous process and operation functionals from an underlying Hamiltonian model. The discrete-time formalism also gives \emph{axiomatic} definitions, grounded in physically motivated properties. Here we present a continuous version of these properties, whose main justification (apart from the formal analogy) rests on the fact that 1) they hold for constructively-defined functionals, proven in Appendix ~\ref{app:properties}, and 2) they reduce to the corresponding multi-time properties in an appropriate limit, Appendix~\ref{app:disc-time}.

\textbf{Positivity.}
Discrete process matrices and testers are required to be positive semidefinite, which means that $\hat{Q}= \hat{Q}^{\dagger}$ and $\expval{\hat{Q}}{\psi}\geq 0$ $\forall \ket{\psi}\in\mathcal{H}^N$, with $\hat{Q}= \hat{W}, \hat{M}$. The functional version of this is $\mathcal{Q}^*[\alpha,\bar{\alpha}] = \mathcal{Q}[\bar{\alpha},\alpha]$ and
\begin{equation} \label{eq:functionalpositivity}
    \int \D[\alpha, \bar{\alpha}] \psi[\alpha]\psi^{*}[\bar{\alpha}] \mathcal{Q}[\alpha,\bar{\alpha}] \geq 0
\end{equation}
for all functionals $\psi[\alpha]$.
Condition \eqref{eq:functionalpositivity} can also equivalently be stated  by including the factor $\ee^{\ii \int_{t_{\text{i}}}^{t_{\text{f}}} \dd t \Im (\dot{\alpha}^*\alpha - \dot{\bar \alpha}^*\bar \alpha  )}$, consistently with the form of the Born rule \eqref{eq:functionalBorn}, as that factor can be absorbed into $\psi[\alpha]$ by redefining $\psi[\alpha]\to \psi[\alpha]\ee^{\ii \int_{t_{\text{i}}}^{t_{\text{f}}} \dd t \Im (\dot{\alpha}^*\alpha )}$.

\textbf{Causality.}
Causally ordered process matrices satisfy a set of linear constraints, often called ``causality conditions'', ensuring that future choices of operations do not affect past measurement statistics \cite{chiribella09b}, as well as an affine constraint ensuring normalization of probabilities; see Appendix~\ref{app:causality}. The functional counterpart is that a (closed-boundary) process functional has to satisfy the following three properties:
\begin{equation}\label{eq:opentoclose}
    \mathcal{W}[\alpha,\bar{\alpha}]{=}\widetilde{\mathcal{W}}[\alpha,\bar{\alpha}] \mathrm{id}(\bar\alpha(\tf), \alpha(\tf));
\end{equation}
\begin{multline}
\label{eq:causalityfunctional}
    \int_{\substack{ \alpha(t') = \alpha' \\ \bar{\alpha}(t') = \bar{\alpha}' }} \D[\alpha_{>},\bar{\alpha}_{>}] \ee^{\ii \int_{t'}^{\tf} \dd t \Im (\dot{\alpha}^*\alpha - \dot{\bar \alpha}^*\bar \alpha  )} \mathcal{W}[\alpha,\bar{\alpha}]  \\
    = \widetilde{\mathcal{W}}[\alpha_{<},\bar{\alpha}_{<}]\, \mathrm{id}(\bar\alpha', \alpha')
\end{multline}
for all $ t'{\in}(\ti,\tf)$, where $\alpha_{>}$, $\bar{\alpha}_{>}$ are functions on $[t',\tf]$ and $\alpha_{<}$, $\bar{\alpha}_{<}$ on $[\ti,t']$; and
\begin{equation} \label{eq:functionalnormalization_main}
\int \D[\alpha,\bar{\alpha}]  \ee^{\ii \int_{t_{\text{i}}}^{t_{\text{f}}} \dd t \Im (\dot{\alpha}^*\alpha - \dot{\bar \alpha}^*\bar \alpha )} {\mathcal{W}}[\alpha,\bar{\alpha}]=1.
\end{equation}
These condition can equivalently be applied to open-future process functionals $\widetilde{\mathcal{W}}$, simply substituting in Eq.~\eqref{eq:opentoclose}.

For an open-past processes $\oW$, Eq.~\eqref{eq:openpastW}, imposing normalization of $\widetilde{\mathcal{W}}[\alpha,\bar{\alpha}] = \lowtilde{\mathcal{W}}[\alpha,\bar{\alpha}]\rho^{\mathrm{S}}(\alpha(t_{\text{i}}),\bar{\alpha}(t_{\text{i}}))$ for all trace-one initial states $\hat{\rho}^S$, and recalling that the trace can be expressed as $\Tr\hat{\rho}^S = \int \dd \alpha \dd \bar\alpha \rho^{\mathrm{S}}(\alpha,\bar{\alpha})\mathrm{id}(\bar\alpha, \alpha)$, yields the open-past normalization condition
\begin{multline} \label{eq:opennormalization}
\int_{\substack{\alpha(\ti) = \alpha_{\mathrm{i}} \\ \bar\alpha(\ti) = \bar\alpha_{\mathrm{i}}}}\!\! \D[\alpha,\bar{\alpha}]  \ee^{\ii \int_{t_{\text{i}}}^{t_{\text{f}}} \dd t \Im (\dot{\alpha}^*\alpha - \dot{\bar \alpha}^*\bar \alpha )} \\
\times {\oW}[\alpha,\bar{\alpha}]\mathrm{id}(\bar\alpha(\tf), \alpha(\tf)) 
= \mathrm{id}(\bar\alpha_{\mathrm{i}}, \alpha_{\mathrm{i}}).
\end{multline}
As discussed in Appendix~\ref{app:openboundarycausality}, this condition ensures that the propagators generated by $\oW$ are trace preserving.

Operation functionals are not required to satisfy causality, because they are conditioned on an outcome $r$. However, like testers in the discrete case, they should define a valid, causally ordered process if the outcome is ignored. Therefore, recalling that operation functionals have open boundary, we require that
$ \int \dd r \mathcal{M}_r[\alpha,\bar{\alpha}] $ satisfies Eqs.~\eqref{eq:causalityfunctional} and \eqref{eq:opennormalization}, where the integral extends over all possible outcome values, and should be substitute by a sum for discrete outcomes or by a functional integral for function-valued outcomes.

\textbf{Markovianity.} For discrete times, a (causally ordered) process matrix is defined to be Markovian if it satisfies  \cite{costa2016, Pollock2018}
\begin{equation} \label{eq:markovprocess}
    \hat W = \hat \rho_{\mathrm{i}} \otimes\left(\bigotimes_{k=1}^{N} \hat{W}_k \right)\otimes \id^{2N+1}, 
\end{equation}
where $\hat \rho_{\mathrm{i}}\in \mathcal{L}(\mathcal{H}^0)$ is the initial state and $\hat{W}_k \in \mathcal{L}(\mathcal{H}^{2k-1} \otimes \mathcal{H}^{2k})$ is the Choi-Jamio{\l}kowski representation of a CPTP map evolving the system from $t_{2k-1}$ to $t_{2k}$.

A Markovian process functional is defined analogously as having the form
\begin{equation}
\mathcal{W}[\alpha,\bar{\alpha}]= \rho(\alpha(\ti),\bar\alpha(\ti)) {\lowtilde{\mathcal{W}}}[\alpha,\bar{\alpha}] \mathrm{id}(\bar{\alpha}(\tf), \alpha(\tf))    
\end{equation}
for some state $\hat{\rho}$, and where the open functional $\lowtilde{\mathcal{W}}$ is \textit{divisible}, that is, for every $t'\in [\ti,\tf]$, 
\begin{equation}\label{eq:functionalmarkovianity}
    {\lowtilde{\mathcal{W}}}[\alpha,\bar{\alpha}] = {\lowtilde{\mathcal{W}}}{}_{<}[\alpha_{<},\bar{\alpha}_{<}]{\lowtilde{\mathcal{W}}}{}_{>}[\alpha_{>},\bar{\alpha}_{>}],
\end{equation}
where $\oW{}_{<}$, $\oW{}_{>}$ are themselves divisible, by recursive application of Eq.~\eqref{eq:causalityfunctional} to the respective domains.
Notably, divisibility for an open-boundary process functional implies that $\oW{}_{<}$ and $\oW{}_{>}$ can be chosen to be valid, open-boundary process functionals. Indeed, if $\oW{}$ is positive over the whole interval $[\ti,\tf]$, Eq.~\eqref{eq:functionalpositivity} has to hold in particular for arbitrary functionals of the form $\psi[\alpha] = \psi_<[\alpha_<]\psi_>[\alpha_>]$. Then, for $\oW$ of the product form \eqref{eq:functionalmarkovianity}, the functional integral in Eq.~\eqref{eq:functionalpositivity} decomposes as a product of integrals over $[\ti,t']$ and $[t',\tf]$, so the inequality implies that $\oW{}_{<}$ and $\oW{}_{>}$ are either both positive semidefinite or both negative semidefinite. In the latter case, $\oW$ can still be written as a product of the positive semidefinite functionals, simply by changing the sign of $\oW{}_{<}$, $\oW{}_{>}$. Furthermore, it is straightforward to verify that applying the causality and open-boundary normalization conditions, Eqs.~\eqref{eq:causalityfunctional} and \eqref{eq:opennormalization}, to a product $\oW$ implies that the same condition holds for $\oW{}_{<}$ and $\oW{}_{>}$. Hence, a Markovian process functional can be interpreted as being implemented as a temporal sequence of valid, independent processes.

To further connect with established notions of Markovianity, note that the recursive formulation of divisibility, Eq.~\eqref{eq:functionalmarkovianity}, is equivalent to requiring that, for every division of $[\ti,\tf]$ in subintervals, the process functional decomposes as a corresponding product of valid process functionals, each defined on one subinterval. This means that $\oW$ can be understood as a continuous product, which can be expressed formally as the exponential of an integral:
\begin{equation}\label{eq:timelocalW}
    \oW[\alpha,\bar\alpha] =  \ee^{\ii \int_{\ti}^{\tf} \dd t L(\alpha(t),\bar{\alpha}(t))}
\end{equation}
for some function $L(\alpha,\bar{\alpha})$. This characterization agrees with the common identification of Markovianity with time-local FV influence functionals \cite{breuer2002theory}. However, our definition has the additional requirement that $\oW[\alpha,\bar\alpha]$ must be a valid (positive and causal) process functional, which is not implied by the time-local form \eqref{eq:timelocalW} alone. Furthermore, divisibility acquires an additional operational meaning within our framework, because the local nature of the process functional can be probed through arbitrary continuous operations, extending to the continuum the process-matrix notion of Markovianity.

\section{Recovering the discrete-time formalism}
We expect the continuous-time formalism to reduce to the discrete-time one in the limit of instantaneous operations separated by finite time intervals. Let the operations act on the intervals $[t_{2k},t_{2k+1}]$, which means that the functionals $\mathcal{M}_r$ are constant over the complementary intervals. Conversely, if the operation intervals $[t_{2k},t_{2k+1}]$ are short enough, the process functional $\mathcal{W}$ can be approximated to be constant over them.
Hence, when we evaluate the functional Born rule, Eq.~\eqref{eq:functionalBorn}, we can split the functional integral into intervals, where only either $\mathcal{M}_r$ or $\mathcal{W}$ contribute, with boundary conditions at the joining times that are integrated at the last step. In this way, the functional integral for $\mathcal{W}$ (resp., $\mathcal{M}_r$), defines the matrix elements (in the coherent-state basis) of a process matrix $\hat W$ (resp., of a tester $\hat M_r$). The integration of the matrix elements then recovers the discrete Born rule, Eq.~\eqref{eq:discreteBorn}, see Appendix~\ref{app:disc-time} for details. Furthermore, in this construction, functional properties, including positive semidefiniteness, causality, normalization, and Markovianity, imply the corresponding properties for the reconstructed discrete process matrix and tester.

\section{Position-space representation}

Upon making the change of variables $\alpha{=}{(x+\mathrm{i}p)}/{\sqrt{2}}$, 
and integrating out the momentum degree of freedom in Eq.~\eqref{eq:functionalBorn}, one may hope that a position-space version of the functional Born rule can be obtained:
$\text{Pr}[r]=\int\mathscr{D}[x,\bar{x}]\mathcal{M}_{r}[x,\,\bar{x}]\mathcal{W}[x,\,\bar{x}]$. 
This is not always the case because, in general, the functional momentum integration does not decompose into separate integrations for $\mathcal{M}_{r}$ and $\mathcal{W}$. However, a convenient special case is when the operation functional does not depend on momentum, $\mathcal{M}_r[x,\bar{x}, p, \bar{p}] = \mathcal{M}^0_r[x,\bar{x}]$, in which case the functional Born rule reduces to 
\begin{equation}\label{eq:positionspaceBorn}
    \Pr[r] 
    = \int\mathscr{D}[x,\bar{x}] \mathcal{M}^0_r[x,\bar{x}]\mathcal{W}[x,\bar{x}],
\end{equation}
where the position-space process functional is given by $\mathcal{W}[x,\bar{x}] = \int \D[p,\bar{p}] \mathcal{W}[x,\bar{x},p,\bar{p}]\ee^{\ii \int_{\ti}^{\tf}\dd t \left(\dot{x}p - \dot{\bar{x}}\bar{p}\right)}
$. We prove this in Appendix~\ref{app:Reduction-to-position-PI}, where we further provide a more general expression for the functional Born rule upon integrating momenta. We also show that a position-space representation is possible if $H^{\tSE}$ and $H^{\tSM}$ satisfy suitable conditions, namely if the system's momentum does not couple to both meter and environment.

\section{Pure, Markovian measurements}
We derive now a functional characterization of the  ``continuous weak measurements'' typically considered in quantum optics~\cite{Jacobs2006,Wiseman2009,jacobs2014quantum,jordan2024quantum}. In the most common scenarios, these are pure measurements, namely the conditional (non-normalized) state transformation is defined by a single Kraus operator $\hat{K}_{r}(\tfti)$ that sends pure states to pure states,
$
|{\tilde{\psi}_{r}(\tf)}\rangle=\hat{K}_{r}(\tfti)|{\tilde{\psi}(t_{i})}\rangle
$
Markovianity means that the Kraus operators satisfy the divisibility property 
\begin{equation}
    \label{eq:divisibilityK}
    \hat{K}_{r}(\tfti) = \hat{K}_{r}(\tfti[\tf][t'])\hat{K}_{r}(\tfti[t'])
\end{equation}
for all $\ti<t'<\tf$, and we shall further assume continuity, $\hat{K}_{r}(\tfti[\ti]) = \id$. Normalization of the Kraus operators is expressed as $
\int\mathscr{D}[r]\hat{K}_{r}^{\dagger}(t_{\mathrm{f}}\!\!\shortleftarrow\!\!t_{\text{i}})\hat{K}_{r}(t_{\mathrm{f}}\!\!\shortleftarrow\!\!t_{\text{i}}) = \id
$, where, as usual in functional integration, a diverging normalization constant is absorbed in the measure $\D[r]$.

Upon defining the non-Hermitian dynamics $
\hat{H}_{r}(t)=\mathrm{i}\lim_{\Delta t\to0}[{\hat{K}_{r}(t+\Delta t\!\!\shortleftarrow\!\!t)-\id}]/{\Delta t}
$, 
these assumptions ensure that the measurement-induced dynamics can be written in terms of the non-Hermitian Schrodinger equation:
$
\text{i}\partial_{t}|{\tilde{\psi}_{r}(t)}\rangle\!=\!\hat{H}_{r}(t)|{\tilde{\psi}_{r}(t)}\rangle
$.
Therefore, the Kraus operator is  
\begin{equation}\label{eq:TorderedK}
\hat{K}_{r}(\tfti)=\mathcal{T}\mathrm{e}^{-\text{i}\int_{\ti}^{\tf}\mathrm{d}t \hat{H}_{r}(t)}.    
\end{equation}
In this case, one can show that the operation functional---up to a normalization constant---is 
\begin{equation}
{\mathcal{M}}_{r}[\alpha,\,\bar{\alpha}]=\mathcal{K}_{r}[\alpha]\mathcal{K}_{r}^{*}[\bar{\alpha}],
\label{eq:Markovian-M-alpha}
\end{equation} 
where the Kraus functional is defined as
$\mathcal{K}_{r}[\alpha]{\coloneqq }\mathrm{e}^{-\text{i}\int_{t_{\text{i}}}^{t_{\text{f}}}\mathrm{d}t H_{r}(\alpha)}$, where $H_{r}(\alpha)$ is the Husimi $Q$-function of $\hat{H}_{r}$ as defined previously. A particular example is continuous weak position measurement \cite{Wiseman2009,jacobs2014quantum,jordan2024quantum}, modeled by a Kraus operator of the form \eqref{eq:TorderedK}, with the anti-Hermitian Hamiltonian $\hat{H}_r\!\coloneqq \!{-\ii(r-\hat{x})^{2}}/({4\tau_{m}})$, where $\tau_m$, called the measurement time, characterizes the strength of the measurement. Noting that this Hamiltonian is normally ordered up to a readout-independent constant, we can readily calculate its $Q$ function and obtain the Kraus functional
\begin{equation} \label{eq:positionKraus}
  \mathcal{K}[x,p] = \mathrm{e}^{-\frac{1}{4 \tau_m}\int_{t_{\text{i}}}^{t_{\text{f}}}\mathrm{d}t (r(t)-x(t))^{2}}.
\end{equation}
As it does not depend on momentum, this can be combined directly with the position-space representation of the process. 

\section{Continuous monitoring of a Generalized Caldeira-Leggett process}

As a concrete application of our formalism, we consider a generalized Caldeira-Leggett processes under continuous-time position measurements. We present here the main results, leaving the details of the calculations to Appendix~\ref{app:GCL-Processes}.
The original Caldeira-Leggett model describes a harmonic oscillator (system) that interacts linearly with a harmonic environment through a time-independent position-position coupling, and where system and environment are initially in a product state, with the environment in a thermal (hence, Gaussian) state~\cite{caldeira1983pathintegral,breuer2002theory}. This model results in a Gaussian FV functional; here we will also consider a more general scenario, characterised by a gaussian functional without a specific reference to the underlying model.

As we have seen in Eq.~\eqref{eq:positionKraus}, the functional representation of continuous position measurement does not depend on momentum, so we only need the position representation of the process functional. The original CL process reads~\cite{caldeira1983pathintegral,breuer2002theory}
\begin{equation} \label{eq:CLprocess}
\mathcal{\widetilde{W}}^{\mathrm{CL}}[x,\bar{x}]=\mathrm{e}^{\ii S^{\mathrm{eff}}[x,\,\bar{x}]}\rho(x_{\mathrm{i}},,\bar{x}_{\mathrm{i}}),
\end{equation}
where 
\begin{equation}
S^{\mathrm{eff}}[x,\,\bar{x}]\coloneqq S_{0}[x,\,\bar{x}]+S^{\mathrm{FV}}[x,\,\bar{x}]
\end{equation}
is the sum of the system's free action, defined as 
\begin{multline}
    S_{0}[x,\,\bar{x}] 
    =\frac{m}{2}\int_{t_{\text{i}}}^{t_{\text{f}}}\dd t\left[\left(\dot{x}^{2}(t)-\dot{\bar{x}}^{2}(t)\right) \right. \\
\left.    -\omega_{0}^{2}\left(x^{2}(t)-\bar{x}^{2}(t)\right)\right],
\end{multline}
and the FV influence action due to the position-coupling to the environment defined as 
\begin{multline}\label{eq:SFV-CL-t-indept}
S^{\mathrm{FV}}[x,\,\bar{x}]=\frac{\mathrm{i}}{2}\int_{t_{\text{i}}}^{t_{\text{f}}}\int_{t_{\text{i}}}^{t_{\text{f}}}\dd t\dd s[x(t)-\bar{x}(t)] \\
\times
[\vartheta(t-s)x(s)-\vartheta^{*}(t-s)\bar{x}(s)]\Theta(t-s)  ,  
\end{multline}
where $\Theta$ is the Heaviside function,
\begin{multline}
\vartheta(t) :=
\int_{0}^{\infty}d\omega I(\omega)\biggl\{ \coth\left(\frac{\beta\omega}{2}\right)\cos(\omega t)  \\ 
 -\ii\sin(\omega t)\biggr\}, 
\end{multline}
and $I(\omega)$ is the spectral density of the thermal bath.

Upon introducing the compact notation $\bm{x}(t)=\left(x(t),\bar{x}(t)\right)^T$, the actions can be expressed as
\begin{equation}
S_{0}[\bm{x}]=\frac{m}{2}\int_{t_{\text{i}}}^{t_{\text{f}}}\dd t\left[\dot{\bm{x}}^{T}(t)\sigma_{z}\dot{\bm{x}}(t)-\omega_{0}^{2}\bm{x}^{T}(t)\sigma_{z}\bm{x}(t)\right]
\end{equation}
and 
\begin{equation}
S^{\mathrm{FV}}[\bm{x}]  =-\frac{1}{2}\int_{t_{\text{i}}}^{t_{\text{f}}}\dd t\int_{t_{\text{i}}}^{t_{\mathrm{f}}}\dd s\bm{x}^{T}(t)\bm{A}(t,s)\bm{x}(s),
\end{equation} where $\sigma_z$ is the third Pauli matrix and the matrix element of $\bm{A}(t,s)$ can be read off from Eq.~(\ref{eq:SFV-CL-t-indept})
as
\begin{multline}
\bm{A}(t-s) \\
=-\mathrm{i}\Theta(t-s)\begin{pmatrix}\vartheta(t-s) & -\vartheta^{*}(t-s)\\
-\vartheta(t-s) & \vartheta^{*}(t-s)
\end{pmatrix}.
\end{multline} 

For the steps below, we can work with a generalized model, where we do not impose particular constraints on the structure  of $\bm{A}(t,s)$. Upon noting the observation 
\begin{multline}
\int_{t_{\text{i}}}^{t_{\text{f}}}\dd t\int_{t_{\text{i}}}^{t_{\mathrm{f}}}\dd s\bm{x}^{T}(t)\bm{A}(t,s)\bm{x}(s) \\
=\int_{t_{\text{i}}}^{t_{\text{f}}}\dd t\int_{t_{\text{i}}}^{t_{\mathrm{f}}}\dd s\,\bm{x}^{T}(t)\bm{A}^{T}(s,t)\bm{x}(s)
\end{multline}
the generic influence action can be written as 
\begin{equation}
S^{\mathrm{FV}}[\bm{x}]  =-\frac{1}{2}\int_{t_{\text{i}}}^{t_{\text{f}}}\dd t\int_{t_{\text{i}}}^{t_{\mathrm{f}}}\dd s\bm{x}^{T}(t)\tilde{\bm{A}}(t,s)\bm{x}(s),
\end{equation} 
where $\tilde{\bm{A}}(t,s):=[\bm{A}(t,s)+\bm{A}^{T}(s,t)]/2$. That is, $\tilde{\bm{A}}(t,s)$ is invariant
under the joint operation of matrix transpose and the swapping between
$t$ and $s$, satisfying $\tilde{\bm{A}}^T(s,t)= \tilde{\bm{A}}(t,s)$. 

When the initial state of the system is Gaussian, the functional Born rule becomes a Gaussian functional integral, which can be solved with standard methods. The result is that the probability to observe a continuous record $r(t)$ is also given by a Gaussian functional:
\begin{equation}
\mathrm{Pr}[r]\propto\mathrm{e}^{-\frac{1}{2\tau_{\mathrm{m}}}\int_{t_{\text{i}}}^{t_{\text{f}}}\dd t\int_{t_{\text{i}}}^{t_{\text{f}}}\dd sr(t)R(t,\,s)r(s)+\int_{t_{\text{i}}}^{t_{\text{f}}}b(s)r(s)},
\end{equation}
where 
\begin{equation}
R(t,\,s)\coloneqq \delta(t-s)-\frac{1}{4}\sum_{ij}\tilde{G}_{ij}(t,\,s)+g(t)h(s)
\end{equation}
is the inverse of the two-time connected correlation function of the readout;
$\tilde{\bm{G}}(t,s)\equiv[\bm{G}(t,s)+\bm{G}^{T}(s,t)]/2$ and $\bm{G}(t,s)$
is an intrinsic Green's function of the damped system, satisfying the following ordinary differential equation:
\begin{multline}
\sigma_{z}\partial_{t}^{2}\bm{G}(t,\,s)+\left(\omega_{0}^{2}\sigma_{z}-\frac{\mathrm{i}}{2m\tau_{\mathrm{m}}}\right)\bm{G}(t,\,s)\\
+\frac{1}{m}\int_{t_{\text{i}}}^{t_{\mathrm{f}}}\dd\tau\tilde{\bm{A}}(t,\tau)\bm{G}(\tau,\,s)=-\frac{\mathrm{i}}{2m\tau_{\mathrm{m}}}\delta(t-s)\id;
\end{multline}
$g(t)$, $h(t)$ and $b(t)$ are determined by the
system's intrinsic Green's function as well as the initial state
$\hat{\rho}$.

In the limit $\tau_m\to 0$, the continuous position measurement approaches a strong, projective measurement, and the probability becomes
\begin{multline} \label{eq:CLprojectiveprobs}
\mathrm{Pr}[r]  
\propto \rho\left(r(t_{\mathrm{i}}),\,r(t_{\mathrm{i}})\right) \\
\times \mathrm{e}^{-\frac{\mathrm{i}}{2}\int_{t_{\text{i}}}^{t_{\text{f}}}\dd t\int_{t_{\text{i}}}^{t_{\mathrm{f}}}\dd s\,r(t)\sum_{ij}\tilde{A}_{ij}(t,s)r(s)},
\end{multline}
which holds for an arbitrary initial state $\hat{\rho}$ of the system, see Appendix~\ref{app:GCL-Processes} for more details. 

Eq.~\eqref{eq:CLprojectiveprobs} can also be understood by taking the formal limit of the Kraus functionals for continuous position measurement, Eq.~\eqref{eq:positionKraus}:
\begin{equation}\label{eq:Krausprojectivelimit}
    \lim_{\tau_{m}\rightarrow 0}\mathrm{e}^{-\frac{1}{4 \tau_m}\int_{t_{\text{i}}}^{t_{\text{f}}}\mathrm{d}t (r(t)-x(t))^{2}}  \propto \delta[r(t)-x(t)].
\end{equation}
In this limit, the functional Born rule results in replacing $x(t)\mapsto r(t)$ and $\bar{x}(t) \mapsto r(t)$ in the process functional. In other words, the probabilities become  proportional to the diagonal elements of the process functional in the position representation. Applying this substitution to Eq.~\eqref{eq:CLprocess} directly gives us Eq.~\eqref{eq:CLprojectiveprobs}.

\section{Conclusion}
Our formalism extends the framework of multi-time processes to continuous time, with process and operation functionals representing the continuum version of process matrices and multi-time operations. It provides an operational description of continuous monitoring in non-Markovian dissipative systems, treating noisy dynamics and measurement operations independently and without relying on a shared microscopic model, thus overcoming limitations of previous approaches. The result is particularly relevant for continuous sensing~\cite{rossi2020noisyquantum,yang2023efficient,yang2024quantum}, which requires measurements to be modeled independently of the dynamics encoding the parameters to be estimated. Further envisioned applications include measurement-induced phase transitions~\cite{li2018quantum,li2019measurementdriven} and continuous error correction and feedback~\cite{mabuchi2009continuous,mohseninia2020alwayson}. By connecting process matrices with the Feynman–Vernon formalism, our work bridges information-theoretic and continuous-time approaches to open quantum systems, offering a springboard for an organic integration of the respective methods and concepts. While we expressed the framework in a path-integral language, we expect that a rigorous algebraic formulation should be possible, building on existing formal treatments of time-ordered products in quantum theory~\cite{Fredenhagen2015, Rejzner2016}.

Beyond open quantum system applications, our work extends to continuous time the notions of \textit{Markovianity} and \textit{causality}, which are at the heart of recent developments in quantum foundations and quantum information. Markovianity is central in quantum causal models \cite{costa2016, Barrett2019}, which underpin modern analysis of causal structure in quantum theory, while causality is the reference constraint to introduce more general quantum causal structures, including indefinite causal order \cite{oreshkov12}. Our functional characterization of Markovianity and causality is therefore a stepping stone towards exploring quantum causality in continuous spacetime, with potential implications for measurements and causality in relativistic quantum field theory~\cite{Papageorgiou2024} and quantum causal structure in quantum-gravity scenarios \cite{Zych2019, SMoller2024gravitational}.

\paragraph{Note added.} After the first version of this work was made public, an independent approach to the continuum limit of quantum combs appeared in Ref.~\cite{Wassner2026}.

\begin{acknowledgments}
We acknowledge support from the Knut and Alice Wallenberg Foundation through the Wallenberg Initiative for Network and Quantum Information (WINQ), Zhejiang University Start-up Grants, Zhejiang Key Laboratory of R\&D and Application of Cutting-edge Scientific Instruments. This work is supported from COST Action CA23115: Relativistic Quantum Information, funded by COST (European Cooperation in Science and Technology).
\end{acknowledgments}

\bibliographystyle{quantum}
\bibliography{Process}

%%%%%%%%%%%%%%%%%%%%%%%%%%%%%%%%%%%%%%%
%%%%%%%%%%%%%%%%%%%%%%%%%%%%%%%%%%%%%%%

\onecolumngrid
\appendix

\section{Conventions}
Here is a summary of conventions and notation used in this work.
\begin{itemize}
	\item Complex phase-space variables (coherent state representation) are denoted with Greek letters: $\alpha,\beta,\gamma,\dots\in \mathbb{C}$.
    \item Position variables are denoted as $x,y,z,\dots \in \mathbb{R}$. Variables associated with their conjugate momenta are $p_x, p_y, p_z,\dots \in \mathbb{R}$. If a single pair of position-momentum variables is used, they are denoted as $(x,p)$.
    \item Coherent-state and phase-space variables are related by $\alpha = \frac{x + \ii p_x}{\sqrt{2}}$, $\beta = \frac{y + \ii p_y}{\sqrt{2}}$, $\gamma = \frac{z + \ii p_z}{\sqrt{2}}, \dots$
	\item Rows vs columns: $\matrixel{x}{\hat\rho}{\bar{x}}, \matrixel{\alpha}{\hat\rho}{\bar{\alpha} }, \dots$ This only for $x$ and $\bar{x}$ ``at the same time'', not e.g., elements of unitary or Kraus operators (where matrix elements are instead distinguished as initial and final, e.g. $\matrixel{x_{\mathrm{f}}}{\hat U}{x_{\mathrm{i}}}$).
	\item Operators always come with hats and their matrix elements contains no hats, e.g., $\rho(x,\bar{x}) = \matrixel{x}{\hat\rho}{\bar{x}}$.
    \item For observables and quantities usually comes with superscript S, SE, SM, or TOT, indicating the degress of freedom that they act on. 
	\item Functional integration: $\int \D[\alpha]$ or $\int \D[x]$; Functionals: calligraphic font and square brackets, as in $\mathcal{K}[\alpha]$ or $\mathcal{K}[x]$ .
	\item Initial and final time: $\ti$ and $\tf$; Intermediate time: $t^{\prime}$; time integration variables: $t$ and then $s$.
	\item Coherent-state notation: functions of coherent states, typically denoted $H(\alpha,\alpha^*)$ in the literature, are simply denoted $H(\alpha)$.
	\item Measure in the overcompleteness relation of coherent state $\int \dd\alpha \coloneqq \int \frac{\dd \text{Re} \alpha \dd \text{Im} \alpha }{\pi}$
\end{itemize}

\section{\label{app:Review-Coherent-PI}Review of coherent-state and phase-space path integrals}

\subsection{Coherent-state representation}

For the sake for simplicity we consider a single-mode Bosonic Hamiltonian
that is normally ordered, i.e.,
\begin{equation}
H(\hat a^{\dagger},\,\hat a)\equiv \; :\!H(\hat a^{\dagger},\,\hat a)\!:
\end{equation}
The overcompleteness relation is 
\begin{equation}
\int\mathrm{d}\alpha\ket{\alpha}\bra{\alpha}=1,\label{eq:closure-boson}
\end{equation}
where the measure is defined as
\begin{equation}
\mathrm{d}\alpha:=\frac{\mathrm{d}\text{Re}\alpha\mathrm{d}\text{Im}\alpha}{\pi}
\end{equation}
We note that 
\begin{equation}
    \langle\alpha|\alpha^{\prime}\rangle=\mathrm{e}^{-\frac{1}{2}(|\alpha|^{2}+|\alpha^{\prime}|^{2})+\alpha^{*}\alpha^{\prime}}\label{eq:overlap}
\end{equation}
Therefore we find

\begin{equation}
\langle\alpha_{\text{f}}|\hat{U}(t_{\text{f}}\!\!\shortleftarrow\!\!t_{\text{i}})|\alpha_{\text{i}}\rangle=\int\mathrm{d}\alpha_{1}\cdots\mathrm{d}\alpha_{N-1}\langle\alpha_{N}|e^{-\text{i}\hat{H}\epsilon}|\alpha_{N-1}\rangle \cdots\langle{\alpha_{1}|e^{-\text{i}\hat{H}\epsilon}|\alpha_{0}}\rangle,
\end{equation}
where $\alpha_{N}:=\alpha_{\text{f}}$, $\alpha_{0}:=\alpha_{\text{i}}$
and $\epsilon=(t_{\text{f}}-t_{\text{i}})/N$. It is straightforward to show that
\begin{equation}
\langle{\alpha_{k}|e^{-\text{i}\hat{H}\epsilon}|\alpha_{k-1}}\rangle\approx\mathrm{e}^{-\frac{1}{2}[\alpha_{k}^{*}(\alpha_{k}-\alpha_{k-1})-(\alpha_{k}^{*}-\alpha_{k-1}^{*})\alpha_{k-1}]-\text{i}H(\alpha_{k}^{*},\,\alpha_{k-1})\epsilon}
\end{equation}
Therefore, we find 
\begin{equation}
\langle{\alpha_{\text{f}}|\hat{U}(t_{\text{f}}\!\!\shortleftarrow\!\!t_{\text{i}})|\alpha_{\text{i}}}\rangle\approx\int\prod_{k=1}^{N-1}\mathrm{d}\alpha_{k}\mathrm{e}^{\left\{ \sum_{k=1}^{M}-\frac{1}{2}[\alpha_{k}^{*}(\frac{\alpha_{k}-\alpha_{k-1}}{\epsilon})-(\frac{\alpha_{k}^{*}-\alpha_{k-1}^{*}}{\epsilon})\alpha_{k}]-\text{i}H(\alpha_{k}^{*},\,\alpha_{k-1})\right\} \epsilon}\label{eq:discrete-phase-PI-propagator}
\end{equation}
In the continuum limit $N\to\infty$, we find 
\begin{equation}
\langle{\alpha_{\text{f}}|\hat{U}(t_{\text{f}}\!\!\shortleftarrow\!\!t_{\text{i}})|\alpha_{\text{i}}}\rangle=\int_{\alpha(t_{\text{i}})=\alpha_{\text{i}}}^{\alpha(t_{\text{f}})=\alpha_{\text{f}}}\mathscr{D}[\alpha]\mathrm{e}^{\frac{1}{2}\int_{t_{\text{i}}}^{t_{\text{f}}}(\dot{\alpha}^{*}\alpha-\alpha^{*}\dot{\alpha})}\mathcal{U}[\alpha]
\end{equation}
where 
\begin{equation}
\mathscr{D}[\alpha]\equiv\lim_{N\to\infty}\prod_{k=1}^{N-1}\mathrm{d}\alpha_{k}:= \lim_{N\to\infty}\prod_{k=1}^{N-1}\frac{\mathrm{d}\text{Re}\alpha_k\mathrm{d}\text{Im}\alpha_k}{\pi}
\end{equation}
and 
\begin{equation}
\mathcal{U}[\alpha]:=\mathrm{e}^{-\text{i}\int_{t_{\text{i}}}^{t_{\text{f}}}H(\alpha) \mathrm{d}t}
\end{equation}
Alternatively, we can rewrite 
\begin{equation}\label{app:coherentpropagator}
\langle\alpha_{\text{f}}|\hat{U}(t_{\text{i}}\!\!\shortleftarrow\!\!t_{\text{i}})|\alpha_{\text{i}}\rangle=\int_{\alpha(t_{\text{i}})=\alpha_{\text{i}}}^{\alpha(t_{\text{f}})=\alpha_{\text{f}}}\mathscr{D}[\alpha]\ee^{\ii \int_{t_{\text{i}}}^{t_{\text{f}}} \dd t \Im (\dot{\alpha}^{*}\alpha)} \mathcal{U}[\alpha]
\end{equation}
We conclude by emphasizing that the boundary condition $\alpha(t_{\text{i}})=\alpha_{\text{i}}$
indicates that neither $\alpha_{\text{i}}$ nor $\alpha_{\text{i}}^{*}$
is integrated over while $\alpha(t_{\text{f}})=\alpha_{\text{f}}$
indicates neither $\alpha_{\text{f}}$ nor $\alpha_{\text{f}}^{*}$
is integrated over. Indeed, from the discrete definition~(\ref{eq:discrete-phase-PI-propagator}),
it is clear that $\langle\alpha_{\text{f}}|\hat{U}(t_{\text{f}}\!\!\shortleftarrow\!\!t_{\text{i}})|\alpha_{\text{i}}\rangle$
depends on $\alpha_{\text{i}}$, $\alpha_{\text{i}}^{*}$ , $\alpha_{\text{f}}$,
and $\alpha_{\text{f}}^{*}$.

\subsection{Connection with the phase-space representation}
\label{app:phasespacecoherent}

The coherent-state path integral reviewed above is tightly connected, but not identical, to the phase-space path integral. The latter is a representation of the \emph{position-space} propagator, obtained by inserting into it alternating position and momentum resolutions of identity. The result is 
\begin{equation}\label{app:phasespacefunctional}
    \matrixel{x_{\text{f}}}{\hat{U}(\tfti)}{x_{\text{i}}} = \int_{x(\ti)=x_{\text{i}}}^{x(\tf)=x_{\text{f}}} \D[x,p]  \ee^{\ii \int_{t_{\text{i}}}^{t_{\text{f}}} \dd t \dot{x} p }\mathcal{U}^{\textrm{PS}}[x,p],
\end{equation}
where $\mathscr{D}[x,p] \coloneqq \lim_{N\to\infty}\prod_{k=1}^{N-1}\frac{\mathrm{d}x_k\mathrm{d}p_k}{2\pi}$,
\begin{equation} \label{app:phasespaceU}
    \mathcal{U}^{\textrm{PS}}[x,p] \coloneqq \ee^{-\ii \int \; \dd t h(x,p)},
\end{equation}
and $h(x,p)$ is a phase-representation of the Hamiltonian $\hat{H}$, whose precise definition depends on the time-slicing prescription used to define the path integral. Usually, a mid-point prescription is used, which results in $h(x,p)$ being equal to the Weyl symbol (i.e., the Wigner function) of $\hat{H}$. However, here we do not necessarily make this identification, since we want to connect this form with the coherent-state formulation, which typically uses a different definition of the path integral and corresponding ``classical Hamiltonian'' (the $Q$-representation in the rest of this work, but the discussion remains valid for different choices).

To connect the two representations, we rewrite the ``Berry phase'' term from the coherent-state representation as
\begin{equation}
    \int_{\ti}^{\tf} \dd t \Im{\dot{\alpha}^*\alpha} = \frac{1}{2}\int_{\ti}^{\tf} \dd t (\dot{x}p - x \dot p) = \int_{\ti}^{\tf} \dd t \left[\dot{x}p - \frac{1}{2} \frac{\dd }{\dd t }(x p)\right] = \int_{\ti}^{\tf} \dd t \dot{x}p - \frac{1}{2} (x_{\text{f}}p_{\text{f}} - x_{\text{i}}p_{\text{i}}),
\end{equation}
where we use $\alpha = \frac{x + \ii p}{\sqrt{2}}$, as usual. So we can rewrite the coherent-state propagator, Eq.~\eqref{app:coherentpropagator}, as
\begin{equation}\label{app:partialconversion}
    \matrixel{\alpha_{\text{f}}}{\hat{U}(\tfti)}{\alpha_{\text{i}}} = 
    \int_{\alpha(\ti)=\alpha_{\text{i}}}^{\alpha(\tf)=\alpha_{\text{f}}} \D[x,p]  \ee^{\ii \int_{t_{\text{i}}}^{t_{\text{f}}} \dd t \dot{x} p }\mathcal{U}\left[\frac{x + \ii p}{\sqrt{2}}\right]
    \ee^{\frac{\ii}{2} (x_{\text{i}}p_{\text{i}} - x_{\text{f}}p_{\text{f}})},
\end{equation}
To complete the identification between the two path integrals, we need to convert the coherent-state propagator into the position-space one. Note that, in the coherent-state path integral, both real and imaginary parts of $\alpha(\ti)$ and $\alpha(\tf)$ are fixed (corresponding to initial and final positions and momenta), while in the phase-space path integral only initial and final positions are fixed. This means that we need to integrate out $p_{\text{i}}$, $p_{\text{f}}$ in \eqref{app:partialconversion} in order to recover the phase-space path integral. In order to also recover the position-space propagator, we use the relation
\begin{equation}
    \int \frac{\dd p}{\pi} \ee^{-\frac{\ii}{2} x p} \ket{\alpha} \propto \ket{x},
\end{equation}
 which can be proven using the position representation of the coherent state:
 \begin{equation}
     \psi_{\alpha}(y) \coloneqq \braket{y}{\alpha} = \frac{1}{\pi^{\frac{1}{4}}}\ee^{-\frac{1}{2}\left(y-x\right)^2+\ii y p - \frac{\ii}{2}x p}.
 \end{equation}
Indeed, we find
\begin{equation}
    \int  \frac{\dd p}{\pi} \ee^{-\frac{\ii}{2} x p} \psi_{\alpha}(y) 
    =\frac{1}{\pi^{\frac{1}{4}}}  \int \frac{\dd p}{\pi} \ee^{-\frac{1}{2}\left(y-x\right)^2+\ii (y-x) p }
    = \frac{2}{\pi^{\frac{1}{4}}}\delta(y-x).
\end{equation}
Therefore, multiplying left and right of Eq.~\eqref{app:partialconversion} by the factor $\ee^{-\frac{\ii}{2} (x_{\text{i}}p_{\text{i}} - x_{\text{f}}p_{\text{f}})}$, and integrating initial and final momenta, we obtain
\begin{equation}
     \int \frac{\dd p_{\text{i}}}{\pi} \frac{\dd p_{\text{f}}}{\pi}
     \ee^{-\frac{\ii}{2} (x_{\text{i}}p_{\text{i}} - x_{\text{f}}p_{\text{f}})} \matrixel{\alpha_{\text{f}}}{\hat{U}(\tfti)}{\alpha_{\text{i}}} 
     \propto \matrixel{x_{\text{f}}}{\hat{U}(\tfti)}{x_{\text{i}}} 
     =  \int_{x(\ti)=x_{\text{i}}}^{x(\tf)=x_{\text{f}}} \D[x,p]  \ee^{\ii \int_{t_{\text{i}}}^{t_{\text{f}}} \dd t \dot{x} p }\mathcal{U}\left[\frac{x + \ii p}{\sqrt{2}}\right].
\end{equation}
Comparing this with Eq.~\eqref{app:phasespacefunctional}, we conclude that, up to a multiplicative factor (omitted as in most of this work), the functional in the coherent-state path integral coincides with the phase-space one, up to the usual change of variables:
\begin{equation} \label{app:coherenttophasespace}
    \mathcal{U}^{\textrm{PS}}[x,p] = \mathcal{U}\left[\frac{x + \ii p}{\sqrt{2}}\right].
\end{equation}
In the following, we will omit the ${}^{\textrm{PS}}$ superscript, and identify the functionals by the variables on which they depend.

Let us remark again that, in the above conversion, a prescription for the definition of the path integrals is understood, corresponding to a particular choice of the classical Hamiltonian, and the conversion holds only as long as the same prescription for the different path integrals is taken.

\section{Derivation of the coherent-state functional Born rule.}
\label{app:Derivation-of-Pr-Phase}

We derive two equivalent versions of the functional Born rule, which differ in the final boundary conditions.
The Born rule Eq.~\eqref{eq:coherentpathintegral} in the main text can be expanded as follows:

\begin{multline}
\text{Pr}[r] =\int\mathrm{d}\alpha_{\mathrm{f}}\mathrm{d}\beta_{\mathrm{f}}\mathrm{d}\gamma_{\mathrm{f}}\mathrm{d}\tilde{\gamma}_{\mathrm{f}}\mathrm{d}\alpha_{\mathrm{i}}\mathrm{d}\beta_{\mathrm{i}}\mathrm{d}\gamma_{\mathrm{i}}\mathrm{d}\bar{\alpha}_{\mathrm{i}}\mathrm{d}\bar{\beta}_{\mathrm{i}}\mathrm{d}\bar{\gamma}_{\mathrm{i}} \;
\langle\alpha_{\mathrm{f}},\beta_{\mathrm{f}},\gamma_{\mathrm{f}}|\hat{E}_{r}^{\mathrm{M}}|\tilde{\gamma}_{\mathrm{f}}\rangle 
\langle\tilde{\gamma}_{\mathrm{f}}|\hat{U}^{\mathrm{TOT}}(t_{{\rm f}}\!\!\shortleftarrow\!\!t_{{\rm i}})|\alpha_{\mathrm{i}},\beta_{\mathrm{i}},\gamma_{\mathrm{i}}\rangle \\
 \times\langle\alpha_{\mathrm{i}},\beta_{\mathrm{i}},\gamma_{\mathrm{i}}|\hat{\sigma}^{\mathrm{M}}\otimes\hat{\rho}^{\mathrm{SE}}|\bar{\alpha}_{\mathrm{i}},\bar{\beta}_{\mathrm{i}},\,\bar{\gamma}_{\mathrm{i}}\rangle\langle\bar{\alpha}_{\mathrm{i}},\bar{\beta}_{\mathrm{i}},\bar{\gamma}_{\mathrm{i}}|\hat{U}^{\mathrm{TOT}\dagger}(t_{\text{i}}\!\!\shortleftarrow\!\!t_{\text{f}})|\alpha_{\mathrm{f}},\beta_{\mathrm{f}},\gamma_{\mathrm{f}}\rangle.
 \label{eq:Pr-resol-config-strat1}
\end{multline}
Upon introducing $O(\alpha,\,\bar{\alpha}):=\langle{\alpha|\hat{O}|\bar{\alpha}}\rangle$
and similar definitions for the meter's or environmental degrees of
freedom, we obtain
\begin{multline}
\text{Pr}[r] =\int\mathrm{d}\alpha_{\mathrm{f}}\mathrm{d}\beta_{\mathrm{f}}\mathrm{d}\gamma_{\mathrm{f}}\mathrm{d}\tilde{\gamma}_{\mathrm{f}}\mathrm{d}\alpha_{\mathrm{i}}\mathrm{d}\beta_{\mathrm{i}}\mathrm{d}\gamma_{\mathrm{i}}\mathrm{d}\bar{\alpha}_{\mathrm{i}}\mathrm{d}\bar{\beta}_{\mathrm{i}}\mathrm{d}\bar{\gamma}_{\mathrm{i}} \;
E_{r}^{\mathrm{M}}(\gamma_{\mathrm{f}},\tilde{\gamma}_{\mathrm{f}})\langle{\alpha_{\mathrm{f}},\,\beta{}_{\mathrm{f}},\,\tilde{\gamma}_{\mathrm{f}}|\hat{U}_{\text{}}^{\mathrm{TOT}}(t_{{\rm f}}\!\!\shortleftarrow\!\!t_{{\rm i}})|\alpha_{\mathrm{i}},\,\beta_{\mathrm{i}},\gamma_{\mathrm{i}}}\rangle  \\
 \times \sigma^{\mathrm{M}}(\gamma_{\mathrm{i}},\,\bar{\gamma}_{\mathrm{i}})\rho^{\mathrm{SE}}(\alpha_{\mathrm{i}},\bar{\alpha}_{\mathrm{i}};\beta_{\mathrm{i}},\bar{\beta}_{\text{i}})\langle{\bar{\alpha}_{\mathrm{i}},\,\bar{\beta}_{\mathrm{i}},\bar{\gamma}_{\mathrm{i}}|\hat{U}_{\text{}}^{\mathrm{TOT}\dagger}(t_{\mathrm{i}}\!\!\shortleftarrow\!\!t_{\mathrm{f}})|\alpha_{\mathrm{f}},\,\beta_{\mathrm{f}},\,\gamma_{\mathrm{f}}}\rangle.\label{eq:Pr-resol-phase}
\end{multline}
Inserting the path integral representation Eq.~\eqref{eq:continuousprob_operators} in the main text, we obtain

\begin{align}
\text{Pr}[r] & =\int\mathrm{d}\alpha_{\text{f}}\mathrm{d}\alpha_{\text{i}}\mathrm{d}\bar{\alpha}_{\text{i}}\int_{\alpha(t_{\text{i}})=\alpha_{\text{i}}}^{\alpha(t_{\text{f}})=\alpha_{\text{f}}}\mathscr{D}[\alpha]\int_{\bar{\alpha}(t_{\text{i}})=\bar{\alpha}_{i}}^{\bar{\alpha}(t_{\mathrm{f}})=\alpha_{\mathrm{f}}}\mathscr{D}[\bar{\alpha}]\mathrm{e}^{\text{i}\int_{t_{\text{i}}}^{t_{\text{f}}}{ \mathrm{Im}(\dot{\alpha}^{*}\alpha-\dot{\bar{\alpha}}^{*}\bar{\alpha})} \mathrm{d}t}\nonumber \\
 & \times\int\mathrm{d}\beta_{\text{f}}\mathrm{d}\beta_{\text{i}}\mathrm{d}\bar{\beta}_{\text{i}}\int_{\beta(t_{\text{i}})=\beta_{\text{i}}}^{\beta(t_{\text{f}})=\beta_{\text{f}}}\mathscr{D}[\beta]\int_{\bar{\beta}(t_{\text{i}})=\bar{\beta}_{\text{i}}}^{\bar{\beta}(t_{\text{f}})=\beta_{\text{f}}}\mathscr{D}[\bar{\beta}]\mathrm{e}^{\text{i}\int_{t_{\text{i}}}^{t_{\text{f}}}{ \mathrm{Im}(\dot{\beta}^{*}\beta-\dot{\bar{\beta}}^{*}\bar{\beta})} \mathrm{d}t}\mathcal{U}^{\mathrm{SE}}[\alpha,\,\beta]\mathcal{U}^{\mathrm{SE}*}[\bar{\alpha},\,\bar{\beta}]\rho(\alpha_{\text{i}},\beta_{\text{i}};\bar{\alpha}_{\text{i}},\bar{\beta}_{\text{i}})\nonumber \\
 & \times\int\mathrm{d}\gamma_{\text{f}}\mathrm{d}\tilde{\gamma}_{\mathrm{f}}\mathrm{d}\gamma_{\text{i}}\mathrm{d}\bar{\gamma}_{\text{i}}E_{r}^{\mathrm{M}}(\gamma_{\text{f}},\tilde{\gamma}_{\mathrm{f}})\sigma(\gamma_{\text{i}},\,\bar{\gamma}_{\text{i}}) \nonumber \\
 & \times \int_{\gamma(t_{\text{i}})=\gamma_{\text{i}}}^{\gamma(t_{\text{f}})=\tilde{\gamma}_{\mathrm{f}}}\mathscr{D}[\gamma]\int_{\bar{\gamma}(t_{\text{i}})=\bar{\gamma}_{\mathrm{i}}}^{\bar{\gamma}(t_{f})=\gamma_{\mathrm{f}}}\mathscr{D}[\bar{\gamma}]\mathrm{e}^{\text{i}\int_{t_{\text{i}}}^{t_{\text{f}}}{ \mathrm{Im}(\dot{\gamma}^{*}\gamma-\dot{\bar{\gamma}}^{*}\bar{\gamma})} \mathrm{d}t}\mathcal{U}^{\mathrm{SM}}[\alpha,\gamma,]\mathcal{U}^{\mathrm{SM}*}[\bar{\alpha},\bar{\gamma}]
\end{align}
Upon noticing the boundary conditions, 
\begin{equation}
    \alpha(t_{\text{f}})=\bar{\alpha}(t_{\mathrm{f}})=\alpha_{\mathrm{f}},\:\beta(t_{\text{f}})=\bar{\beta}(t_{\mathrm{f}})=\beta_{\mathrm{f}},\:\gamma(t_{\text{f}})=\tilde{\gamma}_{\text{f}},\,\bar{\gamma}(t_{\text{f}})=\gamma_{\text{f}}
\end{equation}
we immediately obtain

\begin{equation}
\mathrm{Pr}[r]=\!\!\int_{\alpha(\tf)=\bar{\alpha}(\tf)}\!\D[\alpha,\bar{\alpha}]\ee^{\ii\int_{t_{\text{i}}}^{t_{\text{f}}}\dd t\mathrm{Im}(\dot{\alpha}^{*}\alpha-\dot{\bar{\alpha}}^{*}\bar{\alpha})}\mathcal{M}_{r}[\alpha,\bar{\alpha}]\widetilde{\mathcal{W}}[\alpha,\bar{\alpha}],\label{eq:Pr-Rep-tilde}
\end{equation}

To derive the Born rule $\mathrm{Pr}[r]$ in the form given in the main text, Eq.~\eqref{eq:functionalBorn}, we use a slightly different technique of inserting the resolutions of identity as follows:
\begin{align}
\text{Pr}[r] & =\int\mathrm{d}\alpha_{\mathrm{f}}\mathrm{d}\beta_{\mathrm{f}}\mathrm{d}\gamma_{\mathrm{f}}\mathrm{d}\bar{\alpha}_{\mathrm{f}}\mathrm{d}\bar{\beta}_{\mathrm{f}}\mathrm{d}\bar{\gamma}_{\mathrm{f}}\mathrm{d}\tilde{\gamma}_{\mathrm{f}}\mathrm{d}\alpha_{\mathrm{i}}\mathrm{d}\beta_{\mathrm{i}}\mathrm{d}\gamma_{\mathrm{i}}\mathrm{d}\bar{\alpha}_{\mathrm{i}}\mathrm{d}\bar{\beta}_{\mathrm{i}}\mathrm{d}\bar{\gamma}_{\mathrm{i}}\nonumber \\
 & \times\mathrm{Tr}\left\{ |\alpha_{\mathrm{f}},\beta_{\mathrm{f}},\gamma_{\mathrm{f}}\rangle\langle\alpha_{\mathrm{f}},\beta_{\mathrm{f}},\gamma_{\mathrm{f}}|\hat{E}_{r}^{\mathrm{M}}|\tilde{\gamma}_{\mathrm{f}}\rangle\langle\tilde{\gamma}_{\mathrm{f}}|\hat{U}^{\mathrm{TOT}}(t_{{\rm f}}\!\!\shortleftarrow\!\!t_{{\rm i}})|\alpha_{\mathrm{i}},\beta_{\mathrm{i}},\gamma_{\mathrm{i}}\rangle\right. \nonumber \\
 & \times\left.\langle\alpha_{\mathrm{i}},\beta_{\mathrm{i}},\gamma_{\mathrm{i}}|\hat{\sigma}^{\mathrm{M}}\otimes\hat{\rho}^{\mathrm{SE}}|\bar{\alpha}_{\mathrm{i}},\bar{\beta}_{\mathrm{i}},\,\bar{\gamma}_{\mathrm{i}}\rangle\langle\bar{\alpha}_{\mathrm{i}},\bar{\beta}_{\mathrm{i}},\bar{\gamma}_{\mathrm{i}}|\hat{U}^{\mathrm{TOT}\dagger}(t_{\text{i}}\!\!\shortleftarrow\!\!t_{\text{f}})|\bar{\alpha}_{\mathrm{f}},\bar{\beta}_{\mathrm{f}},\bar{\gamma}_{\mathrm{f}}\rangle\langle\bar{\alpha}_{\mathrm{f}},\bar{\beta}_{\mathrm{f}},\bar{\gamma}_{f}|\right\} .\label{eq:Pr-resol-config-strat2}
\end{align}
This will leads to the same expression as above, but with
the following different boundary conditions
\begin{align}
\int\mathscr{D}[\alpha,\,\bar{\alpha},]\mathrm{id}(\bar{\alpha}(t_{\mathrm{f}}),\,\alpha(t_{\mathrm{f}})) F[\alpha,\,\bar{\alpha}] & :=\int\mathrm{d}\alpha_{\mathrm{f}}\int\mathrm{d}\bar{\alpha}_{\mathrm{f}}\int^{\alpha(t_{\text{f}})=\alpha_{\text{f}},\,\bar{\alpha}(t_{\mathrm{f}})=\bar{\alpha}_{\mathrm{f}}}\mathscr{D}[\alpha,\,\bar{\alpha},]\langle{\bar{\alpha}_{\mathrm{f}}|\alpha_{\mathrm{f}}}\rangle\\
\int\mathscr{D}[\beta,\,\bar{\beta}]\mathrm{id}(\bar{\beta}(t_{\mathrm{f}}),\,\beta(t_{\mathrm{f}})) F[\beta,\,\bar{\beta}] & :=\int\mathrm{d}\beta_{\mathrm{f}}\int\mathrm{d}\bar{\beta}_{\mathrm{f}}\int^{\beta(t_{\text{f}})=\beta_{\text{f}},\,\bar{\beta}(t_{\mathrm{f}})=\bar{\beta}_{\mathrm{f}}}\mathscr{D}[\beta,\,\bar{\beta}]\langle{\bar{\beta}_{\mathrm{f}}|\beta{\mathrm{f}}}\rangle
\end{align}
These conditions imply that 
\begin{align} 
\int\mathscr{D}[\alpha,\,\bar{\alpha}]  \mathrm{id}(\bar{\alpha}(t_{\mathrm{f}}),\,\alpha(t_{\mathrm{f}})) F[\alpha,\,\bar{\alpha}] & =\int_{\alpha(t_{\text{f}})=\bar{\alpha}(t_{\mathrm{f}})}\mathscr{D}[\alpha,\,\bar{\alpha}] F[\alpha,\,\bar{\alpha}]\label{eq:id-funcint-alpha}\\
\int\mathscr{D}[\beta,\,\bar{\beta}]  \mathrm{id}(\bar{\beta}(t_{\mathrm{f}}),\,\beta(t_{\mathrm{f}})) F[\beta,\,\bar{\beta}]  & =\int_{\beta(t_{\text{f}})=\bar{\beta}(t_{\mathrm{f}})}\mathscr{D}[\beta,\,\bar{\beta}] F[\beta,\,\bar{\beta}]\label{eq:id-funcint-beta}
\end{align}
In other words, $\mathrm{id}(\bar{\alpha(\tf)},\alpha(\tf))$ is a function that identifies the final boundary condition of the functional integral over $\bar{\alpha}(t)$ and $\alpha(t)$ and similar observation also holds for the functional integration over $\bar{\beta}(t)$ and $\beta(t)$. With these observations, we obtain
Eq.~\eqref{eq:functionalBorn} in the main text from Eq.~\eqref{eq:Pr-Rep-tilde}.

Eqs.~(\ref{eq:id-funcint-alpha},~\ref{eq:id-funcint-beta}) also
imply the following useful identity
\begin{equation}
\int\dd\bar{\alpha}_{\mathrm{f}}\dd\alpha_{\mathrm{f}}\mathrm{id}(\bar{\alpha}_{\mathrm{f}},\,\alpha_{\mathrm{f}})f(\alpha_{\mathrm{f}},\,\bar{\alpha}_{\mathrm{f}})=\int\dd\alpha_{\mathrm{f}}f(\alpha_{\mathrm{f}},\,\alpha_{\mathrm{f}}),\label{eq:id-property-int}
\end{equation}
where
\begin{equation}
f(\alpha_{\mathrm{f}},\,\bar{\alpha}_{\mathrm{f}}):=\int^{\alpha(t_{\mathrm{f}})=\alpha_{\mathrm{f}},\,\bar{\alpha}(t_{\mathrm{f}})=\bar{\alpha}_{\mathrm{f}}}\mathscr{D}[\alpha,\,\bar{\alpha}]F[\alpha,\,\bar{\alpha}].
\end{equation}

The post-measurement conditional state after the measurement $\hat{K}_{r}^{\mathrm{M}}$, where $\hat{K}_{r}^{\mathrm{M}}$ is the Kraus operator satisfying $\hat{E_r}=\hat{K}_{r}^{\mathrm{M}\dagger}\hat{K}_{r}^{\mathrm{M}}$, given by 
\begin{align}
\langle\alpha_{f}|\tilde{\rho}_{r}|\bar{\alpha}_{f}\rangle 
& =  \mathrm{Tr}_{\mathrm{EM}}[\hat{K}_{r}^{\mathrm{M}}\hat{U}^{\mathrm{TOT}}(t_{{\rm f}}\!\!\shortleftarrow\!\!t_{{\rm i}})\hat{\sigma}^{\mathrm{M}}\otimes\hat{\rho}^{\mathrm{SE}}\hat{U}^{\mathrm{TOT}\dagger}(t_{{\rm f}}\!\!\shortleftarrow\!\!t_{{\rm i}})\hat{K}_{r}^{\mathrm{M}\dagger}] \nonumber \\
 & = \int\mathrm{d}\alpha_{\mathrm{f}}\int\mathrm{d}\bar{\alpha}_{\mathrm{f}}\mathrm{d}\beta_{\mathrm{f}}\mathrm{d}\gamma_{\mathrm{f}}\mathrm{d}\tilde{\gamma}_{\mathrm{f}}\mathrm{d}\alpha_{\mathrm{i}}\mathrm{d}\beta_{\mathrm{i}}\mathrm{d}\gamma_{\mathrm{i}}\mathrm{d}\bar{\alpha}_{\mathrm{i}}\mathrm{d}\bar{\beta}_{\mathrm{i}}\mathrm{d}\bar{\gamma}_{\mathrm{i}} \langle\alpha_{\mathrm{f}},\beta_{\mathrm{f}},\gamma_{\mathrm{f}}|\hat{E}_{r}^{\mathrm{M}}|\tilde{\gamma}_{\mathrm{f}}\rangle  \nonumber \\
 \times \langle\tilde{\gamma}_{\mathrm{f}}|\hat{U}^{\mathrm{TOT}} & (t_{{\rm f}}\!\!\shortleftarrow\!\!t_{{\rm i}})|\alpha_{\mathrm{i}},\beta_{\mathrm{i}},\gamma_{\mathrm{i}}\rangle\langle\alpha_{\mathrm{i}},\beta_{\mathrm{i}},\gamma_{\mathrm{i}}|\hat{\sigma}^{\mathrm{M}}\otimes\hat{\rho}^{\mathrm{SE}}|\bar{\alpha}_{\mathrm{i}},\bar{\beta}_{\mathrm{i}},\,\bar{\gamma}_{\mathrm{i}}\rangle\langle\bar{\alpha}_{\mathrm{i}},\bar{\beta}_{\mathrm{i}},\bar{\gamma}_{\mathrm{i}}|\hat{U}^{\mathrm{TOT}}(t_{\text{i}}\!\!\shortleftarrow\!\!t_{\text{f}})|\bar{\alpha}_{\mathrm{f}},\beta_{\mathrm{f}},\gamma_{\mathrm{f}}\rangle.
\end{align}
Then following similar calculations above, we obtain Eq.~\eqref{eq:Post-Meas-state} in the main text.

Following similar steps we can also arrive at the equivalent phase-space form of the Born rule: 
\begin{equation}
    P[r] = \int_{x(\tf) = \bar{x}(\tf)} \D[x, p, \bar{x}, \bar{p}] \ee^{\ii\int_{\ti}^{\tf} (\dot{x}p - \dot{\bar{x}}\bar{p})} \mathcal{M}_r[x,p,\bar{x},\bar{p}]\mathcal{W}[x,p,\bar{x},\bar{p}],
\end{equation}
with 
\begin{equation}
   \mathcal{W}[x,p,\bar{x},\bar{p}] \coloneqq \widetilde{\mathcal{W}}[x,p,\bar{x},\bar{p}]\,\mathrm{id}(\bar{x}(\tf),x(\tf)),
\end{equation}
$\mathrm{id}(\bar{x},x)\coloneqq\langle\bar{x}|x \rangle = \delta(\bar{x}-x)$, 
\begin{multline}
    \widetilde{\mathcal{W}}[x,p,\bar{x},\bar{p}]
    \coloneqq \int\!\!\!\D[y,p_y,\bar{y},\bar{p}_y] 
    \ee^{\ii\int_{\ti}^{\tf} (\dot{y}p_y - \dot{\bar{y}}\bar{p}_y)} \mathrm{id}(\bar{y}(\tf),y(\tf)) \\
\times \mathcal{U}^{\tSE}(x,p_x,y,p_y)\mathcal{U}^{\tSE*}(\bar{x},\bar{p}_x,\bar{y},\bar{p}_y) \rho\left(x(\ti),y(\ti), \bar{x}(\ti),\bar{y}(\ti)\right),
\end{multline}
$\rho\left(x,y,\bar{x},\bar{y}\right)\coloneqq \matrixel{x,y}{\hat{\rho}}{\bar{x},\bar{y}}$; and
\begin{multline}
    \mathcal{M}_r[x,p,\bar{x},\bar{p}]
    \coloneqq \int\D[z,p_z,\bar{z},\bar{p}_z] \ee^{\ii\int_{\ti}^{\tf} (\dot{z}p_z - \dot{\bar{z}}\bar{p}_z)} \\
\times \mathcal{U}^{\tSM}(x,p_x,z,p_z)\mathcal{U}^{\tSM*}(\bar{x},\bar{p}_x,\bar{z},\bar{p}_z) \sigma(z(\ti), \bar{z}(\ti)) E_r(\bar{z}(\tf),z(\tf)),
\end{multline}
with $\sigma(z, \bar{z})\coloneqq  \matrixel{z}{\hat{\sigma}}{\bar{z}} $, $E_r(\bar{z},z)\coloneqq  \matrixel{\bar{z}}{\hat{E}_r}{z}$.

Here, $\mathcal{U}^{\tSE}(x,p_x,y,p_y)$ and  $\mathcal{U}^{\tSM}(x,p_x,z,p_z)$ are the phase-space functionals, Eq.~\eqref{app:phasespaceU}, associated with the respective Hamiltonians. Using the conversion from coherent-state to phase-space representations discussed in Sec.~\ref{app:phasespacecoherent}, we see that the phase space functionals $\mathcal{W}$ and $\mathcal{M}_r$ coincide with the corresponding coherent-state ones, up to the usual change of variables and as long as the same path-integral prescriptions are used.

\section{Properties of the process and operation functionals}
\label{app:properties}

Here we prove the properties of positivity and causality for process and operation functionals defined constructively through a Hamiltonian model, Eqs.~\eqref{eq:Wfunctional} and \eqref{eq:Mfunctional} respectively.

\subsection{Positivity\label{app:positivity}}
Let us first consider the positivity of the process functional $\mathcal{W}$
and $\widetilde{W}$ for pure unitary system and initial pure states,
where $\mathcal{\widetilde{W}}[\alpha,\,\bar{\alpha}]=\mathcal{U}_{0}[\alpha]\mathcal{U}_{0}^{*}[\bar{\alpha}]\phi(\alpha(t_{\text{i}}))\phi^{*}(\bar{\alpha}(t_{\text{i}}))$.
Clearly the functional integral can be broken into 
\begin{equation}
\int\mathscr{D}[\alpha,\bar{\alpha}]=\int\mathrm{d}\alpha_{\text{i}}\mathrm{d}\bar{\alpha}_{\text{i}}\int\mathrm{d}\alpha_{\text{f}}\mathrm{d}\bar{\alpha}_{\text{f}}\int_{\alpha(t_{\text{i}})=\alpha_{\text{i}}}^{\alpha(t_{\text{f}})=\alpha_{\text{f}}}\mathscr{D}[\alpha]\int_{\bar{\alpha}(t_{\text{i}})=\bar{\alpha}_{i}}^{\bar{\alpha}(t_{\mathrm{f}})=\bar{\alpha}_{\mathrm{f}}}\mathscr{D}[\bar{\alpha}]
\end{equation}
Upon defining $R(\alpha_{\mathrm{i}},\alpha_{\mathrm{f}}):=\int_{\alpha(t_{\text{i}})=\alpha_{\text{i}}}^{\alpha(t_{\text{f}})=\alpha_{\text{f}}}\mathscr{D}[\alpha]\mathcal{U}_{0}[\alpha]\psi[\alpha]$,
then we find 
\begin{align}
\int\mathscr{D}[\alpha,\bar{\alpha}]\psi^{*}[\alpha]\widetilde{\mathcal{W}}[\alpha,\,\bar{\alpha}]\psi[\bar{\alpha}] & =\int\mathrm{d}\alpha_{\text{f}}\mathrm{d}\bar{\alpha}_{\text{f}}\mathrm{d}\alpha_{\text{i}}\mathrm{d}\bar{\alpha}_{\text{i}}R(\alpha_{\mathrm{i}},\alpha_{\mathrm{f}})R^{*}(\bar{\alpha}_{\mathrm{i}},\bar{\alpha}_{\mathrm{f}})\phi(\alpha_{\mathrm{i}})\phi^{*}(\bar{\alpha}_{\mathrm{i}})\nonumber \\
 & =\big|\int\mathrm{d}\alpha_{\text{f}}\mathrm{d}\alpha_{\text{i}}R(\alpha_{\mathrm{i}},\alpha_{\mathrm{f}})\phi(\alpha_{\mathrm{i}})\big|^{2}\ge0
\end{align}
Similarly, we find
\begin{equation} \label{app:mixedspositivity}
\int\mathscr{D}[\alpha,\bar{\alpha}]\psi^{*}[\alpha]\mathcal{W}[\alpha,\,\bar{\alpha}]\psi[\bar{\alpha}]=\int\mathrm{d}\alpha_{\text{f}}\mathrm{d}\bar{\alpha}_{\text{f}}\mathrm{d}\alpha_{\text{i}}\mathrm{d}\bar{\alpha}_{\text{i}}R(\alpha_{\mathrm{i}},\alpha_{\mathrm{f}})R^{*}(\bar{\alpha}_{\mathrm{i}},\bar{\alpha}_{\mathrm{f}})\phi(\alpha_{\mathrm{i}})\phi^{*}(\bar{\alpha}_{\mathrm{i}})\mathrm{id}(\bar{\alpha}_{\text{f}},\,\alpha_{\text{f}})
\end{equation}
Using Eq.~\eqref{eq:id-property-int}, we immediately conclude that
\begin{equation}
\int\mathscr{D}[\alpha,\bar{\alpha}]\psi^{*}[\alpha]\mathcal{W}[\alpha,\bar{\alpha}]\psi[\bar{\alpha}]=\int\mathrm{d}\alpha_{\text{f}}\,\left|\int\mathrm{d}\alpha_{\text{i}}R(\alpha_{\mathrm{i}},\alpha_{\mathrm{f}})\phi(\alpha_{\mathrm{i}})\right|^{2}\ge0 
\end{equation}

The idea of generalizing to the generic case is to regard the system-ancilla (``ancilla'' refers to either environment or meter) 
as a joint system, so that the above argument applies again. Without loss of generality, we can take initial system-environment state $\rho^{\tSE}$ to be pure denoted as $|\phi_k\rangle\langle\phi_k|$ and final measurement $E_{r}$ to be a rank-one denoted as $|\xi_{sr}\rangle\langle\xi_{sr}|$. Since both $\rho^{\tSE}$ and $E_{r}$ are positive semidefinite operators, upon spectral decomposition, it is straightforward to see that if positivity holds for rank one initial system-environment state and rank-one final measurement, it will also hold for the high-rank case.  

Let us recall
\begin{align} 
\mathcal{\widetilde{W}}[\alpha,\bar{\alpha}] & =\int\mathscr{D}[\beta,\bar{\beta}]\ee^{\ii\int\dd t\mathrm{Im}(\dot{\beta}^{*}\beta-\dot{\bar{\beta}}^{*}\bar{\beta})}\mathcal{U}^{\tSE}(\alpha,\beta)\mathcal{U}^{\tSE*}(\bar{\alpha},\bar{\beta})\phi_{}(\alpha(\ti),\beta(\ti))\phi_{}^{*}(\bar{\alpha}(\ti),\bar{\beta}(\ti))\mathrm{id}(\bar{\beta}(t_{\mathrm{f}}),\,\beta (t_{\mathrm{f}})),\\
\mathcal{M}_{r}[\alpha,\bar{\alpha}] & =\int\mathscr{D}[\gamma,\bar{\gamma}]\ee^{\ii\int\dd t\mathrm{Im}(\dot{\gamma}^{*}\gamma-\dot{\bar{\gamma}}^{*}\bar{\gamma})}\mathcal{U}^{\tSM}(\alpha,\gamma)\mathcal{U}^{\tSM*}(\bar{\alpha},\bar{\gamma})\xi_{r}(\bar{\gamma}(\tf))\xi_{r}^{*}(\gamma(\tf)),
\end{align}
We define 
\begin{align}
R(\alpha_{\mathrm{i}},\beta_{\mathrm{i}};\,\alpha_{\mathrm{f}},\beta_{\mathrm{f}}) & :=\int_{\alpha(t_{\text{i}})=\alpha_{\text{i}},\,\beta(t_{\text{i}})=\alpha_{\text{i}}}^{\alpha(t_{\text{f}})=\alpha_{\text{f}},\,\beta(t_{\mathrm{f}})=\beta_{\mathrm{f}}}\mathscr{D}[\alpha,\,\beta]\mathcal{U}^{\tSE}(\alpha,\beta)\psi[\alpha]\\
T(\alpha_{\mathrm{i}},\gamma_{\mathrm{i}};\,\alpha_{\mathrm{f}},\gamma_{\mathrm{f}}) & :=\int_{\alpha(t_{\text{i}})=\alpha_{\text{i}},\,\gamma(t_{\text{i}})=\alpha_{\text{i}}}^{\alpha(t_{\text{f}})=\alpha_{\text{f}},\,\gamma(t_{\mathrm{f}})=\gamma_{\mathrm{f}}}\mathscr{D}[\alpha,\,\beta]\mathcal{U}^{\tSM}(\alpha,\beta)\psi[\alpha]
\end{align}
It is then straightforward to check, upon using Eq.~(\ref{eq:id-property-int}),
\begin{align} 
\int\mathscr{D}[\alpha,\bar{\alpha}]\psi^{*}[\alpha]\widetilde{\mathcal{W}}[\alpha,\,\bar{\alpha}]\psi[\bar{\alpha}] & =\int\mathrm{d}\alpha_{\text{i}}\mathrm{d}\beta_{\text{i}}\mathrm{d}\alpha_{\text{f}}\mathrm{d}\beta_{\text{f}}\mathrm{d}\bar{\alpha}_{\text{i}}\mathrm{d}\bar{\beta}_{\text{i}}\mathrm{d}\bar{\alpha}_{\text{f}}\mathrm{d}\bar{\beta}_{\text{f}}\nonumber \\
 & \times R(\alpha_{\mathrm{i}},\beta_{\mathrm{i}};\,\alpha_{\mathrm{f}},\beta_{\mathrm{f}})R^{*}(\bar{\alpha}_{\mathrm{i}},\bar{\beta}_{\mathrm{i}};\,\bar{\alpha}_{\mathrm{f}},\bar{\beta}_{\mathrm{f}})\phi(\alpha_{\mathrm{i}},\beta_{\mathrm{i}})\phi^{*}(\bar{\alpha}_{\mathrm{i}},\bar{\beta}_{\mathrm{i}})\mathrm{id}(\bar{\beta}_{\mathrm{f}},\,\beta_{\mathrm{f}})\nonumber \\
 & =\int d\beta_{\mathrm{f}}\big|\int\mathrm{d}\alpha_{\text{i}}\mathrm{d}\beta_{\text{i}}\mathrm{d}\alpha_{\text{f}}R(\alpha_{\mathrm{i}},\beta_{\mathrm{i}};\,\alpha_{\mathrm{f}},\beta_{\mathrm{f}})\phi(\alpha_{\mathrm{i}},\beta_{\mathrm{i}})\big|{}^{2}\ge0
\end{align}
Note that $\mathcal{W}[\alpha,\,\bar{\alpha}]$ is related to $\widetilde{\mathcal{W}}[\alpha,\,\bar{\alpha}]$
up to $\mathrm{id}(\bar{\alpha}(t_{\mathrm{f}}),\alpha(t_{\mathrm{f}}))$,
using Eq.~(\ref{eq:id-property-int}), one can show that 
\begin{equation}
\int\mathscr{D}[\alpha,\bar{\alpha}]\psi^{*}[\alpha]\mathcal{W}[\alpha,\,\bar{\alpha}]\psi[\bar{\alpha}]=\int d\beta_{\mathrm{f}}\mathrm{d}\alpha_{\text{f}}\big|\int\mathrm{d}\alpha_{\text{i}}\mathrm{d}\beta_{\text{i}}R(\alpha_{\mathrm{i}},\beta_{\mathrm{i}};\,\alpha_{\mathrm{f}},\beta_{\mathrm{f}})\phi(\alpha_{\mathrm{i}},\beta_{\mathrm{i}})\big|{}^{2}\ge0
\end{equation}
On the other hand, we note that there is no final-time boundary condition
on the functional integration over $\gamma$, we can show that 
\begin{align}
\int\mathscr{D}[\alpha,\bar{\alpha}]\psi^{*}[\alpha]\mathcal{M}_{r}[\alpha,\,\bar{\alpha}]\psi[\bar{\alpha}] & =\int\mathrm{d}\alpha_{\text{i}}\mathrm{d}\gamma_{\text{i}}\mathrm{d}\alpha_{\text{f}}\mathrm{d}\gamma_{\text{f}}\mathrm{d}\bar{\alpha}_{\text{i}}\mathrm{d}\bar{\gamma}_{\text{i}}\mathrm{d}\bar{\alpha}_{\text{f}}\mathrm{d}\bar{\gamma}_{\text{f}}\nonumber \\
 & \times{T}(\alpha_{\mathrm{i}},\gamma_{\mathrm{i}};\,\alpha_{\mathrm{f}},\gamma_{\mathrm{f}}){T}^{*}(\bar{\alpha}_{\mathrm{i}},\bar{\gamma}_{\mathrm{i}};\,\bar{\alpha}_{\mathrm{f}},\bar{\gamma}_{\mathrm{f}})\xi_{r}(\bar{\gamma}_{\mathrm{f}})\xi_{r}^{*}(\gamma_{\mathrm{f}})\nonumber \\
 & =\big|\int\mathrm{d}\alpha_{\text{i}}\mathrm{d}\gamma_{\text{i}}\mathrm{d}\alpha_{\text{f}}\mathrm{d}\gamma_{\text{f}}{T}(\alpha_{\mathrm{i}},\gamma_{\mathrm{i}};\,\alpha_{\mathrm{f}},\gamma_{\mathrm{f}})\xi_{r}^{*}(\gamma_{\mathrm{f}})\big|^{2}\ge0
\end{align}

\subsection{Causality}
\label{app:causality}
Here we derive the causality conditions for the closed-future process functional $\mathcal{W}$ and for the open-future process functional $\widetilde{\mathcal{W}}$.
\subsubsection{Causality for discrete-time process} 

    Let us first recall the causality condition for the discrete-time process matrices \cite{chiribella09b}: A positive semidefinite operator $\hat{W}\in \mathcal{L}(\otimes_{i=0}^{2N+1}\mathcal{H}^i)$ is causally ordered (or satisfies causality) if, for $k=0,\dots,N$, there exists a set of operators $\hat{\widetilde{W}}^{0: 2k} \in \mathcal{L}(\otimes_{i=0}^{2k}\mathcal{H}^i) $ such that
\begin{align} 
\label{eq:discreteclosedfuture}
\hat{W} &= \hat{\widetilde{W}}^{0: 2N}\!\!\!\otimes \id^{2N+1}, \\ \label{eq:discretecausality}
\Tr_{2k}\hat{\widetilde{W}}^{0: 2k} & = \hat{\widetilde{W}}^{0: 2k-2}\!\!\!\otimes \id^{2k-1},\quad k=1,\dots, N ,  \\ \label{eq:discretenormalization}
    \Tr \hat{\widetilde{W}}^{0:0} & = 1.
\end{align}

In words, calling even times ``inputs'' and odd times ``outputs'', a causally ordered process matrix has to be identity on the last output. This defines a process matrix, $\hat{\widetilde{W}}^{0: 2N}$, which ends with an input space (and can be called ``open future'', in alignment with the terminology introduced for functionals). Tracing out the last input space yield identity on the next-to-last output, leaving again an open-future process ending on the next-to-last input. Such a condition iterates, identifying a set of causally-ordered process matrices that end on an input space, which yields identity on the previous output when traced out. When the iteration reaches the first input, the remaining process matrix $\hat{\widetilde{W}}^{0:0}$ is required to be a normalized density matrix. 

\subsubsection{Causality for the continuous-time processes}

The continuous-time analog of Eq.~\eqref{eq:discretecausality} is given by
\begin{equation}
\label{app:causalityfunctional}
    \int^{\alpha(\tf) = \bar{\alpha}(\tf)}_{\substack{ \alpha(t') = \alpha' \\ \bar{\alpha}(t') = \bar{\alpha}' }} \D[\alpha_{>},\bar{\alpha}_{>}] \ee^{\ii \int_{\ti}^{\tf} \dd t \Im (\dot{\alpha}^*\alpha - \dot{\bar \alpha}^*\bar \alpha  )} \mathcal{W}[\alpha,\bar{\alpha}]  \\
    = \ee^{\ii \int_{\ti}^{t'} \dd t \Im (\dot{\alpha}^*\alpha - \dot{\bar \alpha}^*\bar \alpha )} \widetilde{\mathcal{W}}[\alpha_{<},\bar{\alpha}_{<}]\, \mathrm{id}(\bar\alpha', \alpha'),
\end{equation} 
which we now prove.

First, let us consider a simple yet insightful case where there is
no environmental degrees of freedom. In this case, the process functional
is divisible (Markovian), i.e.,
\begin{equation}
\mathcal{W}[\alpha,\,\bar{\alpha}]={\rho(\alpha(\ti),\bar{\alpha}(\ti))}{\lowtilde{\mathcal{W}}}[\alpha_{>},\,\bar{\alpha}_{>}]{\lowtilde{\mathcal{W}}}[\alpha_{<},\,\bar{\alpha}_{<}]\,\mathrm{id}(\bar{\alpha}(t_{\mathrm{f}}),\,\alpha(t_{\mathrm{f}})).
\end{equation}
where 
\begin{align}
{\lowtilde{\mathcal{W}}}[\alpha_{>},\,\bar{\alpha}_{>}] & :=\mathcal{U}_{0}[\alpha_{>}]\mathcal{U}_{0}^{*}[\bar{\alpha}_{>}] \qquad \text{for } \alpha_{>}:[t',\tf]\rightarrow \mathbb{C},\\
{\lowtilde{\mathcal{W}}}[\alpha_{<},\,\bar{\alpha}_{<}] & :=\mathcal{U}_{0}[\alpha_{<}]\mathcal{U}_{0}^{*}[\bar{\alpha}_{<}] \qquad \text{for } \alpha_{<}:[\ti,t']\rightarrow \mathbb{C}
\end{align}
We note that 
\begin{align}
 & \int_{\alpha_{>}(t^{\prime})=\alpha^{\prime};\,\bar{\alpha}_{>}(t^{\prime})=\bar{\alpha}^{\prime}}^{\alpha(t_{\mathrm{f}})=\bar{\alpha}(t_{\mathrm{f}})}\mathscr{D}[\alpha_{>},\,\bar{\alpha}_{>}]\ee^{\ii\int_{t_{\text{i}}}^{t_{\text{f}}}\dd t\mathrm{Im}(\dot{\alpha}_{>}^*\alpha_{>}-\dot{\bar{\alpha}}_{>}^*\bar{\alpha}_{>})}{\lowtilde{\mathcal{W}}}[\alpha_{>},\,\bar{\alpha}_{>}]\nonumber \\
= & \int\mathrm{d}\alpha_{\text{f}}\int_{\bar{\alpha}_{>}(t^{\prime})=\bar{\alpha}^{\prime}}^{\bar{\alpha}(t_{\text{f}})=\alpha_{\text{f}}}\D[\bar{\alpha}_{>}]\ee^{-\ii\int_{t_{\text{i}}}^{t_{\text{f}}}\dd t\mathrm{Im}(\dot{\bar{\alpha}}_{>}^*\bar{\alpha}_{>})}\mathcal{U}_{0}^{*}[\bar{\alpha}_{>}]\int_{\alpha_{>}(t^{\prime})=\alpha^{\prime}}^{\alpha_{>}(t_{\text{f}})=\alpha_{\text{f}}}\D[\alpha_{>}]\ee^{\ii\int_{t_{\text{i}}}^{t_{\text{f}}}\dd t\mathrm{Im}(\dot{\alpha}_{>}^*\alpha_{>})}\mathcal{U}_{0}[\alpha_{>}]\nonumber \\
= & \int\mathrm{d}\alpha_{\text{f}}\langle{\bar{\alpha}^{\prime}|\hat{U}_{0}^{\dagger}(t_{\text{f}}\leftarrow t^{\prime})|\alpha_{\text{f}}}\rangle\langle{\alpha_{\text{f}}|\hat{U}_{0}(t_{\text{f}}\leftarrow t^{\prime})|\alpha^{\prime}}\rangle\nonumber \\
= & \mathrm{id}(\bar{\alpha}^{\prime},\,\alpha^{\prime})\label{eq:PI-Identity-env-free}
\end{align}
With above formula, we can easily check
\begin{align}
 & \int_{\alpha_{>}(t^{\prime})=\alpha^{\prime};\,\bar{\alpha}_{>}(t^{\prime})=\bar{\alpha}^{\prime}}\mathscr{D}[\alpha_{>},\bar{\alpha}_{>}]\ee^{\ii\int_{t_{\text{i}}}^{t_{\text{f}}}\dd t\mathrm{Im}(\dot{\alpha}^{*}\alpha-\dot{\bar{\alpha}}^{*}\bar{\alpha})}\mathcal{W}[\alpha,\,\bar{\alpha}]\nonumber \\
= & \ee^{\ii\int_{t_{\text{i}}}^{t_{\text{f}}}\dd t\mathrm{Im}(\dot{\alpha}_{<}^*\alpha_{<}-\dot{\bar{\alpha}}_{<}^*\bar{\alpha}_{<})}{\lowtilde{\mathcal{W}}}[\alpha_{<},\,\bar{\alpha}_{<}]\int_{\alpha_{>}(t^{\prime})=\alpha^{\prime};\,\bar{\alpha}_{>}(t^{\prime})=\bar{\alpha}^{\prime}}^{\alpha_{>}(t_{\text{f}})=\bar{\alpha}_{>}(t_{\text{f}})}\mathscr{D}[\alpha_{>},\bar{\alpha}_{>}]\ee^{\ii\int_{t_{\text{i}}}^{t_{\text{f}}}\dd t\mathrm{Im}(\dot{\alpha}_{>}^*\alpha_{>}-\dot{\bar{\alpha}}_{>}^*\bar{\alpha}_{>})}{\lowtilde{\mathcal{W}}}[\alpha_{>},\,\bar{\alpha}_{>}]\nonumber \\
= & \ee^{\ii\int_{t_{\text{i}}}^{t_{\text{f}}}\dd t\mathrm{Im}(\dot{\alpha}_{<}^*\alpha_{<}-\dot{\bar{\alpha}}_{<}^*\bar{\alpha}_{<})}{\lowtilde{\mathcal{W}}}[\alpha_{<},\,\bar{\alpha}_{<}]\mathrm{id}(\bar{\alpha}^{\prime},\,\alpha^{\prime})\label{eq:Causality}
\end{align}
Similar with the proof of positivity, the idea of generalizing to
the generic case is to regard the system-environment as a joint system,
so that the above argument applies again. More precisely,\\
\begin{align}
\widetilde{\mathcal{W}}[\alpha,\,\bar{\alpha}] & =\int\mathrm{d}\beta^{\prime}\mathrm{d}\bar{\beta}^{\prime}\int_{\beta_{<}(t^{\prime})=\beta^{\prime},\,\bar{\beta}_{<}(t^{\prime})=\bar{\beta}^{\prime}}\mathscr{D}[\beta_{<},\,\bar{\beta}_{<}]\ee^{\ii\int_{t_{\text{i}}}^{t_{\text{f}}}\dd t\mathrm{Im}(\dot{\beta}_{<}^*\beta_{<}-\dot{\bar{\beta}}_{<}^*\bar{\beta}_{<})}\nonumber \\
 & \times\mathcal{U}^{\tSE}[\alpha_{<},\,\beta_{<}]\mathcal{U}^{\tSE*}[\bar{\alpha}_{<},\,\bar{\beta}_{<}]\rho_{\text{i}}(\alpha_{\text{i}},\beta_{\text{i}};\bar{\alpha}_{\text{i}},\bar{\beta}_{\text{i}})\nonumber \\
 & \times\int_{\substack{\beta_{>}(t^{\prime})=\beta^{\prime},\,\bar{\beta}_{>}(t^{\prime})=\bar{\beta}^{\prime}}
}^{\beta_{>}(t_{\text{f}})=\bar{\beta}_{>}(t_{\text{f}})}\mathscr{D}[\beta_{>},\,\bar{\beta}_{>}]\ee^{\ii\int_{t_{\text{i}}}^{t_{\text{f}}}\dd t\mathrm{Im}(\dot{\beta}_{>}^*\beta_{>}-\dot{\bar{\beta}}_{>}^*\bar{\beta}_{>})}\mathcal{U}^{\tSE}[\alpha_{>},\,\beta_{>}]\mathcal{U}^{\tSE*}[\bar{\alpha}_{>},\,\bar{\beta}_{>}]\label{eq:W-causal-structure}
\end{align}
where we have use the identity 
\begin{equation}
\int_{\beta(t_{\text{f}})=\bar{\beta}(t_{\text{f}})}\mathscr{D}[\beta,\,\bar{\beta}]=\int\mathrm{d}\beta^{\prime}\mathrm{d}\bar{\beta}^{\prime}\int_{\substack{\beta_{<}(t^{\prime})=\beta^{\prime}\\
\bar{\beta}_{<}(t^{\prime})=\bar{\beta}^{\prime}
}
}\mathscr{D}[\beta_{<},\,\bar{\beta}_{<}]\int_{\beta_{>}(t^{\prime})=\beta^{\prime},\,\bar{\beta}_{>}(t^{\prime})=\bar{\beta}^{\prime}}^{\beta_{>}(t_{\text{f}})=\bar{\beta}_{>}(t_{\text{f}})}\mathscr{D}[\beta_{>},\,\bar{\beta}_{>}]
\end{equation}
Using a similar trick with Eq.~(\ref{eq:PI-Identity-env-free}),
we can show that 
\begin{align}
 & \int_{\alpha_{>}(t^{\prime})=\alpha^{\prime};\,\bar{\alpha}_{>}(t^{\prime})=\bar{\alpha}^{\prime}}^{\alpha_{>}(t_{\text{f}})=\bar{\alpha}_{>}(t_{\text{f}})}\mathscr{D}[\alpha_{>},\bar{\alpha}_{>}]\int_{\substack{\beta_{>}(t^{\prime})=\beta^{\prime},\,\bar{\beta}_{>}(t^{\prime})=\bar{\beta}^{\prime}}
}^{\beta_{>}(t_{\text{f}})=\bar{\beta}_{>}(t_{\text{f}})}\mathscr{D}[\beta_{>},\,\bar{\beta}_{>}]\nonumber \\
\times & \ee^{\ii\int_{t_{\text{i}}}^{t_{\text{f}}}\dd t\mathrm{Im}(\dot{\alpha}_{>}^*\alpha_{>}-\dot{\bar{\alpha}}_{>}^*\bar{\alpha}_{>}+\dot{\beta}_{>}^*\beta_{>}-\dot{\bar{\beta}}_{>}^*\bar{\beta}_{>})}\mathcal{U}^{\tSE}[\alpha_{>},\,\beta_{>}]\mathcal{U}^{\tSE*}[\bar{\alpha}_{>},\,\bar{\beta}_{>}]\nonumber \\
= & \int\mathrm{d}\alpha_{\text{f}}\mathrm{d}\beta_{\text{f}}\int_{\alpha_{>}(t^{\prime})=\alpha^{\prime},\,\beta_{>}(t^{\prime})=\beta^{\prime}}^{\alpha_{>}(t_{\text{f}})=\alpha_{\text{f}},\,\beta_{>}(t_{\text{f}})=\beta_{\text{f}}}\mathscr{D}[\alpha_{>},\,\beta_{>}]\ee^{\ii\int_{t_{\text{i}}}^{t_{\text{f}}}\dd t\mathrm{Im}(\dot{\alpha}_{>}^*\alpha_{>}+\dot{\beta}_{>}^*\beta_{>})}\mathcal{U}^{\tSE}[\alpha_{>},\,\beta_{>}]\nonumber \\
\times & \int_{\bar{\alpha}_{>}(t^{\prime})=\bar{\alpha}^{\prime},\,\bar{\beta}_{>}(t^{\prime})=\bar{\beta}^{\prime}}^{\bar{\alpha}_{>}(t_{\text{f}})=\alpha_{\text{f}},\,\bar{\beta}_{>}(t_{\text{f}})=\bar{\beta}_{\text{f}}}\mathscr{D}[\alpha_{>},\,\beta_{>}]\ee^{-\ii\int_{t_{\text{i}}}^{t_{\text{f}}}\dd t\mathrm{Im}(\dot{\bar{\alpha}}_{>}^*\bar{\alpha}_{>}+\dot{\bar{\beta}}_{>}^*\bar{\beta}_{>})}\mathcal{U}^{\tSE*}[\bar{\alpha}_{>},\,\bar{\beta}_{>}]\nonumber \\
= & \int\mathrm{d}\alpha_{\text{f}}\mathrm{d}\beta_{\text{f}}
\langle{\alpha_{\text{f}},\,\beta_{\text{f}}|\hat{U}_{\text{SE}}(t_{\text{f}}\leftarrow t^{\prime})|\alpha^{\prime},\,\beta^{\prime}}\rangle
\matrixel{\bar{\alpha}^{\prime},\,\bar{\beta}^{\prime}}{\hat{U}_{\text{SE}}^{\dagger}(t_{\text{f}}\leftarrow t^{\prime})}{\alpha_{\text{f}},\,\beta_{\text{f}}}\nonumber \\
= & \langle{\bar{\alpha}^{\prime},\,\bar{\beta}^{\prime}|\hat{U}_{\text{SE}}^{\dagger}(t_{\text{f}}\leftarrow t^{\prime})\hat{U}_{\text{SE}}(t_{\text{f}}\leftarrow t^{\prime})|\alpha^{\prime},\,\beta^{\prime}}\rangle\nonumber \\
= & \mathrm{id}(\bar{\alpha}^{\prime},\,\alpha^{\prime})\mathrm{id}(\bar{\beta}^{\prime},\,\beta^{\prime})\label{eq:PI-idetity}
\end{align}
Integrating both sides of Eq.~(\ref{eq:W-causal-structure}) 
\begin{align}
 & \int_{\alpha_{>}(t^{\prime})=\alpha^{\prime};\,\bar{\alpha}_{>}(t^{\prime})=\bar{\alpha}^{\prime}}\mathscr{D}[\alpha_{>},\bar{\alpha}_{>}]\ee^{\ii\int_{t_{\text{i}}}^{t_{\text{f}}}\dd t\mathrm{Im}(\dot{\alpha}^{*}\alpha-\dot{\bar{\alpha}}^{*}\bar{\alpha})}\mathcal{W}[\alpha,\,\bar{\alpha}]\nonumber \\
= & \ee^{\ii\int_{t_{\text{i}}}^{t_{\text{f}}}\dd t\mathrm{Im}(\dot{\alpha}_{<}^*\alpha_{<}-\dot{\bar{\alpha}}_{<}^*\bar{\alpha}_{<})}\int\mathrm{d}\beta^{\prime}\mathrm{d}\bar{\beta}^{\prime}\int_{\beta_{<}(t^{\prime})=\beta^{\prime},\,\bar{\beta}_{<}(t^{\prime})=\bar{\beta}^{\prime}}\mathscr{D}[\beta_{<},\,\bar{\beta}_{<}]\ee^{\ii\int_{t_{\text{i}}}^{t_{\text{f}}}\dd t\mathrm{Im}(\dot{\beta}_{<}^*\beta_{<}-\dot{\bar{\beta}}_{<}^*\bar{\beta}_{<})}\nonumber \\
\times & \mathcal{U}^{\tSE}[\alpha_{<},\,\beta_{<}]\mathcal{U}^{\tSE*}[\bar{\alpha}_{<},\,\bar{\beta}_{<}]\rho_{\text{i}}(\alpha_{\text{i}},\beta_{\text{i}};\bar{\alpha}_{\text{i}},\bar{\beta}_{\text{i}})\nonumber \\
\times & \int_{\alpha_{>}(t^{\prime})=\alpha^{\prime};\,\bar{\alpha}_{>}(t^{\prime})=\bar{\alpha}^{\prime}}^{\alpha_{>}(t_{\text{f}})=\bar{\alpha}_{>}(t_{\text{f}})}\mathscr{D}[\alpha_{>},\bar{\alpha}_{>}]\int_{\substack{\beta_{>}(t^{\prime})=\beta^{\prime},\,\bar{\beta}_{>}(t^{\prime})=\bar{\beta}^{\prime}}
}^{\beta_{>}(t_{\text{f}})=\bar{\beta}_{>}(t_{\text{f}})}\mathscr{D}[\beta_{>},\,\bar{\beta}_{>}]\ee^{\ii\int_{t_{\text{i}}}^{t_{\text{f}}}\dd t\mathrm{Im}(\dot{\alpha}_{>}^*\alpha_{>}-\dot{\bar{\alpha}}_{>}^*\bar{\alpha}_{>}+\dot{\beta}_{>}^*\beta_{>}-\dot{\bar{\beta}}_{>}^*\bar{\beta}_{>})}\nonumber\\
\times & \mathcal{U}^{\tSE}[\alpha_{>},\,\beta_{>}]\mathcal{U}^{\tSE*}[\bar{\alpha}_{>},\,\bar{\beta}_{>}]
\end{align}
The last two lines can be evaluated using Eq.~(\ref{eq:PI-idetity}),
which leads to 
\begin{align}
 & \int_{\alpha_{>}(t^{\prime})=\alpha^{\prime};\,\bar{\alpha}_{>}(t^{\prime})=\bar{\alpha}^{\prime}}\mathscr{D}[\alpha_{>},\bar{\alpha}_{>}]\ee^{\ii\int_{t_{\text{i}}}^{t_{\text{f}}}\dd t\mathrm{Im}(\dot{\alpha}^{*}\alpha-\dot{\bar{\alpha}}^{*}\bar{\alpha})}\mathcal{W}[\alpha,\,\bar{\alpha}]\nonumber \\
= & \ee^{\ii\int_{t_{\text{i}}}^{t_{\text{f}}}\dd t\mathrm{Im}(\dot{\alpha}_{<}^*\alpha_{<}-\dot{\bar{\alpha}}_{<}^*\bar{\alpha}_{<})}\mathrm{id}(\bar{\alpha}^{\prime},\,\alpha^{\prime})\int\mathrm{d}\beta^{\prime}\mathrm{d}\bar{\beta}^{\prime}\mathrm{id}(\bar{\beta}^{\prime},\,\beta^{\prime})\int_{\beta_{<}(t^{\prime})=\beta^{\prime},\,\bar{\beta}_{<}(t^{\prime})=\bar{\beta}^{\prime}}\mathscr{D}[\beta_{<},\,\bar{\beta}_{<}]\ee^{\ii\int_{t_{\text{i}}}^{t_{\text{f}}}\dd t\mathrm{Im}(\dot{\beta}_{<}^*\beta_{<}-\dot{\bar{\beta}}_{<}^*\bar{\beta}_{<})}\nonumber \\
\times & \mathcal{U}^{\tSE}[\alpha_{<},\,\beta_{<}]\mathcal{U}^{\tSE*}[\bar{\alpha}_{<},\,\bar{\beta}_{<}]\rho_{\text{i}}(\alpha_{\text{i}},\beta_{\text{i}};\bar{\alpha}_{\text{i}},\bar{\beta}_{\text{i}})\nonumber \\
= & \ee^{\ii\int_{t_{\text{i}}}^{t_{\text{f}}}\dd t\mathrm{Im}(\dot{\alpha}_{<}^*\alpha_{<}-\dot{\bar{\alpha}}_{<}^*\bar{\alpha}_{<})}\widetilde{\mathcal{W}}[\alpha_{<},\,\bar{\alpha}_{<}]\mathrm{id}(\bar{\alpha}^{\prime},\,\alpha^{\prime})
\end{align}

Let us now discuss the normalization condition. In the discrete-time case, the normalization condition~\eqref{eq:discretenormalization} can be relaxed since it can be always satisfied upon redefining $\hat{\widetilde{W}}^{0:0}/\mathrm{Tr}\hat{\widetilde{W}}^{0:0}$.  To see the analog of Eq.~\eqref{eq:discretenormalization} in the continuous-time case, we set $t'=t_\mathrm{i}$ in the causality condition and obtain
\begin{equation}
\widetilde{\mathcal{W}}(\alpha_\mathrm{i},\,\bar{\alpha}_\mathrm{i})\mathrm{id}(\bar{\alpha}_\mathrm{i},\,\alpha_\mathrm{i})=\int_{\alpha(t_\mathrm{i})=\alpha_\mathrm{i};\,\bar{\alpha}(t_\mathrm{i})=\bar{\alpha}_\mathrm{i}} \D[\alpha,\bar{\alpha}]  \ee^{\ii \int_{t_{\text{i}}}^{t_{\text{f}}} \dd t \Im (\dot{\alpha}^*\alpha - \dot{\bar \alpha}^*\bar \alpha )} {\mathcal{W}}[\alpha,\bar{\alpha}],
\label{eq:W-alpha-i-def}
\end{equation}
One can view Eq.~\eqref{eq:W-alpha-i-def} as the definition of $\widetilde{\mathcal{W}}(\alpha_\mathrm{i},\,\bar{\alpha}_\mathrm{i})$. Since $\mathrm{id}(\bar\alpha_i, \alpha_i)$ vanishes nowhere, $\widetilde{\mathcal{W}}(\alpha_\mathrm{i},\,\bar{\alpha}_\mathrm{i})$ is well-defined in the sense that it can be extracted from Eq.~\eqref{eq:W-alpha-i-def} dividing both sides by $\mathrm{id}(\bar\alpha_i, \alpha_i)$.

From Eq.~\eqref{eq:W-alpha-i-def}, it is clear that the analog of Eq.~\eqref{eq:discretenormalization} becomes 
\begin{equation} \label{eq:functionalnormalization}
    \int \dd \alpha_\mathrm{i} \dd \bar{\alpha}_\mathrm{i}\widetilde{\mathcal{W}}[\alpha_\mathrm{i},\,\bar{\alpha}_\mathrm{i}]\mathrm{id}(\bar{\alpha}_\mathrm{i},\,\alpha_\mathrm{i})=\int \D[\alpha,\bar{\alpha}]  \ee^{\ii \int_{t_{\text{i}}}^{t_{\text{f}}} \dd t \Im (\dot{\alpha}^*\alpha - \dot{\bar \alpha}^*\bar \alpha )} {\mathcal{W}}[\alpha,\bar{\alpha}]=1
\end{equation}
In the Hamiltonian-based model, we note that the state at the final time $\tf$ can be represented as
\begin{equation}
    \varrho^{\tS}_\mathrm{f}(\alpha_\mathrm{f},\bar{\alpha}_\mathrm{f}):=\int_{\alpha(\tf) = \alpha_\mathrm{f}, \bar \alpha(\tf) = \bar \alpha_\mathrm{f}}\D[\alpha,\bar{\alpha}]  \ee^{\ii \int_{t_{\text{i}}}^{t_{\text{f}}} \dd t \Im (\dot{\alpha}^*\alpha - \dot{\bar \alpha}^*\bar \alpha )} \widetilde{{\mathcal{W}}}[\alpha,\bar{\alpha}]
\end{equation}
Clearly, the normalization condition for the continuous-time case is physically equivalent to the normalization of the final state. In fact, integrating both sides of the causality condition $\int \dd \alpha^{\prime} \dd \bar{\alpha}^{\prime} \int_{\alpha_{<}(t^{\prime})=\alpha^{\prime};\,\bar{\alpha}_{<}(t^{\prime})=\bar{\alpha}^{\prime}} \mathscr{D}[\alpha_{<},\bar{\alpha}_{<}]$, we find 
\begin{equation}
    \mathrm{Tr}\hat{\varrho}^{\tS}_{t^{'}}:=\int \D[\alpha_<,\bar{\alpha_<}]  \ee^{\ii \int_{t_{\text{i}}}^{t_{\text{f}}} \dd t \Im (\dot{\alpha_<}^*\alpha_< - \dot{\bar \alpha}_<^*\bar \alpha_< )} {\mathcal{W}}[\alpha_<,\bar{\alpha}_<]=\int \D[\alpha,\bar{\alpha}]  \ee^{\ii \int_{t_{\text{i}}}^{t_{\text{f}}} \dd t \Im (\dot{\alpha}^*\alpha - \dot{\bar \alpha}^*\bar \alpha )} {\mathcal{W}}[\alpha,\bar{\alpha}]=1,
\end{equation}
that is, the normalization condition also fixes the normalization of any state at any intermediate time.

\subsection{Causality, trace-preserving property for the open-boundary process functional and the definition of Markovianity} \label{app:openboundarycausality}
We have introduced open-boundary processes $\oW$ for the particular case in which the initial system state is uncorrelated from the environment, so that $\widetilde{W}[\alpha,\bar{\alpha}] =  \oW[\alpha,\bar{\alpha}]\rho^{\tS}(\alpha(\ti),\bar\alpha(\ti))$ 
and the open-boundary process functional is constructed from the Hamiltonian model as
\begin{equation} \label{app:openW}
    \oW[\alpha,\bar{\alpha}] 
\coloneqq \int\!\!\!\D[\beta,\bar{\beta}] \ee^{\ii \int_{t_{\text{i}}}^{t_{\text{f}}} \dd t \Im(\dot{\beta}^*\beta - \dot{\bar{\beta}}^*\bar{\beta})} \mathrm{id}(\bar{\beta}(\tf),\beta(\tf)) \mathcal{U}^{\tSE}[\alpha,\beta]\mathcal{U}^{\tSE*}[\bar{\alpha},\bar{\beta}] \rho^{\tE}(\beta(\ti), \bar{\beta}(\ti)).
\end{equation}
Imposing that $\widetilde{W}[\alpha,\bar{\alpha}] =  \oW[\alpha,\bar{\alpha}]\rho^{\tS}(\alpha(\ti),\bar\alpha(\ti))$ satisfies causality for all $\hat\rho_\mathrm{i}$ immediately yields the causality condition for open-boundary functionals:
\begin{equation}\label{app:opencausality}
     \int_{\substack{ \alpha(t') = \alpha' \\ \bar{\alpha}(t') = \bar{\alpha}' }} \D[\alpha_{>},\bar{\alpha}_{>}] \ee^{\ii \int_{t'}^{\tf} \dd t \Im (\dot{\alpha}^*\alpha - \dot{\bar \alpha}^*\bar \alpha  )} \oW[\alpha,\bar{\alpha}]  \mathrm{id}(\bar\alpha(\tf), \alpha(\tf)) = \oW[\alpha_{<},\bar{\alpha}_{<}]\, \mathrm{id}(\bar\alpha', \alpha'),\,\forall t^{'}\in[\ti,\tf].
\end{equation}

Let us now consider normalization for the open-boundary process functional. Recall that, as discussed in the main text, the open-boundary functional relates directly to the Feynman-Vernon influence functional, Eq.~\eqref{eq:W-FV-Phase}, so it generates the propagators for density matrices:
\begin{align} \label{app:rhoitorhof}
    \varrho^{\tS}_\mathrm{f}(\alpha_\mathrm{f},\bar{\alpha}_\mathrm{f}) &= \int \dd \alpha_\mathrm{i} \dd \bar\alpha_\mathrm{i}
    \Gamma(\alpha_\mathrm{f},\bar{\alpha}_\mathrm{f}; \alpha_\mathrm{i},\bar{\alpha}_\mathrm{i} ) \rho_\mathrm{i}(\alpha_\mathrm{i},\bar{\alpha}_\mathrm{i} ) \\ \label{app:propagator}
    \Gamma(\alpha_\mathrm{f},\bar{\alpha}_\mathrm{f}; \alpha_\mathrm{i},\bar{\alpha}_\mathrm{i}) & \coloneqq  \int^{\alpha(\tf) = \alpha_\mathrm{f}, \bar \alpha(\ti) = \bar \alpha_\mathrm{i}}_{\alpha(\ti)=\alpha_\mathrm{i}, \bar \alpha(\ti) = \bar \alpha_\mathrm{i}}\
    \!\D[\alpha,\bar{\alpha}] \oW[\alpha,\bar\alpha],
\end{align}
where
\begin{equation}
    \hat\varrho^{\tS}_\mathrm{f} = \Tr_E\left[\hat{U}^{\tSE}(\tfti)\hat \rho^{\tS}_\mathrm{i}\otimes \hat{\rho}^{\tE} \hat{U}^{\tSE \dagger}(\tfti) \right], \qquad \hat{U}^{\tSE}(\tfti) \coloneqq \mathcal{T}\ee^{-\ii \int_{\ti}^{\tf} \dd t \, \hat H^{\tSE}}
\end{equation}
This implies the trace-preserving condition, $\Tr\hat\varrho^{\tS}_\mathrm{f} = \Tr\hat\rho^{\tS}_\mathrm{i}$,  $\forall \hat\rho_\mathrm{i}$, which, in coherent state representation, reads
\begin{equation}
    \int \dd\alpha_\mathrm{f} \dd \bar\alpha_\mathrm{f} \,\varrho^{\tS}_\mathrm{f}(\alpha_\mathrm{f},\bar{\alpha}_\mathrm{f}) \mathrm{id}(\bar{\alpha}_\mathrm{f},\alpha_\mathrm{f}) 
    = \int \dd \alpha_\mathrm{i} \dd \bar\alpha_\mathrm{i} \,
     \rho^{\tS}(\alpha_\mathrm{i},\bar{\alpha}_\mathrm{i} ) \mathrm{id}(\bar{\alpha}_\mathrm{i},\alpha_\mathrm{i}) \qquad \forall \hat\rho.
\end{equation}
This, substituted into Eq.~\eqref{app:rhoitorhof} and then Eq.~\eqref{app:propagator}, becomes
\begin{align} \label{eq:TP}
    \int \dd \alpha_\mathrm{f} \dd \bar\alpha_\mathrm{f}
    \Gamma (\alpha_\mathrm{f},\bar{\alpha}_\mathrm{f}; \alpha_\mathrm{i},\bar{\alpha}_\mathrm{i} ) \mathrm{id}(\bar{\alpha}_\mathrm{f},\alpha_\mathrm{f}) & = \mathrm{id}(\bar{\alpha}_\mathrm{i},\alpha_\mathrm{i}) ,\\ \label{eq:functionalTP}
    \int_{\substack{ \alpha(\ti) = \alpha_\mathrm{i} \\ \bar{\alpha}(\ti) = \bar{\alpha}_\mathrm{i} }} 
   \!\!\! \D[\alpha,\bar{\alpha}]  \oW[\alpha,\bar\alpha] \mathrm{id}(\bar{\alpha}_\mathrm{f},\alpha_\mathrm{f})&= \mathrm{id}(\bar{\alpha}_\mathrm{i},\alpha_\mathrm{i}).
\end{align}

Eq.~\eqref{eq:functionalTP} characterises the open-boundary process functionals that generate trace-preserving propagators. To see how this relates to causality and normalisation, take the limit $t'\rightarrow \ti$  in the causality condition Eq.~\eqref{app:opencausality} to obtain
\begin{equation}\label{app:initialboundary}
     \int_{\substack{ \alpha(\ti) = \alpha_\mathrm{i} \\ \bar{\alpha}(\ti) = \bar{\alpha}_\mathrm{i} }} \D[\alpha,\bar{\alpha}] \ee^{\ii \int_{t_\mathrm{i}}^{\tf} \dd t \Im (\dot{\alpha}^*\alpha - \dot{\bar \alpha}^*\bar \alpha  )} \oW[\alpha,\bar{\alpha}]  \mathrm{id}(\bar\alpha(\tf), \alpha(\tf)) = \oW(\alpha_\mathrm{i},\bar{\alpha}_\mathrm{i})\, \mathrm{id}(\bar\alpha_\mathrm{i}, \alpha_\mathrm{i}).
\end{equation}
Similar to $\widetilde{\mathcal{W}}(\alpha_\mathrm{i},\,\bar{\alpha}_\mathrm{i})$, the above equation can be viewed as the definition of $\oW(\alpha_i,\bar{\alpha}_i)$. Since $\mathrm{id}(\bar\alpha_i, \alpha_i)$ vanishes nowhere, $\oW(\alpha_i,\bar{\alpha}_i)$ is well-defined in the sense that it can be extracted from Eq.~\eqref{app:initialboundary} dividing both sides by $\mathrm{id}(\bar\alpha_i, \alpha_i)$. Then, the trace-preserving condition~\eqref{eq:functionalTP} can be obtained from the causality condition upon imposing the following condition
\begin{equation} \label{app:initialTP}
\oW(\alpha_\mathrm{i},\bar{\alpha}_\mathrm{i}) = 1.
\end{equation}

In the main text, we define Markovianity on the open-boundary process functionals satisfying the following recursive properties $\forall t^{'}\in[\ti,\tf]$: (a) Divisibility, i.e., ${\lowtilde{\mathcal{W}}}[\alpha,\bar{\alpha}] = {\lowtilde{\mathcal{W}}}{}_{<}[\alpha_{<},\bar{\alpha}_{<}]{\lowtilde{\mathcal{W}}}{}_{>}[\alpha_{>},\bar{\alpha}_{>}]$.
(b) The open-boundary causality condition: ${\lowtilde{\mathcal{W}}}{}_{<}[\alpha_{<},\bar{\alpha}_{<}]$ and ${\lowtilde{\mathcal{W}}}{}_{>}[\alpha_{>},\bar{\alpha}_{>}]$ are positive-semi-definite and  satisfy  Eq.~\eqref{app:opencausality}. (c) The trace-preserving properties, i.e.,
\begin{equation}
    \oW{}_{<}(\alpha(\ti), \bar \alpha(\ti)) = \oW{}_{> }(\alpha(t'), \bar \alpha(t')) = 1.
\end{equation}
(d) Recursive divisibility: $\oW{}_{<}$ and $\oW{}_{>}$ can be further divided, namely they satisfy properties (a)--(d) in the respective domain intervals.

Note that the above recursive definition is equivalent to stating that, for any sequence of times $\ti<t_1<\dots<t_{N-1}<\tf$, the open-boundary process functional can be divided as
\begin{equation}
    \oW[\alpha,\bar\alpha] = \prod_{j=1}^N \oW{}_{j}[\alpha_{(t_{j-1},t_j)},\bar\alpha_{(t_{j-1},t_j)}],
\end{equation}
where $\alpha_{(t_{j-1},t_j)}$, $\bar\alpha_{(t_{j-1},t_j)}$ are functions over the interval $(t_{j-1},t_j)$ and $\oW{}_{j}$ are valid open-boundary process functionals.

\section{Recover the discrete-time formalism\label{app:disc-time}}
\subsection{The Born rule}
{Let us note first that the identity evolution is represented by a unit open-boundary process functional, $\lowtilde{\mathcal{W}}[\alpha,\bar{\alpha}] = 1$. This is simply because the identity unitary is generated by the null Hamiltonian, $\id = \hat{U}(\tfti) = \ee^{-\ii \hat{H}_0(\tf-\ti)}$ for $\hat{H}_0 = 0$, which corresponds to $\lowtilde{\mathcal{W}}[\alpha,\bar{\alpha}]=\mathcal{U}_0[\alpha]\mathcal{U}^*_0[\bar\alpha]$, with $\mathcal{U}_0[\alpha] = \ee^{-\ii\int_{\ti}^{\tf} \dd t H_0(\alpha)} = 1$. (More generally, any fully-degenerate Hamiltonian, $\hat{H}_0 = E_0 \id$ for some $E_0\in \mathbb{R}$, results in $\lowtilde{\mathcal{W}}[\alpha,\bar{\alpha}] = 1$.) 

Formally, the discrete-time formalism is recovered when operations and process act on complementary intervals. Having in mind the above characterization of trivial evolution, this is realized by operation and process functionals that are constant on interleaving intervals:
\begin{align} \label{app:Mintermittent}
\mathcal{M}_r[\alpha,\bar\alpha]
& =\mathcal{M}_r[\{\alpha_{(t_{2k},\,t_{2k+1})}\}_{k=0}^{N},\,\{\bar{\alpha}_{(t_{2k},\,t_{2k+1})}\}_{k=0}^{N}]\\ \label{app:Wintermittent}
\widetilde{\mathcal{W}}[\alpha,\bar{\alpha}]
& =\widetilde{\mathcal{W}}[\alpha(\ti),\{\alpha_{(t_{2k+1},\,t_{2k+2})}\}_{k=0}^{N-1},\,\bar\alpha(\ti), \{\bar{\alpha}_{(t_{2k+1},\,t_{2k+2})}\}_{k=0}^{N-1}],
\end{align}
with $\ti=t_0$, $\tf=t_{2N+1}$, and where $\alpha_{(t_{\ell},\,t_{\ell+1})}$, $\bar\alpha_{(t_{\ell},\,t_{\ell+1})}$, denote functions defined only
on the time interval $(t_{\ell},\,t_{\ell+1})$. The dependence of the process functional on the initial-time variables $\alpha(\ti),\bar\alpha(\ti)$ accounts for possible initial system-environment correlations, but we will suppress it in the following.

Physically, the above condition arises when the operations act only on time intervals that are much shorter than the relevant time scales of the process.  At the level of the Hamiltonian model, we can enforce operations that only act on the desired intervals by setting $\hat{H}^\tSM(t) = 0$ for $t\in [t_{2k+1}, t_{2k+2}]$, $k=0,\dots,N-1$. Let us denote with $\Delta_k \coloneqq |t_{2k+1}-t_{2k}|$ the length of the operation intervals. A sufficient condition for system-environment dynamics to be negligible over those interval is when, in operator norm, $\|\hat{H}^\tSE(t)\|\Delta_k \ll 1$ for $t\in [t_{2k+1}, t_{2k+2}]$. The evolution operator over those intervals can then be approximated as\footnote{For unbounded Hamiltonians, we can relax this by requiring that only the projection of $\hat{H}^\tSE$ on a finite-energy subspace satisfies this condition, which still leads to approximately trivial dynamics as long as states remain within that subspace. Note that these are sufficient, not necessary, conditions for Eq.~\eqref{app:Wintermittent}. An environment can also have fast internal evolution that is irrelevant for the reduced process dynamics.}
\begin{equation}
    \hat{U}(\tfti[t_{2k+2}][t_{2k+2}]) = \mathcal{T}\ee^{- \ii \int_{t_{2k+1}}^{t_{2k+2}}\hat{H}^\tSE(t)} \approx \id.
\end{equation}

Under the above conditions, } it can be readily recognized that
\begin{align}
\mathcal{U}^{\tSM}(\alpha,\,\beta) & =\prod_{k=0}^{N}\mathcal{U}^{\tSM}(\alpha_{(t_{2k},\,t_{2k+1})},\,\gamma_{(t_{2k},\,t_{2k+1})}),\\
\mathcal{U}^{\tSE}(\alpha,\,\beta) & =\prod_{k=0}^{N-1}\mathcal{U}^{\tSE}(\alpha_{(t_{2k+1},\,t_{2k+2})},\,\beta_{(t_{2k+1},\,t_{2k+2})}).
\end{align}
where
\begin{align}
\mathcal{U}^{\tSM}(\alpha_{(t_{2k},\,t_{2k+1})},\,\gamma_{(t_{2k},\,t_{2k+1})}) & :=\ee{}^{-\ii\int_{t_{2k}}^{t_{2k+1}}\dd tH^{\tSM}(\alpha,\gamma)}\\
\mathcal{U}^{\tSE}(\alpha_{(t_{2k+1},\,t_{2k+2})},\,\beta_{(t_{2k+1},\,t_{2k+2})}) & :=\ee{}^{-\ii\int_{t_{2k+1}}^{t_{2k+2}}\dd tH^{\tSE}(\alpha,\,\beta)}
\end{align}
Upon substituting into the definitions of the operation and process
functional, Eqs.~(\ref{eq:Wfunctional},\ref{eq:Mfunctional}) in the main text, 
we obtain Eqs.~\eqref{app:Mintermittent} and \eqref{app:Wintermittent} above.

Introducing the notation
\begin{equation}
    \mathcal{E}[\alpha,\,\bar{\alpha}] \coloneqq 
    \ee^{\ii\int_{t_{\text{i}}}^{t_{\text{f}}}\dd t\mathrm{Im}(\dot{\alpha}^{*}\alpha-\dot{\bar{\alpha}}^{*}\bar{\alpha})},
\end{equation}
we now analyze the behavior
of the functionals $\mathcal{M}[\alpha,\,\bar{\alpha}]$, $\widetilde{\mathcal{W}}[\alpha,\,\bar{\alpha}]$
and $\mathcal{E}[\alpha,\,\bar{\alpha}]$ when the evolution is frozen
alternately, as described above.
Observe that the exponential factor has the product structure
\begin{equation}
\mathcal{E}[\alpha,\,\bar{\alpha}]=\mathcal{E}[\{\alpha_{(t_{2k},\,t_{2k+1})}\}_{k=0}^{N-1},\,\{\bar{\alpha}_{(t_{2k},\,t_{2k+1})}\}_{k=0}^{N-1}]\mathcal{E}[\{\alpha_{(t_{2k+1},\,t_{2k+2})}\}_{k=0}^{N-1},\,\{\bar{\alpha}_{(t_{2k+1},\,t_{2k+2})}\}_{k=0}^{N-1}]
\end{equation}
where 
\begin{align}
\mathcal{E}[\{\alpha_{(t_{2k},\,t_{2k+1})}\}_{k=0}^{N},\,\{\bar{\alpha}_{(t_{2k},\,t_{2k+1})}\}_{k=0}^{N}] & :=\prod_{k=0}^{N}\mathcal{E}[\alpha_{(t_{2k},\,t_{2k+1})},\,\bar{\alpha}_{(t_{2k},\,t_{2k+1})}]\\
\mathcal{E}[\{\alpha_{(t_{2k+1},\,t_{2k+2})}\}_{k=0}^{N-1},\,\{\bar{\alpha}_{(t_{2k+1},\,t_{2k+2})}\}_{k=0}^{N-1}] & :=\prod_{k=0}^{N-1}\mathcal{E}[\alpha_{(t_{2k+1},\,t_{2k+2})},\,\bar{\alpha}_{(t_{2k+1},\,t_{2k+2})}]
\end{align}
and 
\begin{equation}
\mathcal{E}[\alpha_{(t_{\ell},\,t_{\ell+1})},\,\bar{\alpha}_{(t_{\ell},\,t_{\ell+1})}]:=\ee^{\ii\int_{t_{\ell}}^{t_{\ell+1}}\dd t\mathrm{Im}(\dot{\alpha}^{*}\alpha-\dot{\bar{\alpha}}^{*}\bar{\alpha})}
\end{equation}
On the other hand, the path integral can be decomposed into 
\begin{align}
\int \mathscr{D}[\alpha,\,\bar{\alpha}] 
 & = \int \dd^{2N+2} \boldsymbol{\alpha} \; \dd^{2N+2} \boldsymbol{\bar{\alpha}}
 \\
 & \times \left(\prod_{k=0}^{N}\int_{\alpha(t_{2k})=\alpha_{2k},\,\bar{\alpha}(t_{2k})=\bar{\alpha}_{2k}}^{\alpha(t_{2k+1})=\alpha_{2k+1},\,\bar{\alpha}(t_{2k+1})=\bar{\alpha}_{2k+1}}\mathscr{D}[\alpha_{(t_{2k},\,t_{2k+1})},\,\bar{\alpha}_{(t_{2k},\,t_{2k+1})}]\right)\nonumber \\
 & \times\left(\prod_{k=0}^{N-1}\int_{\alpha(t_{2k+1})=\alpha_{2k+1},\,\bar{\alpha}(t_{2k+1})=\bar{\alpha}_{2k+1}}^{\alpha(t_{2k+2})=\alpha_{2k+2},\,\bar{\alpha}(t_{2k+2})=\bar{\alpha}_{2k+2}}\mathscr{D}[\alpha_{(t_{2k+1},\,t_{2k+2})},\,\bar{\alpha}_{(t_{2k+1},\,t_{2k+2})}]\right)\nonumber .
\end{align} 
with $\boldsymbol{\alpha}\coloneqq \{\alpha_{j}\}_{j=0}^{2N+1}$ and $\bar{\boldsymbol{\alpha}}\coloneqq \{\bar\alpha_{j}\}_{j=0}^{2N+1}$.
With above observations, the functional Born rule~\eqref{eq:functionalBorn} in the main text can be rewritten
as
\begin{align}
\mathrm{Pr}[r] & =\int \dd^{2N+2} \boldsymbol{\alpha} \; \dd^{2N+2} \boldsymbol{\bar{\alpha}} \nonumber \\
 & \times\left(\prod_{k=0}^{N}\int_{\alpha(t_{2k})=\alpha_{2k},\,\bar{\alpha}(t_{2k})=\bar{\alpha}_{2k}}^{\alpha(t_{2k+1})=\alpha_{2k+1},\,\bar{\alpha}(t_{2k+1})=\bar{\alpha}_{2k+1}}\mathscr{D}[\alpha_{(t_{2k},\,t_{2k+1})},\,\bar{\alpha}_{(t_{2k},\,t_{2k+1})}]\right)\mathcal{E}[\{\alpha_{(t_{2k},\,t_{2k+1})}\}_{k=0}^{N},\,\{\bar{\alpha}_{(t_{2k},\,t_{2k+1})}\}_{k=0}^{N}]\nonumber \\
 & \times\mathcal{M}_r[\{\alpha_{(t_{2k},\,t_{2k+1})}\}_{k=0}^{N},\,\{\bar{\alpha}_{(t_{2k},\,t_{2k+1})}\}_{k=0}^{N}]\nonumber \\
 & \times\left(\prod_{k=0}^{N-1}\int_{\alpha(t_{2k+1})=\alpha_{2k+1},\,\bar{\alpha}(t_{2k+1})=\bar{\alpha}_{2k+1}}^{\alpha(t_{2k+2})=\alpha_{2k+2},\,\bar{\alpha}(t_{2k+2})=\bar{\alpha}_{2k+2}}\mathscr{D}[\alpha_{(t_{2k+1},\,t_{2k+2})},\,\bar{\alpha}_{(t_{2k+1},\,t_{2k+2})}]\right) \nonumber \\
 & \times
 \mathcal{E}[\{\alpha_{(t_{2k+1},\,t_{2k+2})}\}_{k=0}^{N-1},\,\{\bar{\alpha}_{(t_{2k+1},\,t_{2k+2})}\}_{k=0}^{N-1}] \widetilde{\mathcal{W}}[\{\alpha_{(t_{2k+1},\,t_{2k+2})}\}_{k=0}^{N-1},\{\bar{\alpha}_{(t_{2k+1},\,t_{2k+2})}\}_{k=0}^{N-1}] \nonumber \\
 & \times
 \mathrm{id}(\bar{\alpha}_{2N+1},\,\alpha_{2N+1})
\end{align}
We define 
\begin{multline}
 M(\boldsymbol{\alpha},\,\bar{\boldsymbol{\alpha}})
:= \left(\prod_{k=0}^{N}\int_{\alpha(t_{2k})=\alpha_{2k},\,\bar{\alpha}(t_{2k})=\bar{\alpha}_{2k}}^{\alpha(t_{2k+1})=\alpha_{2k+1},\,\bar{\alpha}(t_{2k+1})=\bar{\alpha}_{2k+1}}\mathscr{D}[\alpha_{(t_{2k},\,t_{2k+1})},\,\bar{\alpha}_{(t_{2k},\,t_{2k+1})}]\right) \\
\times \mathcal{E}[\{\alpha_{(t_{2k},\,t_{2k+1})}\}_{k=0}^{N},\,\{\bar{\alpha}_{(t_{2k},\,t_{2k+1})}\}_{k=0}^{N}]  \mathcal{M}_r[\{\alpha_{(t_{2k},\,t_{2k+1})}\}_{k=0}^{N},\,\{\bar{\alpha}_{(t_{2k},\,t_{2k+1})}\}_{k=0}^{N}],
 \label{app:Mreconstruction}
\end{multline}

\begin{multline}
 W(\boldsymbol{\alpha},\,\bar{\boldsymbol{\alpha}})
 :=  \left(\prod_{k=0}^{N-1}\int_{\alpha(t_{2k+1})=\alpha_{2k+1},\,\bar{\alpha}(t_{2k+1})=\bar{\alpha}_{2k+1}}^{\alpha(t_{2k+2})=\alpha_{2k+2},\,\bar{\alpha}(t_{2k+2})=\bar{\alpha}_{2k+2}}\mathscr{D}[\alpha_{(t_{2k+1},\,t_{2k+2})},\,\bar{\alpha}_{(t_{2k+1},\,t_{2k+2})}]\right) \\ 
\times \mathcal{E}[\{\alpha_{(t_{2k+1},\,t_{2k+2})}\}_{k=0}^{N-1},\,\{\bar{\alpha}_{(t_{2k+1},\,t_{2k+2})}\}_{k=0}^{N-1}]
 \widetilde{\mathcal{W}}[\{\alpha_{(t_{2k+1},\,t_{2k+2})}\}_{k=0}^{N-1},\,\{\bar{\alpha}_{(t_{2k+1},\,t_{2k+2})}\}_{k=0}^{N-1}] \\
 \times \mathrm{id}(\bar\alpha_{2N+1},\alpha_{2N+1})
\label{app:Wreconstruction}
\end{multline}

The functional Born rule reduces to 
\begin{equation}
\mathrm{Pr}[r]  =\int \dd^{2N+2} \boldsymbol{\alpha} \; \dd^{2N+2} \boldsymbol{\bar{\alpha}} 
M_r(\boldsymbol{\alpha},\,\bar{\boldsymbol{\alpha}})
W(\boldsymbol{\alpha},\,\bar{\boldsymbol{\alpha}})
\end{equation}
Upon defining 
\begin{align}
\langle\alpha_{0},\,\dots,\alpha_{2N+1}|\hat{M}_r|\bar\alpha_{0},\,\dots,\bar\alpha_{2N+1}\rangle &
:=M_r(\boldsymbol{\alpha},\,\bar{\boldsymbol{\alpha}})\\
\langle\alpha_{0},\,\dots,\alpha_{2N+1}|\hat{W}|\bar\alpha_{0},\,\dots,\bar\alpha_{2N+1}\rangle & 
:=W(\boldsymbol{\alpha},\,\bar{\boldsymbol{\alpha}})
\end{align}
We recover the discrete-time Born rule
\begin{equation}
    \mathrm{Pr}[r]  = \Tr \hat{M}_r^T \hat{W}.
\end{equation}

\subsection{Recovering discrete-time properties}
Let us show that the functional properties of positivity, causality, normalization (closed and open past), and Markoviniaty introduced in the main text imply the corresponding discrete-time properties under the reconstruction presented above.

\textbf{\textit{Positivity}}.
An operator $\hat{Q}$ is positive semidefinite if $\expval{\hat Q}{\psi} \geq 0$ for all $\ket{\psi}$. In coherent-state representation, this reads
\begin{equation}
    \expval{\hat Q}{\psi} = \int \dd^{2N+2} \! \boldsymbol{\alpha} \, \dd^{2N+2} \! \boldsymbol{\bar{\alpha}} \, \psi(\boldsymbol{\alpha}) \psi^*(\bar{\boldsymbol{\alpha}})Q(\boldsymbol{\alpha},\bar{\boldsymbol{\alpha}})\geq 0.
\end{equation}
Substituting one of the reconstruction formulas, say Eq.~\eqref{app:Mreconstruction} for $\hat{M}_r$, we find
\begin{align}
\expval{\hat Q}{\psi} &=  \int \dd^{2N+2} \! \boldsymbol{\alpha} \, \dd^{2N+2} \! \boldsymbol{\bar{\alpha}} \, \psi(\boldsymbol{\alpha}) \psi^*(\bar{\boldsymbol{\alpha}}) \\
& \left(\prod_{k=0}^{N}\int_{\alpha(t_{2k})=\alpha_{2k},\,\bar{\alpha}(t_{2k})=\bar{\alpha}_{2k}}^{\alpha(t_{2k+1})=\alpha_{2k+1},\,\bar{\alpha}(t_{2k+1})=\bar{\alpha}_{2k+1}}\mathscr{D}[\alpha_{(t_{2k},\,t_{2k+1})},\,\bar{\alpha}_{(t_{2k},\,t_{2k+1})}]\right)\mathcal{E}[\{\alpha_{(t_{2k},\,t_{2k+1})}\}_{k=0}^{N},\,\{\bar{\alpha}_{(t_{2k},\,t_{2k+1})}\}_{k=0}^{N}]\nonumber \\
\times & \mathcal{Q}[\{\alpha_{(t_{2k},\,t_{2k+1})}\}_{k=0}^{N},\,\{\bar{\alpha}_{(t_{2k},\,t_{2k+1})}\}_{k=0}^{N}] \nonumber \\ \label{app:reconstructingpositivity}
=& \int \D[\alpha,\,\bar{\alpha}] \psi(\boldsymbol{\alpha}) \psi^*(\bar{\boldsymbol{\alpha}}) \mathcal{E}[\{\alpha_{(t_{2k},\,t_{2k+1})}\}_{k=0}^{N},\,\{\bar{\alpha}_{(t_{2k},\,t_{2k+1})}\}_{k=0}^{N}]  \mathcal{Q}[\{\alpha_{(t_{2k},\,t_{2k+1})}\}_{k=0}^{N},\,\{\bar{\alpha}_{(t_{2k},\,t_{2k+1})}\}_{k=0}^{N}].
\end{align}
Recalling that the exponential factor $\mathcal{E}$ can be absorbed into $\psi$, $\psi^*$, expression \eqref{app:reconstructingpositivity} is by definition non-negative for a positive semidefinite functional If $\mathcal{Q}$. The same conclusion holds if we use the reconstruction formula for $\hat{W}$, Eq.~\eqref{app:Wreconstruction}.

\textit{\bf{Causality}}. Let us recall again, for convenience, the process-matrix causality conditions we need to recover: A positive semidefinite operator $\hat{W}\in \mathcal{L}(\otimes_{i=0}^{2N+1}\mathcal{H}^i)$ is causally ordered (or satisfies causality) if, for $k=0,\dots,N$, there exists a set of operators $\hat{\widetilde{W}}^{0: 2k} \in \mathcal{L}(\otimes_{i=0}^{2k}\mathcal{H}^i) $ such that
\begin{align} 
\label{app:discreteclosedfuture}
\hat{W} &= \hat{\widetilde{W}}^{0: 2N}\!\!\!\otimes \id^{2N+1}, \\ \label{app:discretecausality}
\Tr_{2k}\hat{\widetilde{W}}^{0: 2k} & = \hat{\widetilde{W}}^{0: 2k-2}\!\!\!\otimes \id^{2k-1},\quad k=1,\dots, N ,  \\ \label{app:discretenormalization}
    \Tr \hat{\widetilde{W}}^{0:0} & = 1.
\end{align}
We immediately see that the reconstruction formula for the closed-boundary process $\hat{W}$, Eq.~\eqref{app:Wreconstruction}, directly implies the form \eqref{app:discreteclosedfuture}, because $\mathrm{id}(\bar\alpha_{2N+1},\alpha_{2N+1})=\matrixel{\bar\alpha_{2N+1}}{\id}{\alpha_{2N+1}}$, where the reconstruction formula for the ``open-future'' process matrix $\hat{\widetilde{W}}^{0: 2N}$ is obtained by removing the factor $\mathrm{id}(\bar\alpha_{2N+1},\alpha_{2N+1})$ from Eq.~\eqref{app:Wreconstruction}:
\begin{align} \nonumber
 \widetilde{W}^{0:2N}&(\alpha_0,\dots,\alpha_{2N},\bar\alpha_0,\dots,\bar\alpha_{2N})
  \\ \nonumber
= & \left(\prod_{k=0}^{N-1}\int_{\alpha(t_{2k+1})=\alpha_{2k+1},\,\bar{\alpha}(t_{2k+1})=\bar{\alpha}_{2k+1}}^{\alpha(t_{2k+2})=\alpha_{2k+2},\,\bar{\alpha}(t_{2k+2})=\bar{\alpha}_{2k+2}}\mathscr{D}[\alpha_{(t_{2k+1},\,t_{2k+2})},\,\bar{\alpha}_{(t_{2k+1},\,t_{2k+2})}]\right) \\ \label{app:Wtildereconstruction}
\times & \mathcal{E}[\{\alpha_{(t_{2k+1},\,t_{2k+2})}\}_{k=0}^{N-1},\,\{\bar{\alpha}_{(t_{2k+1},\,t_{2k+2})}\}_{k=0}^{N-1}] \widetilde{\mathcal{W}}[\alpha,\bar{\alpha}],
\end{align}
where we leave implicit the dependency of $\widetilde{W}$ on interleaved intervals, $\widetilde{\mathcal{W}}[\alpha,\bar{\alpha}]{=}\widetilde{\mathcal{W}}[\{\alpha_{(t_{2k+1},\,t_{2k+2})}\}_{k=0}^{N-1},\,\{\bar{\alpha}_{(t_{2k+1},\,t_{2k+2})}\}_{k=0}^{N-1}]$.

Let us now trace out the last tensor factor from $\hat{\widetilde{W}}^{0: 2N}$:\footnote{Recall, a partial trace in coherent-state representation can be taken either as $\int \dd\alpha_{2N}\expval{\hat{\widetilde{W}}^{0: 2N}}{\alpha_{2N}}$ or, as we do here, as $\int \dd\alpha_{2N}\dd\bar\alpha_{2N} \matrixel{\alpha_{2N}}{\hat{\widetilde{W}}^{0: 2N}}{\bar\alpha_{2N}}
\mathrm{id}(\bar\alpha_{2N},\alpha_{2N})$, see Eq.~\eqref{eq:id-property-int} above.}
\begin{align}
&\int \dd\alpha_{2N}\dd\bar\alpha_{2N} 
 \widetilde{W}^{0:2N}(\alpha_0,\dots,\alpha_{2N},\bar\alpha_0,\dots,\bar\alpha_{2N}) \mathrm{id}(\bar\alpha_{2N},\alpha_{2N})
 \nonumber \\
= &\int \dd\alpha_{2N}\dd\bar\alpha_{2N}  \left(\prod_{k=0}^{N-1}\int_{\alpha(t_{2k+1})=\alpha_{2k+1},\,\bar{\alpha}(t_{2k+1})=\bar{\alpha}_{2k+1}}^{\alpha(t_{2k+2})=\alpha_{2k+2},\,\bar{\alpha}(t_{2k+2})=\bar{\alpha}_{2k+2}}\mathscr{D}[\alpha_{(t_{2k+1},\,t_{2k+2})},\,\bar{\alpha}_{(t_{2k+1},\,t_{2k+2})}]\right) \nonumber \\ \nonumber
& \hspace{1.5cm} \times 
\mathcal{E}[\{\alpha_{(t_{2k+1},\,t_{2k+2})}\}_{k=0}^{N-1},\,\{\bar{\alpha}_{(t_{2k+1},\,t_{2k+2})}\}_{k=0}^{N-1}] \widetilde{\mathcal{W}}[\alpha,\bar{\alpha}] \mathrm{id}(\bar\alpha_{2N}, \alpha_{2N}) \\
= &\left(\prod_{k=0}^{N-2}\int_{\alpha(t_{2k+1})=\alpha_{2k+1},\,\bar{\alpha}(t_{2k+1})=\bar{\alpha}_{2k+1}}^{\alpha(t_{2k+2})=\alpha_{2k+2},\,\bar{\alpha}(t_{2k+2})=\bar{\alpha}_{2k+2}}\D[\alpha_{(t_{2k+1},\,t_{2k+2})},\,\bar{\alpha}_{(t_{2k+1},\,t_{2k+2})}]\right)\mathcal{E}[\{\alpha_{(t_{2k+1},\,t_{2k+2})}\}_{k=0}^{N-2},\,\{\bar{\alpha}_{(t_{2k+1},\,t_{2k+2})}\}_{k=0}^{N-2}]\nonumber \\ \label{app:nexttolastinterval}
\times &\int_{\substack{\alpha(t_{2N-1})=\alpha_{2N-1}\\
\bar{\alpha}(t_{2N-1})=\bar{\alpha}_{2N-1}}}
\D[\alpha_{(t_{2N-1},\,t_{2N})},\,\bar{\alpha}_{(t_{2N-1},\,t_{2N})}] 
\ee^{\ii \int_{t_{2N-1}}^{t_{2N}} \dd t \Im(\dot{\alpha}^{*}\alpha-\dot{\bar{\alpha}}^{*}\bar{\alpha})} 
\widetilde{\mathcal{W}}[\alpha,\bar{\alpha}] \mathrm{id}(\bar\alpha(t_{2N}), \alpha(t_{2N})).
\end{align}
To use the functional causality property of $\mathcal{W}$, we need to extend the path integral to include the last time interval, from $t_{2N}$ to $t_{2N+1}=\tf$, where $\widetilde{\mathcal{W}}$ is constant. To this end, we note the identity
\begin{equation}\label{app:lastintervalcomeon}
\int_{\substack{\alpha(t_{2N})=\alpha_{2N}\\
\bar{\alpha}(t_{2N})=\bar{\alpha}_{2N}}}
\D[\alpha_{(t_{2N},\,t_{2N+1})},\,\bar{\alpha}_{(t_{2N},\,t_{2N+1})}] 
\ee^{\ii \int_{t_{2N}}^{t_{2N+1}} \dd t \Im(\dot{\alpha}^{*}\alpha-\dot{\bar{\alpha}}^{*}\bar{\alpha})} \mathrm{id}(\bar\alpha(t_{2N+1}), \alpha(t_{2N+1})) 
= \mathrm{id}(\bar\alpha_{2N}, \alpha_{2N}),
\end{equation}
which holds because the path integral on the left-hand side is the propagator of the identity evolution (Hamiltonian $\hat{H}=0$), traced at its output, and the equality to the right-hand side expresses the trace preserving condition. Substituting Eq.~\eqref{app:lastintervalcomeon} into the last line of Eq.~\eqref{app:nexttolastinterval}, we find
\begin{multline}
    \int_{\substack{\alpha(t_{2N-1})=\alpha_{2N-1} \\ \bar{\alpha}(t_{2N-1})=\bar{\alpha}_{2N-1}}}
\D[\alpha_{(t_{2N-1},\,t_{2N})},\,\bar{\alpha}_{(t_{2N-1},\,t_{2N})}] 
\ee^{\ii \int_{t_{2N-1}}^{t_{2N}} \dd t \Im(\dot{\alpha}^{*}\alpha-\dot{\bar{\alpha}}^{*}\bar{\alpha})} 
\widetilde{\mathcal{W}}[\alpha,\bar{\alpha}] \mathrm{id}(\bar\alpha(t_{2N}), \alpha(t_{2N})) = \\
\int_{\substack{\alpha(t_{2N-1})=\alpha_{2N-1} \\ \bar{\alpha}(t_{2N-1})=\bar{\alpha}_{2N-1}}}
\D[\alpha_{(t_{2N-1},\,t_{2N+1})},\,\bar{\alpha}_{(t_{2N-1},\,t_{2N+1})}] 
\ee^{\ii \int_{t_{2N-1}}^{t_{2N+1}} \dd t \Im(\dot{\alpha}^{*}\alpha-\dot{\bar{\alpha}}^{*}\bar{\alpha})} 
\widetilde{\mathcal{W}}[\alpha,\bar{\alpha}] \mathrm{id}(\bar\alpha(t_{2N+1}), \alpha(t_{2N+1})) \\
= \widetilde{\mathcal{W}}[\alpha_{(\ti,t_{2N-1})},\bar{\alpha}_{(\ti,t_{2N-1})}] \mathrm{id}(\bar\alpha_{2N-1}, \alpha_{2N-1}),
\end{multline}
where in the last step we have used the functional causality property of $\widetilde{W}$, Eq.~\eqref{app:causalityfunctional}. Substituting the result back into Eq.~\eqref{app:nexttolastinterval}, we finally obtain
\begin{multline}
    \int \dd\alpha_{2N}\dd\bar\alpha_{2N} 
 \widetilde{W}^{0:2N}(\alpha_0,\dots,\alpha_{2N},\bar\alpha_0,\dots,\bar\alpha_{2N}) \mathrm{id}(\bar\alpha_{2N},\alpha_{2N}) \\
= \left(\prod_{k=0}^{N-2}\int_{\alpha(t_{2k+1})=\alpha_{2k+1},\,\bar{\alpha}(t_{2k+1})=\bar{\alpha}_{2k+1}}^{\alpha(t_{2k+2})=\alpha_{2k+2},\,\bar{\alpha}(t_{2k+2})=\bar{\alpha}_{2k+2}}\D[\alpha_{(t_{2k+1},\,t_{2k+2})},\,\bar{\alpha}_{(t_{2k+1},\,t_{2k+2})}]\right) \\
\times \mathcal{E}[\{\alpha_{(t_{2k+1},\,t_{2k+2})}\}_{k=0}^{N-2},\,\{\bar{\alpha}_{(t_{2k+1},\,t_{2k+2})}\}_{k=0}^{N-2}] \widetilde{\mathcal{W}}[\alpha_{(\ti,t_{2N-1})},\bar{\alpha}_{(\ti,t_{2N-1})}] \mathrm{id}(\bar\alpha_{2N-1}, \alpha_{2N-1}) \\
= \widetilde{W}^{0:2N-2}(\alpha_0,\dots,\alpha_{2N-2},\bar\alpha_0,\dots,\bar\alpha_{2N-2}) \mathrm{id}(\bar\alpha_{2N-1}, \alpha_{2N-1}),
\end{multline}
which is the first iterative step of condition \eqref{app:discretecausality}. 
This procedure can be iterated to prove Eq.~\eqref{app:discretecausality} for all $k=N-1,\dots,0$.

When working with non-normalized process functionals, the normalization condition for the discrete-time processes Eq.~\eqref{app:discretenormalization} can be recovered upon redefining $\widetilde{W}^{0:0}(\alpha_0,\bar\alpha_0)\to \widetilde{W}^{0:0}(\alpha_0,\bar\alpha_0)/\int \dd \alpha_0 \dd \bar\alpha_0 \widetilde{W}^{0:0}(\alpha_0,\bar\alpha_0)\mathrm{id}(\alpha_0,\bar\alpha_0)$. On the other hand, if we additionally impose process-functional normalization, Eq.~\eqref{eq:functionalnormalization}, we immediately find
\begin{equation}
    \Tr \hat{\widetilde{W}}^{0:0} = \int \dd \alpha_0 \dd \bar\alpha_0 \widetilde{W}^{0:0}(\alpha_0,\bar\alpha_0)\mathrm{id}(\alpha_0,\bar\alpha_0) = 1.
\end{equation}

\textit{\bf{Markovianity}}. 
Consider first the reconstruction formula for a closed-past process, Eq.~\eqref{app:Wtildereconstruction}, for the particular case of initially uncorrelated state, $\widetilde{W}[\alpha,\bar{\alpha}] = \rho(\alpha(\ti), \bar{\alpha}(\ti))\oW[\alpha,\bar{\alpha}] $. The initial state factors out of the path integral and we find:
\begin{equation}
    \widetilde{W}^{0:2N} (\alpha_0,\dots,\alpha_{2N},\bar\alpha_0,\dots,\bar\alpha_{2N}) = \rho(\alpha_0,\bar\alpha_0)\lowtilde{W}^{1:2N}(\alpha_1,\dots,\alpha_{2N},\bar\alpha_1,\dots,\bar\alpha_{2N}),
\end{equation}
where the open-boundary process matrix is defined through the reconstruction formula
\begin{align} \nonumber
 \lowtilde{W}^{1:2N}&(\alpha_1,\dots,\alpha_{2N},\bar\alpha_1,\dots,\bar\alpha_{2N})
  \\ \nonumber
\coloneqq & \left(\prod_{k=0}^{N-1}\int_{\alpha(t_{2k+1})=\alpha_{2k+1},\,\bar{\alpha}(t_{2k+1})=\bar{\alpha}_{2k+1}}^{\alpha(t_{2k+2})=\alpha_{2k+2},\,\bar{\alpha}(t_{2k+2})=\bar{\alpha}_{2k+2}}\mathscr{D}[\alpha_{(t_{2k+1},\,t_{2k+2})},\,\bar{\alpha}_{(t_{2k+1},\,t_{2k+2})}]\right) \\ \label{app:openreconstruction}
\times & \mathcal{E}[\{\alpha_{(t_{2k+1},\,t_{2k+2})}\}_{k=0}^{N-1},\,\{\bar{\alpha}_{(t_{2k+1},\,t_{2k+2})}\}_{k=0}^{N-1}] \oW[\alpha,\bar{\alpha}],
\end{align}
where, recall, we are assuming that the process is constant over the odd time intervals, Eq.~\eqref{app:Wintermittent}.

Now, by iterating the functional definition of Markovianity [and again using the fact that the process is constant over intervals $(t_{2k},t_{2k+1})$], we can decompose $\oW$ as
\begin{equation}
    \oW[\alpha,\bar{\alpha}] = \prod_{k=0}^{N-1}\oW[\alpha_{(t_{2k+1},\,t_{2k+2})},\,\bar{\alpha}_{(t_{2k+1},\,t_{2k+2})}].
\end{equation}
Plugging this expression into the reconstruction formula \eqref{app:openreconstruction}, we recover the discrete Markovianity condition:
\begin{equation}
    \lowtilde{W}^{1:2N}(\alpha_1,\dots,\alpha_{2N},\bar\alpha_1,\dots,\bar\alpha_{2N}) = \prod_{k=0}^{N-1} \lowtilde{W}^{2k+1:2k+2}(\alpha_{2k+1},\alpha_{2k+2},\bar\alpha_{2k+1},\bar\alpha_{2k+2}),
\end{equation}
where
\begin{align}
    \lowtilde{W}^{2k+1:2k+2}(\alpha_{2k+1},\alpha_{2k+2},\bar\alpha_{2k+1},\bar\alpha_{2k+2}) &
    \coloneqq  \int_{\alpha(t_{2k+1})=\alpha_{2k+1},\,\bar{\alpha}(t_{2k+1})=\bar{\alpha}_{2k+1}}^{\alpha(t_{2k+2})=\alpha_{2k+2},\,\bar{\alpha}(t_{2k+2})=\bar{\alpha}_{2k+2}} \D[\alpha_{(t_{2k+1},\,t_{2k+2})},\,\bar{\alpha}_{(t_{2k+1},\,t_{2k+2})}] \\
    \times & \ee^{\ii\int_{t_{2k+1}}^{t_{2k+2}} \Im(\dot{\alpha}^*\alpha - \dot{\bar{\alpha}}^*\bar{\alpha})} \oW[\alpha_{(t_{2k+1},\,t_{2k+2})},\,\bar{\alpha}_{(t_{2k+1},\,t_{2k+2})}].
\end{align}

\section{Position-space path integrals}
\label{app:Reduction-to-position-PI}

\subsection{Generalized position-space functional Born rule}
The purpose of this subsection is to derive a functional Born rule upon integrating momenta, and to prove a particular case in which this reduces to a position-space formula. To be more precise, we first derive a functional Born rule on the \textit{tangent bundle}, namely on the space of positions and velocities, from which we can recover a position-space formula under given conditions.

First, recall from Sec.~\ref{app:phasespacecoherent} that the functional Born rule can be written in terms of phase space variables as 
\begin{equation}\label{eq:phasespaceBorn}
    \Pr[r] = \int \D[x,\bar{x}, p, \bar{p}] \mathcal{M}_r[x,\bar{x}, p, \bar{p}]\mathcal{W}[x,\bar{x}, p, \bar{p}] \ee^{\ii \int_{\ti}^{\tf}\dd t \left(\dot{x}p - \dot{\bar{x}}\bar{p}\right)} .
\end{equation}
%In the following, we will omit the outcome label $r$ 
Let us now introduce the functional Fourier transform over momentum for a generic phase-space functional $\mathcal{Q}$:
\begin{equation}
\check{\mathcal{Q}}[x,\bar{x}, v, \bar{v}] := \int \D[p,\bar{p}]\mathcal{Q}[x,\bar{x}, p, \bar{p}]\ee^{\ii \int_{\ti}^{\tf}\dd t \left(vp - \bar{v}\bar{p}\right)}.
\end{equation}
The \emph{position-space} functional corresponding to $\mathcal{Q}[x,\bar{x}, p, \bar{p}]$  is given by
\begin{equation}\label{eq:positionspacefunctional}
    \mathcal{Q}[x,\bar{x}]:= \check{\mathcal{Q}}[x,\bar{x}, \dot{x}, \dot{\bar{x}}],
\end{equation}
where, with a small abuse of notation, we use the same symbol $\mathcal{Q}$ for phase-space and position-space functionals.

Up to normalization factors (which, as in most of this article, are omitted in functional expressions), the functional Fourier transform is invertible:
\begin{equation}\label{eq:inverseFourier}
 \mathcal{Q}[x,\bar{x}, p, \bar{p}] = \int \D[v,\bar{v}]\check{\mathcal{Q}}[x,\bar{x}, v, \bar{v}] \ee^{-\ii \int_{\ti}^{\tf}\dd t \left(vp - \bar{v}\bar{p}\right)},
\end{equation}
which can be demonstrated using the formula for the functional Dirac delta
\begin{equation}
 \int \D[v]\ee^{\ii \int_{\ti}^{\tf}\dd t v\left(p - p'\right)} = \delta[p-p'].
\end{equation}
Plugging Eq.~\eqref{eq:inverseFourier} into the phase-space Born rule, Eq.~\eqref{eq:phasespaceBorn}, we obtain
\begin{multline}
    \label{eq:tangentbundleBorn}
    \Pr[r] = \int \D[x,\bar{x}, p, \bar{p},v_1,\bar{v}_1,v_2,\bar{v}_2] \check{\mathcal{M}}_r[x,\bar{x}, v_1, \bar{v}_1]\check{\mathcal{W}}[x,\bar{x}, v_2, \bar{v}_2] \ee^{\ii \int_{\ti}^{\tf}\dd t \left[(\dot{x}-v_1-v_2)p - (\dot{\bar{x}}-\bar{v}_1-\bar{v} _2)\bar{p}\right]} \\
    = \int \D[x,\bar{x}, v,\bar{v}] \check{\mathcal{M}}_r[x,\bar{x}, \dot{x}-v, \dot{\bar{x}} -\bar{v}]\check{\mathcal{W}}[x,\bar{x}, v, \bar{v}] .
\end{multline}

We can interpret Eq.~\eqref{eq:tangentbundleBorn} as the \emph{tangent bundle} form of the functional Born rule, where $v, \bar{v}$ represent velocities---treated here as independent from positions---and $\check{\mathcal{M}}_r$, $\check{\mathcal{W}}$ are the tangent-bundle operation and process functionals, respectively. We see that this expression involves a convolution over the velocities that, in general, cannot be simplified to obtain a simple position-space rule. However, simplifications can occur under given conditions. In particular, consider operations that do not depend on momentum:
\begin{equation}
    \mathcal{M}_r[x,\bar{x}, p, \bar{p}] = \mathcal{M}^0_r[x,\bar{x}].
\end{equation}
Taking the Fourier transform, we obtain the tangent-bundle functional
\begin{equation}\label{eq:tangetM}
   \check{\mathcal{M}}_r[x,\bar{x}, v, \bar{v}] = \mathcal{M}^0_r[x,\bar{x}]\delta[v]\delta[\bar{v}],
\end{equation}
In this case, the tangent-bundle expression, Eq.~\eqref{eq:tangentbundleBorn}, becomes
\begin{multline}\label{eq:positionBorn}
    \Pr[r] 
    = \int \D[x,\bar{x}, v,\bar{v}] \mathcal{M}^0_r[x,\bar{x}]\delta[\dot{x}-v]\delta[\dot{\bar{x}} -\bar{v}]\check{\mathcal{W}}[x,\bar{x}, v, \bar{v}] \\
    = \int \D[x,\bar{x}] {\mathcal{M}}^0_r[x,\bar{x}]\check{\mathcal{W}}[x,\bar{x}, \dot{x}, \dot{\bar{x}}] \\
    = \int \D[x,\bar{x}] \mathcal{M}^0_r[x,\bar{x}]\mathcal{W}[x,\bar{x}],
\end{multline}
with the position-space process functional $\mathcal{W}[x,\bar{x}]$ defined as in Eq.~\eqref{eq:positionspacefunctional}.
Note that $\mathcal{M}^0_r[x,\bar{x}]$ \textit{is not} the position-space operation functional, which, as can be seen from Eq.~\eqref{eq:tangetM}, has an additional delta functional in velocity, $\mathcal{M}_r[x,\bar{x}] = \mathcal{M}^0_r[x,\bar{x}]\delta[\dot x]\delta[\dot{\bar{x}}]$.

\subsection{Constraints on system-meter-environment models}
As opposed to the previous subsection, where we derive conditions for operation functionals under which a position Born rule can be written, here we identify conditions on the underlying Hamiltonian model for system-meter-environment coupling for which a position-space formula holds.

Using Hamilton's equation of motion, we find 
\begin{align}
\dot{x} & =\frac{\partial H^{\tSE}}{\partial p_{x}}+\frac{\partial H^{\tSM}}{\partial p_{x}},\label{eq:px-Legendre}\\
\dot{y} & =\frac{\partial H^{\tSE}}{\partial p_{y}},\label{eq:py-Legendre}\\
\dot{z} & =\frac{\partial H^{\tSM}}{\partial p_{z}}.\label{eq:pz-Legendre}
\end{align}
The total Lagrangian can obtained through the Legendre transform.
Upon inserting Eqs.~(\ref{eq:px-Legendre}-\ref{eq:pz-Legendre}),
the total Lagrangian can be split into two parts
\begin{equation}
L^{\mathrm{TOT}}=L^{\mathrm{SE}}+L^{\mathrm{SM}}
\end{equation}
where 

\begin{align}
L^{\mathrm{SE}} & \equiv p_{x}\frac{\partial H^{\mathrm{SE}}}{\partial p_{x}}+p_{y}\frac{\partial H^{\mathrm{SE}}}{\partial p_{y}}-H^{\tSE}\label{eq:LSE-def}\\
L^{\mathrm{SM}} & \equiv p_{x}\frac{\partial H^{\mathrm{SM}}}{\partial p_{x}}+p_{z}\frac{\partial H^{\tSM}}{\partial p_{z}}-H^{\tSM}\label{eq:LSM-def}
\end{align}

This seems to imply that $L^{\tSE}$ is the system-environment Lagrangian
while $L^{\tSM}$ is the system-meter Lagrangian. If this observation
were true in general, we would have smooth transition to write down
the functional Born rule in the position space. 

However, in general $L^{\tSE}$ can still depend on the meter's degrees
of freedom while $L_{\text{ }}^{\tSM}$ can depend on the environmental
degrees of freedom. To see this, let us consider that we solve for
$p_{x}$ from Eq.~(\ref{eq:px-Legendre}). Clearly, in the most general
case, it does not only involve the system's degree of freedom $\dot{x}$
and $x$, but also on $y$, $p_{y}$ and $z$, $p_{z}$. On the other
hand, upon solving Eq.~(\ref{eq:py-Legendre}) and Eq.~(\ref{eq:pz-Legendre})
for $p_{y}$ and $p_{z}$ respectively, one can find $p_{y}$ depends
on $y$, $\dot{y}$, $x$, $p_{x}$ while $p_{z}$ depends on $z$,
$\dot{z}$, $x$, $p_{x}$. As a consequence, if one views $x$,$\dot{x}$,$y$,$\dot{y}$,$z$,$\dot{z}$
as fundamental variables, then $p_{x}$, $p_{y}$ and $p_{z}$ in
general will depend on all of them. 

To get around of this issue, we consider the following two assumptions: {(a) While $H^{\tSE}$ and $H^{\tSM}$ can have arbitrary dependence
on their respective positions, they are at most quadratic these momenta. (b) 
The momentum of the system can only couple to either the environmental degrees of freedom or the meter's degrees of freedom,
but not both}. The first assumption allows us to represent the joint system-environment-meter
propagator in terms of the path integral in the position space.
The second assumption guarantees that the total Lagrangian contains
no direct interaction between the meter's degrees of freedom and the
environment's degrees of freedom so that $L^{\mathrm{TOT}}$ can be
well split into $L^{\tSE}$ and $L^{\tSM}$.

Without loss of generality, under these two assumptions, we consider
the case where the system's momentum only couples to the environment's
degree of freedom at most quadratically. i.e. $H^{\tSM}$ does not
depends on $p_{x}$. Upon solving Eq.~(\ref{eq:pz-Legendre}), this
immediately leads to the fact that $p_{z}$ only depends on $z$,
$\dot{z}$ and $x$. On the other hand, since $H_{\text{SE}}$ is
at most quadratic in $p_{x}$, solve Eq.~(\ref{eq:px-Legendre})
for $p_{x}$ with $\partial H^{\tSM}/\partial p_{x}=0$ will not involve
$z$ and $\dot{z}$ and thus $p_{x}$ is only a function of $x$,
$\dot{x}$, $y$, $p_{y}$. Further solving Eq.~(\ref{eq:py-Legendre})
for $p_{y}$ does not involve $z$ and $\dot{z}$ either. This leads
to a well-defined $L^{\tSE}$ and $L^{\tSM}$. Similar arguments also
hold when the system's momentum only couples to the meter's degree
of freedom.

Once the two assumptions discussed in the main text is satisfied so
that $L^{\tSE}$ and $L_{\text{ }}^{\tSM}$ are well defined , we can
apply similar analysis as the coherent-state path integral formalism
and obtain
\begin{align}
\mathcal{M}_{r}[x,\,\bar{x}] & \equiv\int\mathscr{D}[z,\bar{z}]E_{r}^{\mathrm{M}}(\bar{z}(t_{\text{f}}),\,z(t_{\text{f}}))\sigma(z(t_{\text{i}}),\,\bar{z}(t_{\text{i}}))\nonumber \\
 & \times\mathcal{U}^{\mathrm{SM}}[x,z]\mathcal{U}^{\mathrm{SM}*}[\bar{x},\,\bar{z}],\label{eq:M-config-def}
\end{align}
$\mathcal{W}[x,\,\bar{x}]=\widetilde{\mathcal{W}}[x,\,\bar{x}]\,\mathrm{id}(\bar{x}(t_{\text{f}}),{x}(t_{\text{f}}))$ with $\mathrm{id}(\bar{x}^{\prime},\,x^{\prime}):=\delta(\bar{x}^{\prime}-x^{\prime})$, and
\begin{align}
\widetilde{\mathcal{W}}[x,\,\bar{x}] & \equiv\int_{y(t_{\text{f}})=\bar{y}(t_{\text{f}})}\mathscr{D}[y,\bar{y}]\mathcal{U}^{\tSE}[x,\,y]\mathcal{U}^{\tSE*}[\bar{x},\,\bar{y}]\nonumber \\
 & \times\rho(x(t_{\text{i}}),y(t_{\text{i}});\bar{x}(t_{\text{i}}),\bar{y}(t_{\text{i}}))\label{eq:W-config-def}
\end{align}
where  $x$, $y$, $z$ denote the degrees of freedom of the system,
environment and meter, respectively, and
\begin{align}
\mathcal{U}^{\tSM}[z,\,x] & \equiv\text{e}^{\text{i}\int_{t_{\text{i}}}^{t_{\text{f}}}L^{\tSM}(z,\,\dot{z};x,\,\dot{x})\mathrm{d}t},\label{eq:calUSM-config-def}\\
\mathcal{U}^{\tSE}[x,y] & \equiv\text{e}^{\text{i}\int_{t_{\text{i}}}^{t_{\text{f}}}L^{\tSE}(y,\dot{y};x,\,\dot{x})\mathrm{d}\tau}.\label{eq:calUSE-config-def}
\end{align}
Note that the position-space process and measurement functional satisfy all previously mentioned properties. For example, the position-space process also has similar connection to the Feynman-Vernon formalism in the position space and Eq.~\eqref{eq:W-FV-Phase}  in the main text generalizes straightforward to position space.
The recovery of discrete-time formalism is also possible from the
continuous-time position space representation. Other properties, such
as positivity, causality and Markovianity also hold similarly. In
particular, for causality, the exponential factor in the coherent-state
representation does not appear here and it reads
\begin{equation}
\int_{x_{>}(t^{\prime})=x^{\prime},\,\bar{x}_{>}(t^{\prime})=\bar{x}^{\prime}}\mathscr{D}[x_{>},\bar{x}_{>}]\mathcal{W}[x,\,\bar{x}]=\widetilde{\mathcal{W}}[x_{<},\,\bar{x}_{<}]\mathrm{id}(\bar{x}^{\prime},\,x^{\prime}).
\end{equation}

\section{Measurement functional for continuous-time Markovian measurements}

When the system is subject to intrinsic and stochastic measurement
dynamics, the stochastic propagator for the system and environment
that depends upon the stream of the measurement outcome $r$ has the
following coherent state path integral representation,
\begin{align}
\langle\alpha_{\mathrm{f}},\,\beta_{\mathrm{f}}|\hat{U}_{r}(t_{{\rm f}}\!\!\shortleftarrow\!\!t_{{\rm i}})|\alpha_{\mathrm{i}},\,\beta_{\mathrm{i}}\rangle & =\langle\alpha_{\mathrm{f}},\,\beta_{\mathrm{f}}|\prod_{k=0}^{N-1}\mathrm{e}^{-\text{i}\hat{H}^{\mathrm{SE}}\Delta t_{k}}\mathrm{e}^{-\text{i}\hat{H}_{r}\Delta t_{k}}|\alpha_{\mathrm{i}},\,\beta_{\mathrm{i}}\rangle\nonumber \\
 & =\langle\alpha_{\mathrm{f}},\,\beta_{\mathrm{f}}|\mathcal{T}\mathrm{e}^{-\text{i}\int_{t_{\text{i}}}^{t_{\text{f}}}\mathrm{d}t(\hat{H}^{\mathrm{SE}}+\hat{H}_{r})}|\alpha_{\mathrm{i}},\,\beta_{\mathrm{i}}\rangle\nonumber \\
 & =\int_{\alpha(t_{\text{i}})=\alpha_{\text{i}},\,\beta(t_{\text{i}})=\beta_{\text{i}}}^{\alpha(t_{\text{f}})=\alpha_{\text{f}},\,\beta(t_{\text{f}})=\beta_{\text{f}}}\mathscr{D}[\alpha,\,\beta] \mathrm{e}^{\text{i}\int_{t_{\text{i}}}^{t_{\text{f}}}{ \mathrm{Im}(\dot{\alpha}^{*}\alpha-\dot{\bar{\alpha}}^{*}\bar{\alpha}+\dot{\beta}^{*}\beta-\dot{\bar{\beta}}^{*}\bar{\beta})} \mathrm{d}t} \mathcal{U}_{r}^{\mathrm{SE}}[\alpha,\beta]\label{eq:Stochastic-Propagator}
\end{align}
where $\Delta t_{k}\equiv t_{k+1}-t_{k}$, $t_{0}=t_{\mathrm{i}}$
and $t_{N}=t_{\mathrm{f}}$, $\mathcal{U}_{r}^{\mathrm{SE}}[\alpha,\,\beta]:=\mathcal{K}_{r}[\alpha]\mathcal{U}^{\mathrm{SE}}[\alpha,\beta]$,
\begin{equation}
\mathcal{K}_{r}[\alpha]:=\mathrm{e}^{-\text{i}\int_{t_{\text{i}}}^{t_{\text{f}}}\mathrm{d}tH_{r}(\alpha)}\label{eq:cal-Va}
\end{equation}
is the Kraus functional. It is worth to note that the functional probability
\begin{equation}
\mathrm{Pr}[r]=\text{Tr}_{\text{SE}}[\hat{U}_{r}(t_{{\rm f}}\!\!\shortleftarrow\!\!t_{{\rm i}})\rho_{\mathrm{i}}\hat{U}_{r}(t_{{\rm i}}\!\!\shortleftarrow\!\!t_{\mathrm{f}})]
\end{equation}
is no longer normalized, which can be made to be normalized upon dividing
the factor $\int\mathscr{D}[r]\mathrm{Pr}[r]$. As before,
upon inserting the resolution of identity, we find
\begin{align}
\mathrm{Pr}[r] & =\int\mathrm{d}^{2}\alpha_{\text{f}}\mathrm{d}^{2}\beta_{\text{f}}\mathrm{d}^{2}\alpha_{\text{i}}\mathrm{d}\bar{\alpha}_{\text{i}}\mathrm{d}^{2}\beta_{\text{i}}\mathrm{d}^{2}\bar{\beta}_{\text{i}}\langle{\alpha_{\text{f}},\beta{}_{\text{f}}|\hat{U}_{r}(t_{{\rm f}}\!\!\shortleftarrow\!\!t_{{\rm i}})|\alpha_{\text{i}},\beta_{\text{i}}}\rangle\rho_{\text{i}}(\alpha_{\text{i}},\alpha_{\text{i}};\bar{\beta}_{\text{i}},\bar{\beta}_{\text{i}})\langle{\bar{\alpha}_{\text{i}},\bar{\beta}_{\text{i}}|\hat{U}_{r}(t_{\text{i}}\!\!\shortleftarrow\!\!t_{\text{f}})|\alpha_{\text{f}},\beta_{\text{f}}}\rangle
\end{align}
From which, we observe 
\begin{align}
\mathrm{Pr}[r] & 
=\int_{\alpha(t_{\text{f}})=\bar{\alpha}(t_{\text{f}})}\mathscr{D}[\alpha,\,\bar{\alpha}] \mathrm{e}^{\text{i}\int_{t_{\text{i}}}^{t_{\text{f}}}{\mathrm{Im}(\dot{\alpha}^{*}\alpha-\dot{\bar{\alpha}}^{*}\bar{\alpha})} \mathrm{d}t} \mathcal{M}_{r}[\alpha,\,\bar{\alpha}]\mathcal{\widetilde{W}}[\alpha,\,\bar{\alpha}] \nonumber \\
& =\int\mathscr{D}[\alpha,\,\bar{\alpha}]\mathrm{e}^{\text{i}\int_{t_{\text{i}}}^{t_{\text{f}}}{\mathrm{Im}(\dot{\alpha}^{*}\alpha-\dot{\bar{\alpha}}^{*}\bar{\alpha})} \mathrm{d}t} \mathcal{M}_{r}[\alpha,\,\bar{\alpha}]\mathcal{W}[\alpha,\,\bar{\alpha}],\label{eq:Pra-unnormalized}
\end{align}
where $\mathcal{\widetilde{W}}$ and $\mathcal{W}$ are defined as
in Eq.~\eqref{eq:Wfunctional0} and Eq.~\eqref{eq:Wfunctional} in the main text, the measurement functional is 
\begin{equation}
\mathcal{M}_{r}[\alpha,\,\bar{\alpha}]=\mathcal{K}_{r}[\alpha]\mathcal{K}_{r}^{*}[\bar{\alpha}]
\end{equation}
Furthermore, if $\hat{H}_{r}$ only depends on the system's momentum
and $H^{\mathrm{SE}}$ is at most quadratic in their respective momenta,then
following similar arguments above, we can obtain the measurement functional
in the position space. 

\section{The Generailized Caldeira-Legett Processes }
\label{app:GCL-Processes}
The generic CL process reads
\begin{equation}
\mathcal{\widetilde{W}}^{\mathrm{CL}}[x,\bar{x}]=\mathrm{e}^{\mathrm{i}S^{\mathrm{eff}}[x,\,\bar{x}]}\rho(x_{i},\,\bar{x}_{i})
\end{equation}
where $S^{\mathrm{eff}}[x,\,\bar{x}]=S_{0}[x,\,\bar{x}]+S^{\mathrm{FV}}[x,\,\bar{x}]$
is the effective action due to the coupling with an environment. Here
both $S_{0}$ and $S^{\mathrm{FV}}$ bear the generic Gaussian structure:
\begin{align}
S_{0}[\bm{x}] & =\frac{m}{2}\int_{t_{\text{i}}}^{t_{\text{f}}}\dd t\left[\dot{\bm{x}}^{T}(t)\sigma_{z}\dot{\bm{x}}(t)-\omega_{0}^{2}\bm{x}^{T}(t)\sigma_{z}\bm{x}(t)\right]\\
S^{\mathrm{FV}}[\bm{x}] & =-\frac{1}{2}\int_{t_{\text{i}}}^{t_{\text{f}}}\dd t\int_{t_{\text{i}}}^{t_{\mathrm{f}}}\dd s\bm{x}^{T}(t)\tilde{\bm{A}}(t,s)\bm{x}(s)
\end{align}
where $\tilde{\bm{A}}(t,s)=[\bm{A}(t,s)+\bm{A}^{T}(s,t)]/2$ is invariant
under the joint operation of matrix transpose and the swapping between
$t$ and $s$.

The measurement functional is 
\begin{equation}
\mathcal{M}_{r}^{0}[x,\,\bar{x}]=\mathrm{e}^{-\frac{1}{4\tau_{\mathrm{m}}}\int_{t_{\text{i}}}^{t_{\text{f}}}\dd t\left([r(t)-x(t)]^{2}+[r(t)-\bar{x}(t)]^{2}\right)}
\end{equation}
Therefore the functional probability distribution of the time-continuous
readout is
\begin{align}
\mathrm{Pr}[r] & =\int\mathscr{D}[x,\bar{x}]\mathcal{\widetilde{W}}^{\mathrm{CL}}[x,\bar{x}]\mathrm{id}(\bar{x}(t_{\mathrm{f}}),\,x(t_{\mathrm{f}}))\mathcal{M}_{r}[x,\,\bar{x}]\nonumber \\
 & =\mathrm{e}^{-\frac{1}{2\tau_{\mathrm{m}}}\int_{t_{\text{i}}}^{t_{\text{f}}}\dd tr^{2}(t)}\int\dd x_{\mathrm{i}}\dd\bar{x}_{\mathrm{i}}\dd x_{\mathrm{f}}\rho(x_{\mathrm{i}},\,\bar{x}_{\mathrm{i}})\int_{x(t_{\mathrm{i}})=x_{\mathrm{i}},\,\bar{x}(t_{\mathrm{i}})=\bar{x}_{\mathrm{i}}}^{x(t_{\text{f}})=x_{\mathrm{f}},\bar{x}(t_{\text{f}})=x_{\mathrm{f}}}\mathscr{D}[x,\bar{x}]\mathrm{e}^{\mathrm{i}\left\{ S^{\mathrm{eff}}[x,\bar{x}]+\frac{\mathrm{i}}{4\tau_{\mathrm{m}}}\int_{t_{\text{i}}}^{t_{\text{f}}}\dd t\left(x^{2}(t)+\bar{x}^{2}(t)-2r(t)[x(t)+\bar{x}(t)]\right)\right\} }
\end{align}
Physically, the Markovian Gaussian measurement amounts to shifting
the frequency of the system by a complex frequency as well as driving
it with a complex-valued force. We define
\begin{align}
Z_{r,\,\bar{r}}[x_{\mathrm{i}},\bar{x}_{\mathrm{i}};\,x_{\mathrm{f}},\,\bar{x}_{\mathrm{f}}] & \equiv\int_{x(t_{\mathrm{i}})=x_{\mathrm{i}},\,\bar{x}(t_{\mathrm{i}})=\bar{x}_{\mathrm{i}}}^{x(t_{\text{f}})=x_{\mathrm{f}},\bar{x}(t_{\text{f}})=\bar{x}_{\mathrm{f}}}\mathscr{D}[x,\bar{x}]\mathrm{e}^{\mathrm{i}S_{r,\,\bar{r}}[x,\bar{x}]}
\end{align}
where 
\begin{align}
S_{r,\,\bar{r}}[x,\bar{x}] & \equiv S^{\mathrm{eff}}[x,\bar{x}]+\frac{\mathrm{i}}{4\tau_{\mathrm{m}}}\int_{t_{\text{i}}}^{t_{\text{f}}}\dd t\left(x^{2}(t)+\bar{x}^{2}(t)\right)-\frac{\mathrm{i}}{2\tau_{\mathrm{m}}}\int_{t_{\text{i}}}^{t_{\text{f}}}\dd t[r(t)x(t)+\bar{r}(t)\bar{x}(t)]
\end{align}
Thanks to Gaussianity, $Z_{r,\,\bar{r}}[x_{\mathrm{i}},\bar{x}_{\mathrm{i}};\,x_{\mathrm{f}},\,\bar{x}_{\mathrm{f}}]$
can be evaluated using the saddle point approximation. We define 
\begin{equation}
\bm{r}(t)\equiv\begin{pmatrix}r(t)\\
\bar{r}(t)
\end{pmatrix},\,\bm{x}_{\mathrm{i}}\equiv\begin{pmatrix}x_{\mathrm{i}}\\
\bar{x}_{\mathrm{i}}
\end{pmatrix},\,\bm{x}_{\mathrm{f}}\equiv\begin{pmatrix}x_{\mathrm{f}}\\
\bar{x}_{\mathrm{f}}
\end{pmatrix}
\end{equation}
Then $S_{\bm{r}}[\bm{x}]\equiv S_{r,\,\bar{r}}[x,\bar{x}]$ can be
written in a more compact form:
\begin{multline}
S_{\bm{r}}[\bm{x}] =\frac{m}{2}\int_{t_{\text{i}}}^{t_{\text{f}}}\dd t\left[\dot{\bm{x}}^{T}(t)\sigma_{z}\dot{\bm{x}}(t)-\bm{x}^{T}(t)\left(\omega_{0}^{2}\sigma_{z}-\frac{\mathrm{i}}{2m\tau_{\mathrm{m}}}\right)\bm{x}(t)-\frac{1}{m}\int_{t_{\text{i}}}^{t_{\text{f}}}\dd t\int_{t_{\text{i}}}^{t_{\mathrm{f}}}\dd s\bm{x}^{T}(t)\bm{A}(t,s)\bm{x}(s) \right. \\
\left.
- \frac{\mathrm{i}}{m\tau_{\mathrm{m}}}\int_{t_{\text{i}}}^{t_{\text{f}}}\dd t\bm{x}^{T}(t)\bm{r}(t)\right]
\end{multline}
The equation of motion for the classical trajectory can be read off
immediately from above equation:
\begin{equation}
\sigma_{z}\ddot{\bm{x}}^{\mathrm{cl}}(t)+\left(\omega_{0}^{2}\sigma_{z}-\frac{\mathrm{i}}{2m\tau_{\mathrm{m}}}\right)\bm{x}^{\mathrm{cl}}(t)+\frac{1}{m}\int_{t_{\text{i}}}^{t_{\mathrm{f}}}\dd s\tilde{\bm{A}}(t,s)\bm{x}^{\mathrm{cl}}(s)=-\frac{\mathrm{i}\bm{r}(t)}{2m\tau_{\mathrm{m}}}\label{eq:cl-traj}
\end{equation}
with the boundary condition 
\begin{equation}
\bm{x}^{\mathrm{cl}}(t_{\mathrm{i}})=\bm{x}_{\mathrm{i}},\,\bm{x}^{\mathrm{cl}}(t_{\mathrm{f}})=\bm{x}_{\mathrm{f}}\label{eq:BC}
\end{equation}
We define fours fundamental solutions $\bm{D}_{\mathrm{i}}(t)$, $\bar{\bm{D}}_{\mathrm{i}}(t)$,
$\bm{D}_{\mathrm{f}}(t)$ and $\bar{\bm{D}}_{\mathrm{f}}(t)$ that
satisfies the homogenous version of Eq.~(\ref{eq:cl-traj}) with
the boundary conditions 
\begin{equation}
\bm{D}_{\mathrm{i}}(t_{\mathrm{i}})=\begin{pmatrix}-1\\
0
\end{pmatrix},\,\bm{D}_{\mathrm{i}}(t_{\mathrm{f}})=0
\end{equation}
\begin{equation}
\bar{\bm{D}}_{\mathrm{i}}(t_{\mathrm{i}})=\begin{pmatrix}0\\
-1
\end{pmatrix},\,\bar{\bm{D}}_{\mathrm{i}}(t_{\mathrm{f}})=0
\end{equation}
\begin{equation}
\bm{D}_{\mathrm{f}}(t_{\mathrm{i}})=0,\,\bm{D}_{\mathrm{f}}(t_{\mathrm{f}})=\begin{pmatrix}1\\
0
\end{pmatrix}
\end{equation}
\begin{equation}
\bar{\bm{D}}_{\mathrm{f}}(t_{\mathrm{i}})=0,\,\bar{\bm{D}}_{\mathrm{f}}(t_{\mathrm{f}})=\begin{pmatrix}0\\
1
\end{pmatrix}
\end{equation}
respectively. Furthermore, we define the Green's function to Eq.~(\ref{eq:cl-traj})
as $G_{kl}(t-s)$, which satisfies 
\begin{equation}
\sigma_{z}\partial_{t}^{2}\bm{G}(t,\,s)+\left(\omega_{0}^{2}\sigma_{z}-\frac{\mathrm{i}}{2m\tau_{\mathrm{m}}}\right)\bm{G}(t,\,s)+\frac{1}{m}\int_{t_{\text{i}}}^{t_{\mathrm{f}}}\dd\tau\tilde{\bm{A}}(t,\tau)\bm{G}(\tau,\,s)=-\frac{\mathrm{i}}{2m\tau_{\mathrm{m}}}\delta(t-s)\mathbb{I}\label{eq:EOM-classical}
\end{equation}
and the homogeneous boundary conditions. Therefore, the structure
of the solution to the Eq.~(\ref{eq:cl-traj}) can be represented
as follows:
\begin{align}
\bm{x}^{\mathrm{cl}}(t) & =x_{\mathrm{f}}\bm{D}_{\mathrm{f}}(t)+\bar{x}_{\mathrm{f}}\bar{\bm{D}}_{\mathrm{f}}(t)-x_{\mathrm{i}}\bm{D}_{\mathrm{i}}(t)-\bar{x}_{\mathrm{i}}\bar{\bm{D}}_{\mathrm{i}}(t)+\int_{t_{\text{i}}}^{t_{\mathrm{f}}}\bm{G}(t,s)\bm{r}(s)\nonumber \\
 & =\begin{bmatrix}\bm{D}_{\mathrm{f}}(t) & \bar{\bm{D}}_{\mathrm{f}}(t)\end{bmatrix}\bm{x}_{\mathrm{f}}-\begin{bmatrix}\bm{D}_{\mathrm{i}}(t) & \bar{\bm{D}}_{\mathrm{i}}(t)\end{bmatrix}\bm{x}_{\mathrm{i}}+\int_{t_{\text{i}}}^{t_{\mathrm{f}}}\bm{G}(t,s)\bm{r}(s)
\end{align}

Now we are in a position to evaluate the classical action. Integrating
by parts, it can be readily obtained
\begin{equation}
\int_{t_{\text{i}}}^{t_{\text{f}}}\dd t\dot{\bm{x}}^{T}(t)\sigma_{z}\dot{\bm{x}}(t)=\bm{x}_{\mathrm{f}}^{T}\sigma_{z}\dot{\bm{x}}(t_{\mathrm{f}})-\bm{x}_{\mathrm{i}}^{T}\sigma_{z}\dot{\bm{x}}(t_{\mathrm{i}})-\int_{t_{\text{i}}}^{t_{\text{f}}}\dd t\bm{x}^{T}(t)\sigma_{z}\left(\ddot{\bm{x}}(t)+\omega_{0}^{2}\bm{x}\right)
\end{equation}
from which, we find 
\begin{align}
S_{\bm{r}}[\bm{x}] & =\frac{m}{2}\left[\bm{x}_{\mathrm{f}}^{T}\sigma_{z}\dot{\bm{x}}_{\mathrm{}}(t_{\mathrm{f}})-\bm{x}_{\mathrm{i}}^{T}\sigma_{z}\dot{\bm{x}}(t_{\mathrm{i}})\right]\nonumber \\
 & -\frac{m}{2}\left[\int_{t_{\text{i}}}^{t_{\text{f}}}\dd t\bm{x}^{T}(t)\left(\sigma_{z}\ddot{\bm{x}}(t)+(\omega_{0}^{2}\sigma_{z}-\frac{\mathrm{i}}{2m\tau_{\mathrm{m}}})\bm{x}+\frac{1}{m}\int_{t_{\text{i}}}^{t_{\mathrm{f}}}\dd s\tilde{\bm{A}}(t,s)\bm{x}(s)+\frac{\mathrm{i}}{2m\tau_{\mathrm{m}}}\bm{r}(t)\right)\right]\nonumber \\
 & -\frac{\mathrm{i}}{4\tau_{\mathrm{m}}}\int_{t_{\text{i}}}^{t_{\text{f}}}\dd t\bm{x}^{T}(t)\bm{r}(t)
\end{align}
Upon replacing $\bm{x}(t)$ with the classical trajectory $\bm{x}^{\mathrm{cl}}(t)$,
the classical action can be represented in the following compact form
\begin{align}
S_{\bm{r}}^{\mathrm{cl}}(\bm{x}_{\mathrm{i}},\bm{x}_{\mathrm{f}}) & =\bm{x}_{\mathrm{f}}^{T}\bm{\Sigma}_{\mathrm{f}}\bm{x}_{\mathrm{f}}+\bm{x}_{\mathrm{i}}^{T}\bm{\Sigma}_{\mathrm{i}}\bm{x}_{\mathrm{i}}-\bm{x}_{\mathrm{i}}^{T}\bm{\Sigma}_{\mathrm{if}}\bm{x}_{\mathrm{f}}\nonumber \\
+ & \bm{x}_{\mathrm{f}}^{T}\int_{t_{\text{i}}}^{t_{\mathrm{f}}}\dd s\bm{F}_{\mathrm{f}}(s)\bm{r}(s)-\bm{x}_{\mathrm{i}}^{T}\int_{t_{\text{i}}}^{t_{\mathrm{f}}}\dd s\bm{F}_{\mathrm{i}}(s)\bm{r}(s)-\frac{\mathrm{i}}{4\tau_{\mathrm{m}}}\int_{t_{\text{i}}}^{t_{\text{f}}}\dd t\dd s\bm{r}^{T}(t)\tilde{\bm{G}}(t,s)\bm{r}(s)
\end{align}
where
\begin{align}
\bm{\Lambda}_{\mathrm{f}} & =\frac{m}{2}\sigma_{z}\begin{bmatrix}\dot{\bm{D}}_{\mathrm{f}}(t_{\mathrm{f}}) & \dot{\bar{\bm{D}}}_{\mathrm{f}}(t_{\mathrm{f}})\end{bmatrix}\\
\bm{\Lambda}_{\mathrm{i}} & =\frac{m}{2}\sigma_{z}\begin{bmatrix}\dot{\bm{D}}_{\mathrm{i}}(t_{\mathrm{i}}) & \dot{\bar{\bm{D}}}_{\mathrm{i}}(t_{\mathrm{i}})\end{bmatrix}\\
\bm{\Lambda}_{\mathrm{if}} & =\frac{m}{2}\left(\sigma_{z}\begin{bmatrix}\dot{\bm{D}}_{\mathrm{f}}(t_{\mathrm{i}}) & \dot{\bar{\bm{D}}}_{\mathrm{f}}(t_{\mathrm{i}})\end{bmatrix}+\begin{bmatrix}\dot{\bm{D}}_{\mathrm{i}}(t_{\mathrm{f}}) & \dot{\bar{\bm{D}}}_{\mathrm{i}}(t_{\mathrm{f}})\end{bmatrix}^{T}\sigma_{z}\right)\\
\bm{F}_{\mathrm{f}}(s) & =\frac{m}{2}\sigma_{z}\dot{\bm{G}}(t_{\mathrm{f}}-s)-\frac{\mathrm{i}}{4\tau_{\mathrm{m}}}\begin{bmatrix}\bm{D}_{\mathrm{f}}(s) & \bar{\bm{D}}_{\mathrm{f}}(s)\end{bmatrix}^{T}\\
\bm{F}_{\mathrm{i}}(s) & =\frac{m}{2}\sigma_{z}\dot{\bm{G}}(t_{\mathrm{i}}-s)-\frac{\mathrm{i}}{4\tau_{\mathrm{m}}}\begin{bmatrix}\bm{D}_{\mathrm{i}}(s) & \bar{\bm{D}}_{\mathrm{i}}(s)\end{bmatrix}^{T}
\end{align}
Therefore, we conclude that 
\begin{equation}
Z_{r,\,\bar{r}}(x_{\mathrm{i}},\bar{x}_{\mathrm{i}};\,x_{\mathrm{f}},\,\bar{x}_{\mathrm{f}})=\mathcal{N}(t_{\mathrm{i}},\,t_{\mathrm{f}})\mathrm{e}^{\mathrm{i}S_{r,\,\bar{r}}^{\mathrm{cl}}(x_{\mathrm{i}},\bar{x}_{\mathrm{i}};\,x_{\mathrm{f}},\,\bar{x}_{\mathrm{f}})}
\end{equation}
Note that the normalization constant of the pat can obtained by integrating
over the fluctuation over the classical trajectory. Since the fluctuations
satisfy the homogeneous boundary conditions, the normalization constant
must be independent of $x_{\mathrm{i}},\bar{x}_{\mathrm{i}};\,x_{\mathrm{f}},\,\bar{x}_{\mathrm{f}}$.
The probability distribution of the readouts is
\begin{equation}
\mathrm{Pr}[r]=\mathrm{e}^{-\frac{1}{2\tau_{\mathrm{m}}}\int_{t_{\text{i}}}^{t_{\text{f}}}\dd tr^{2}(t)}\int\dd x_{\mathrm{i}}\dd\bar{x}_{\mathrm{i}}\dd x_{\mathrm{f}}\rho(x_{\mathrm{i}},\,\bar{x}_{\mathrm{i}})Z_{r,\,r}(x_{\mathrm{i}},\bar{x}_{\mathrm{i}};\,x_{\mathrm{f}},\,x_{\mathrm{f}})\label{eq:Pr-Z}
\end{equation}
We further assume the initial state is a Gaussian state
\begin{equation}
\rho(x_{\mathrm{i}},\,\bar{x}_{\mathrm{i}})\propto\mathrm{e}^{-\frac{1}{2}\bm{x}_{\mathrm{i}}^{T}\bm{\Xi}\bm{x}_{i}+\bm{x}_{\mathrm{i}}\bm{c}^{T}},
\end{equation}
where $\bm{c}\equiv(c_{\mathrm{i}},\,\mathrm{\bar{c}}_{\mathrm{i}})$
and evaluate the probability exactly. To this end, we set $x_{\mathrm{f}}=\bar{x}_{\mathrm{f}}$
and $r(t)=\bar{r}(t)$, we obtain
\begin{align}
S_{r,\,r}^{\mathrm{cl}}(x_{\mathrm{i}},\bar{x}_{\mathrm{i}};\,x_{\mathrm{f}},\,x_{\mathrm{f}}) & =-\frac{1}{2}\left(\begin{array}{ccc}
x_{\mathrm{i}} & \bar{x}_{\mathrm{i}} & x_{\mathrm{f}}\end{array}\right)\bm{\Omega}\begin{pmatrix}x_{\mathrm{i}}\\
\bar{x}_{\mathrm{i}}\\
x_{\mathrm{f}}
\end{pmatrix}+\left(\begin{array}{ccc}
x_{\mathrm{i}} & \bar{x}_{\mathrm{i}} & x_{\mathrm{f}}\end{array}\right)\begin{pmatrix}\sum_{k}\int_{t_{\text{i}}}^{t_{\mathrm{f}}}\dd sF_{\mathrm{i},\,1k}(s)r(s)\\
\sum_{k}\int_{t_{\text{i}}}^{t_{\mathrm{f}}}\dd sF_{\mathrm{i},\,2k}(s)r(s)\\
\sum_{kl}\int_{t_{\text{i}}}^{t_{\mathrm{f}}}\dd sF_{\mathrm{f},\,kl}(s)r(s)
\end{pmatrix}\\
 & -\frac{\mathrm{i}}{4\tau_{\mathrm{m}}}\sum_{ij}\int_{t_{\text{i}}}^{t_{\text{f}}}\dd t\dd sr(t)\tilde{G}_{ij}(t,\,s)r(s)
\end{align}
where 
\begin{equation}
\bm{\Omega}=\begin{pmatrix}-2\Lambda_{\mathrm{i},\,11} & -2\Lambda_{\mathrm{i},\,12} & \sum_{k}\Lambda_{\mathrm{if},\,1k}\\
-2\Lambda_{\mathrm{i},\,21} & -2\Lambda_{\mathrm{i},\,22} & \sum_{k}\Lambda_{\mathrm{if},\,2k}\\
\sum_{k}\Lambda_{\mathrm{if},\,1k} & \sum_{k}\Lambda_{\mathrm{if},\,2k} & -2\sum_{kl}\Lambda_{\mathrm{f},kl}
\end{pmatrix}
\end{equation}
Upon integration we find
\begin{align}
\mathrm{Pr}[r] & \propto\exp\left[-\frac{1}{2\tau_{\mathrm{m}}}\int_{t_{\text{i}}}^{t_{\text{f}}}\dd tr^{2}(t)+\frac{\mathrm{1}}{4\tau_{\mathrm{m}}}\sum_{ij}\int_{t_{\text{i}}}^{t_{\text{f}}}\dd t\dd sr(t)\tilde{G}_{ij}(t,\,s)r(s)\right.\nonumber \\
 & \left.+\frac{1}{2}\begin{pmatrix}\sum_{k}\int_{t_{\text{i}}}^{t_{\mathrm{f}}}\dd sF_{\mathrm{i},\,1k}(s)r(s)+c_{\mathrm{i}}\\
\sum_{k}\int_{t_{\text{i}}}^{t_{\mathrm{f}}}\dd sF_{\mathrm{i},\,2k}(s)r(s)+\bar{c}_{\mathrm{i}}\\
\sum_{kl}\int_{t_{\text{i}}}^{t_{\mathrm{f}}}\dd sF_{\mathrm{f},\,kl}(s)r(s)
\end{pmatrix}^{T}{\bm{\check{\Omega}}^{-1}}\begin{pmatrix}\sum_{k}\int_{t_{\text{i}}}^{t_{\mathrm{f}}}\dd sF_{\mathrm{i},\,1k}(s)r(s)+c_{\mathrm{i}}\\
\sum_{k}\int_{t_{\text{i}}}^{t_{\mathrm{f}}}\dd sF_{\mathrm{i},\,2k}(s)r(s)+\bar{c}_{\mathrm{i}}\\
\sum_{kl}\int_{t_{\text{i}}}^{t_{\mathrm{f}}}\dd sF_{\mathrm{f},\,kl}(s)r(s)
\end{pmatrix}\right]\label{eq:Pr-r-formal}
\end{align}
where the normalization constant is independent of the readout $r(t)$
and
\begin{equation}
{\bm{\check{\Omega}}}\equiv\bm{\Omega}+\begin{pmatrix}\bm{\Xi} & 0\\
0 & 0
\end{pmatrix}
\end{equation}
It can be read off from Eq.~\eqref{eq:Pr-r-formal}
\begin{equation}
\mathrm{Pr}[r]\propto\mathrm{e}^{-\frac{1}{2\tau_{\mathrm{m}}}\int_{t_{\text{i}}}^{t_{\text{f}}}\dd t\int_{t_{\text{i}}}^{t_{\text{f}}}\dd sr(t)R(t,\,s)r(s)+\int_{t_{\text{i}}}^{t_{\text{f}}}b(s)r(s)}
\end{equation}
where
\begin{equation}
R(t,\,s)=\delta(t-s)-\frac{1}{4}\sum_{ij}\tilde{G}_{ij}(t,\,s)+g(t)h(s)
\end{equation}
and the functions $g(t)$, $h(t)$ and $b(t)$ are determined by the
system's intrinsic Gaussian function as well as the initial state
$\rho$. Experimentally, one can measure the correlation of the readouts.
$r(t)$: $\langle r(t)r(s)\rangle-\langle r(t)\rangle\langle r(s)\rangle=\tau_{m}/R(t,s)$.

In the singular projective limit $\tau_{m}\to0$, it is clear from
that Eq.~(\ref{eq:cl-traj}) that 
\begin{equation}
\bm{x}^{\mathrm{cl}}(t)=\bm{r}(t)
\end{equation}
where $r(t)$ satisfies the boundary condition~(\ref{eq:BC}). It
is also clear from Eq.~(\ref{eq:cl-traj}) that $\bm{G}(t,s)\to\delta(t-s)\id$.
In this case, one can calculate Eq.~(\ref{eq:Pr-Z})
\begin{equation}
\mathrm{Pr}[r]\propto\int\dd x_{\mathrm{i}}\dd\bar{x}_{\mathrm{i}}\dd x_{\mathrm{f}}\rho(x_{\mathrm{i}},\,\bar{x}_{\mathrm{i}})\delta\left(r(t_{\mathrm{i}})-x_{\mathrm{i}}\right)\delta\left(r(t_{\mathrm{i}})-\bar{x}_{\mathrm{i}}\right)\delta\left(r(t_{\mathrm{f}})-x_{\mathrm{f}}\right)\mathrm{e}^{\mathrm{i}S_{r,\,r}^{\mathrm{eff}}(x_{\mathrm{i}},\bar{x}_{\mathrm{i}};\,x_{\mathrm{f}},\,x_{\mathrm{f}})}
\end{equation}
where the delta functions account for the fact that readouts must
satisfy the boundary condition~(\ref{eq:BC}) and 
\begin{equation}
S_{r,\,r}^{\mathrm{eff}}(x_{\mathrm{i}},\bar{x}_{\mathrm{i}};\,x_{\mathrm{f}},\,x_{\mathrm{f}})
=-\frac{1}{2}\int_{t_{\text{i}}}^{t_{\text{f}}}\dd t\int_{t_{\text{i}}}^{t_{\mathrm{f}}}\dd s\;r(t)\sum_{ij}\tilde{A}_{ij}(t,s)r(s)
\end{equation}
Thus, for generic initial states, we obtain 
\begin{equation}
\mathrm{Pr}[r]\propto\rho\left(r(t_{\mathrm{i}}),\,r(t_{\mathrm{i}})\right)\mathrm{e}^{-\frac{\mathrm{i}}{2}\int_{t_{\text{i}}}^{t_{\text{f}}}\dd t\int_{t_{\text{i}}}^{t_{\mathrm{f}}}\dd s\, r(t)\sum_{ij}\tilde{A}_{ij}(t,s)r(s)}.
\end{equation}

\end{document}